\documentclass[twoside,11pt]{article}
\usepackage{color}
\usepackage{listings}
\usepackage{xcolor}
\usepackage{amsmath}
\usepackage{caption,subcaption}
\usepackage{multirow}
\usepackage{multicol}
\usepackage{bbm}

\usepackage{bm}
\usepackage[ruled]{algorithm2e}

\usepackage{jmlr2e}

\newtheorem{claim}[theorem]{Claim}

\def\EE{\mathbb{E}}

\def\RR{\mathbb{R}}

\def\calF{\mathcal{F}}

\def\calN{\mathcal{N}}

\def\calR{\mathcal{R}}
\def\calS{\mathcal{S}}

\def\bR{\mathbf{R}}

\def\1{\mathbbm{1}}

\newcommand\independent{\protect\mathpalette{\protect\independenT}{\perp}}
\def\independenT#1#2{\mathrel{\rlap{$#1#2$}\mkern2mu{#1#2}}}
\DeclareMathOperator*{\argmax}{arg\,max}
\DeclareMathOperator*{\argmin}{arg\,min}

\def \bR {\mathbb{R}}

\def\independenT#1#2{\mathrel{\rlap{$#1#2$}\mkern2mu{#1#2}}}

\definecolor{codegreen}{rgb}{0,0.6,0}
\definecolor{codegray}{rgb}{0.5,0.5,0.5}
\definecolor{codepurple}{rgb}{0.58,0,0.82}
\definecolor{backcolour}{rgb}{0.95,0.95,0.92}
\lstdefinestyle{mystyle}{
	backgroundcolor=\color{backcolour},   
	commentstyle=\color{codegreen},
	keywordstyle=\color{magenta},
	numberstyle=\tiny\color{codegray},
	stringstyle=\color{codepurple},
	basicstyle=\ttfamily\footnotesize,
	breakatwhitespace=false,         
	breaklines=true,                 
	captionpos=b,                    
	keepspaces=true,                 
	numbers=left,                    
	numbersep=5pt,                  
	showspaces=false,                
	showstringspaces=false,
	showtabs=false,                  
	tabsize=2
}

\usepackage{lastpage}
\jmlrheading{22}{2021}{1-\pageref{LastPage}}{7/21; Revised
	11/21}{12/21}{21-0840}{Zijun Gao and Trevor Hastie}
\ShortHeadings{LinCDE: Conditional Density Estimation via Lindsey's Method}{Gao and Hastie}
\firstpageno{1}

\begin{document}
	\title{LinCDE: Conditional Density Estimation via Lindsey's Method}
		
	\author{\name Zijun Gao \email zijungao@stanford.edu \\
		\addr Department of Statistics \\
		Stanford University\\
		Stanford, CA 94305, USA
		\AND
		\name Trevor Hastie \email hastie@stanford.edu\\
		\addr Department of Statistics and Department of Biomedical Data Science\\
		Stanford University\\
		Stanford, CA 94305, USA}
	
	\editor{Victor Chernozhukov}
	\maketitle

	\begin{abstract}%
          Conditional density estimation is a fundamental problem in
          statistics, with scientific and practical
          applications in biology, economics, finance and
          environmental studies, to name a few.  In this paper,
          we propose a conditional density estimator based on gradient
          boosting and Lindsey's method (LinCDE). LinCDE admits
          flexible modeling of the density family and can capture
          distributional characteristics like modality and shape. In
          particular, when suitably parametrized, LinCDE will produce smooth and non-negative
          density estimates. Furthermore, like boosted regression
          trees, LinCDE does automatic feature selection. We
          demonstrate LinCDE's efficacy through extensive simulations
          and three real data examples.
	\end{abstract}


\section{Introduction}

In statistics, a fundamental problem is characterizing how a response depends on a set of covariates. Numerous methods have been developed for estimating the mean response conditioning on the covariates---the so-called regression problem. However, the conditional mean may not always be sufficient in practice, and various distributional characteristics or even the full conditional distribution are called for, such as the mean-variance analysis of portfolios \citep{markowitz1959portfolio}, the bimodality of gene expression distributions \citep{desantis2014breast, moody2019computational}, and the peak patterns of galaxy redshift densities \citep{ball2008robust}.
Conditional distributions can be used for constructing prediction intervals,  downstream analysis, visualization, and interpretation \citep{arnold1999conditional}.
Therefore, it is worthwhile to take a step forward from the conditional mean to the conditional distribution.

There are several difficulties in estimating conditional distributions. First, distribution estimation is more complicated than mean estimation regardless of the conditioning. 
Second, as with conditional mean estimation, conditioning on a
potentially large number of covariates suffers from the curse of
dimensionality. When only a small subset of the covariates are
relevant, proper variable selection is necessary to mitigate overfitting, reduce computational burden, and identify covariates that may be of interest to the practitioners.

In this paper, we develop a tree-boosted conditional density estimator
based on Lindsey's method, which we call LinCDE (pronounced ``linseed'') boosting.
LinCDE boosting is built on the base learner LinCDE tree. 
A LinCDE tree partitions the covariate space into subregions with homogeneous conditional distributions, estimates a local unconditional density in each subregion, and aggregates the unconditional densities to form the final conditional estimator. 
LinCDE boosting combines LinCDE trees to form a strong ensemble learner.

LinCDE boosting possesses several desirable properties.
LinCDE boosting provides a flexible modeling paradigm and is capable of capturing distributional properties such as heteroscedasticity and multimodality. 
LinCDE boosting also inherits the advantages of  tree and boosting methods, and in particular, LinCDE boosting is able to detect influential covariates. 
Furthermore, the conditional density estimates are automatically
non-negative and smooth, and other useful statistics such as
conditional quantiles or conditional cumulant distribution functions
(CDFs) can be obtained in a straightforward way from the LinCDE boosting estimates.

The organization of the paper is as follows. In Section
\ref{sec:background}, we formulate the problem and discuss related
work. We develop LinCDE in three steps:
\begin{itemize}
\item 
In Section~\ref{sec:Lindsey}, we describe Lindsey's method for
(marginal) density estimation. 
\item In Section~\ref{sec:LinCDETree}, we introduce LinCDE trees for
  conditional density estimation, which combine Lindsey's method with
  recursive partitioning. 
\item In Section~\ref{sec:LinCDEBoosting}, we develop a boosted
ensemble model using  LinCDE trees.
\end{itemize}
In Section~\ref{sec:pretreatment}, we discuss two optional but helpful
preprocessing steps---response transformation and conditional mean centering. In Section~\ref{sec:simulation}, we evaluate the performance of LinCDE boosting on simulated data sets. In Section~\ref{sec:realData}, we apply LinCDE boosting to three real data sets. We conclude the paper with discussions in Section~\ref{sec:discussion} and provide links to the R software and instructions in Section \ref{sec:software}. All proofs are deferred to the appendix.

\section{Background}\label{sec:background}

In this section, we formulate the conditional density estimation problem and discuss related
work.

\subsection{Problem Formulation}
Let $y \in \bR$ be a continuous response.\footnote{The paper will focus on univariate responses, and the generalization to multivariate responses is straightforward and discussed in Section~\ref{sec:discussion}.} Let $x$ be a $d$-dimensional covariate vector and $x^{(j)}$ be its $j$-th coordinate. We assume the covariates are generated from an unknown underlying distribution $f_x(x)$, and the response given the covariates are sampled from an unknown conditional density $f_{y|x}(y\mid x)$. The model is summarized as
\begin{align}\label{eq:model}
\begin{split}
	x_i &\stackrel{i.i.d.}{\sim} f_x,\\
	y_i \mid x_i &\stackrel{ind.}{\sim} f_{y|x}. 
\end{split}
\end{align}
We observe $n$ data pairs $\{(x_i, y_i)\}$ and aim to estimate the conditional density $f_{y|x}(y\mid x)$.

\subsection{Literature}

In general, there are three ways to characterize a conditional distribution: conditional density, conditional quantile, and conditional CDF. 
We categorize related works on conditional density, quantile, and CDF estimation according to the methodology and provide a review below. We conclude by discussing two desired properties of conditional density estimation and how our work complements the existing literature on the two points.

A line of study estimates the conditional distribution by localizing 
unconditional distribution estimators. Localization methods weight 
observations according to the distances between their covariates 
and those at the target point, and solve the unconditional 
distribution estimation problem based on the weighted sample. For conditional density, \citet{fan1996estimation} obtain conditional density estimates by local polynomial regression. For conditional quantile, \citet{chaudhuri1991global, chaudhuri1991nonparametric} partitions the covariate space into bins and fit a quantile model in each bin separately, and \citet{yu1998local} tackle the conditional quantile estimation via local quantile loss minimization. For conditional CDF, \citet{stone1977consistent} proposes a weighted sum of indicator functions, and \citet{hall1999methods} consider a local logistic regression and a locally adjusted Nadaraya-Watson estimator. Localization methods enable systematic extensions of any unconditional estimator. Nevertheless, the weights usually treat covariates as equally important, and variable selection is typically not accommodated. This leaves the methods vulnerable to the curse of dimensionality.

As discussed by \citet{stone1991asymptotics, stone1991multivariate}, another approach making use of unconditional methods, first obtains the joint distribution estimate $\hat{f}_{y,x}(y, x)$ and the covariate distribution estimate $\hat{f}_x(x)$, and then follows 
\begin{align}\label{eq:ratio}
	\hat{f}_{y|x}(y \mid x) = \hat{f}_{y,x}(y, x)/\hat{f}_x(x)
\end{align} 
to derive the conditional density. Nevertheless, the joint distribution estimation is also challenging, if not more. \citet{arnold1999conditional} point out except for special cases like multivariate Gaussian, the estimation of a bivariate joint distribution in a certain exponential form is onerous due to the normalizing constant. Moreover, the approach is inefficient both statistically and computationally if the conditional distribution is comparatively simpler than the joint and covariate distributions; as an example, when the response is independent of the covariates.

A different thread directly models the dependence of the response's distribution on the covariates by a linear combination of a finite or infinite number of bases.
\begin{itemize}
	\item 
	For conditional density, \citet{kooperberg1991study, kooperberg1992logspline}, \citet{stone1991multivariate}, \citet{maacsse1999conditional},  and \citet{barron1991approximation} study the conditional logspline density model: modeling $\log({f}_{y|x}(y \mid x))$ by tensor products of splines or trigonometric series and maximizing the conditional log-likelihood to estimate the parameters. The method is also known as entropy maximization subject to empirical constraints. In addition, \citet{sugiyama2010conditional} model the conditional density in reproducing kernel Hilbert spaces and estimate the loadings by the unconstrained least-squares importance fitting, and \citet{izbicki2016nonparametric} expand the conditional density in the eigenfunctions of a kernel-based operator which adapts to the intrinsic dimension of the covariates.\footnote{We remark that the approach in \citep{izbicki2016nonparametric} adapts to the low intrinsic dimensionality of covariates but not the sparse dependency of the response on the covariates. }
	\item For conditional quantiles,  \citet{koenker1978regression} formulate conditional quantiles as linear functions of covariates and minimize quantile losses to estimate parameters. \citet{koenker1994quantile} explore quantile smoothing splines minimizing for a single covariate, and \citet{he1998bivariate} extend the approach to the bi-variate setting, i.e., two covariates. \citet{li2007quantile} propose kernel quantile regression (KQR) considering quantile regression in reproducing Hilbert kernel spaces. \citet{belloni2019conditional} approximate the conditional quantile function by a growing number of bases as the sample size increases.
	\item For conditional CDF, \citet{foresi1995conditional} and \citet{ chernozhukov2013inference} consider distribution regression: estimating a sequence of conditional logit models over a grid of values of the response variable. \citet{belloni2019conditional} further extend the method to the high-dimensional-sparse-model setting.
\end{itemize}
The performance of parametric models depends on the selected bases or kernels. Covariate-specific bases or kernels require a lot of tuning, and the bases or kernels treat covariates equally, making the approaches less powerful in the presence of many nuisance covariates.

More recently, tree-based estimators arose in conditional distribution estimation. 
The overall idea is partitioning the covariate space recursively and 
fitting an unconditional model at each terminal node.
For conditional quantiles, \citet{chaudhuri2002nonparametric} investigate a tree-structured quantile regression. 
Nevertheless, the estimation of different quantiles requires separate quantile loss minimization, which complicates the full conditional distribution calculation. \citet{meinshausen2006quantile} proposes the quantile regression forest (QRF) that computes all quantiles simultaneously. QRF first builds a standard random forest, then estimates the conditional CDF by a weighted distribution of the observed responses, and finally inverts the CDF to quantiles. 
For conditional CDF, \citet{friedman2019contrast} proposes distribution boosting (DB). DB relies on Friedman's contrast trees---a method to detect the lack-of-fit regions of any conditional distribution estimator. DB estimates the conditional distribution by iteratively transforming the conditional distribution estimator and correcting the errors uncovered by contrast trees.

Beyond trees, neural network-based conditional distribution estimators have also been developed. For conditional quantile, standard neural networks with the quantile losses serve the goal. For conditional density, \citet{bishop1994mixture, bishop2006pattern} introduces the mixture density networks (MDNs) which model the conditional density as a mixture of Gaussian distributions. Neural network-based methods can theoretically approximate any conditional distributions well but are computationally heavy and lack interpretation.

In this paper, we propose LinCDE boosting which complements existing works by producing smooth density curves and performing feature selection. 
\begin{itemize}
	\item  Compared to quantiles and CDF, smooth density curves are more suitable for the following reasons.
	\begin{itemize}
		\item Smooth density curves can give valuable indications of distributional characteristics such as skewness and multimodality. Visualization of smooth density curves is comprehensible to practitioners and can yield self-evident conclusions or point the way to further analysis.
		\item Conditional densities can be used to compute class-posterior probabilities using the Bayes' rule. The class-posterior probabilities can be further used for classification.
		\item Densities can be used to detect outliers: if an observation lies in a very low-density region, the data point is likely to be an outlier. 
	\end{itemize}
	LinCDE boosting generates smooth density curves, while directly transforming estimates of conditional quantiles or CDF to conditional density estimates usually produces bumpy results.
	\item LinCDE trees and LinCDE boosting calculate feature importances and automatically focus on influential features in estimation. In contrast, methods like localization and modeling conditional densities in a linear space or RKHS often treat covariates equally, and covariate-specific neighborhoods, bases, and kernels may require a lot of tuning. 
	Therefore, LinCDE boosting is more scalable to a large number of features and less sensitive to the presence of nuisance covariates.
\end{itemize}

\section{Lindsey's Method}\label{sec:Lindsey}
In this section, we first introduce the density estimation problem ---
an intermediate step towards the conditional density estimation. We
then discuss how to solve the density estimation problem by Lindsey's
method \citep{lindsey1974comparison}---a stepping stone of
LinCDE. Lindsey's method cleverly avoids the normalizing issue by
discretization and solves the problem by fitting a simple Poisson
regression. It can be thought of as a method for fitting a smooth
histogram with a large number of bins.

We consider the density family
\begin{align}\label{eq:g(y)}
	f(y) = \kappa(y) e^{g(y)},
\end{align}
where $\kappa: \RR \to \RR$ is some carrying density, and $g(y)$ is
known as a \emph{tilting} function.
The idea is that $\kappa(y)$ is known or assumed (such as Gaussian or
uniform), and $g(y)$ is represented by a model. We represent $g$ as a
linear expansion 
\begin{align} \label{eq:g(y)2}
	g(y) = z(y)^\top\beta + \beta_0,
\end{align}
where $z(y)$ is a basis of $k$ smooth functions. As a simple example, if
we use standard Gaussian  as the carrying measure and choose 
$z(y)^\top=(y, y^2)$, the resulting density family corresponds to all
possible Gaussian distributions. More generally we use a basis of
natural cubic
splines in $z(y)$ with knots spread over the domain of $y$, to achieve
a flexible representation \citep{wahba1975smoothing, wahba1990spline}.

Our goal is to find the density that maximizes the log-likelihood
\begin{align}\label{eq:loglikelihood}
	\max_{\beta, \beta_0} \frac{1}{n}\sum_{i=1}^n \log(\kappa(y_i)) + z(y_i)^\top\beta + \beta_0, \quad \text{s.t.}	\int \kappa(y) e^{z(y)^\top\beta + \beta_0} dy = 1.
\end{align}
The constrained optimization problem
\eqref{eq:loglikelihood} can be simplified to the unconstrained
counterpart below by the method of Lagrange multipliers \citep{silverman1986density},
\begin{align}\label{eq:loglikelihoodPenalty2}
	\frac{1}{n}\sum_{i=1}^n \left(\log(\kappa(y_i)) + z(y_i)^\top\beta  + \beta_0\right) - \int \kappa(y) e^{z(y)^\top\beta  + \beta_0 } dy,
\end{align}
where the multiplier turns out to be one.

The optimization problem \eqref{eq:loglikelihoodPenalty2} is difficult since the integral $\int \kappa(y) e^{z(y)^\top\beta  + \beta_0 } dy$ is generally unavailable in closed form. One way to avoid the integral is by discretization---the key idea underlying Lindsey's method. 
One divides the response range into $B$ equal bins of width $\Delta$ with mid-points $y_b$. The integral is approximated by the finite sum
\begin{align*}
	\int \kappa(y) e^{z(y)^\top\beta  + \beta_0 } dy ~\approx~ \sum_{b=1}^B  \kappa(y_b) e^{z(y_b)^\top\beta  + \beta_0} \Delta.
\end{align*}
As for the first part of \eqref{eq:loglikelihoodPenalty2}, one replaces $y_i$ by its bin  midpoint $y_{b(i)}$ and groups the observations,
\begin{align*}
	\sum_{i=1}^n \log(\kappa(y_i)) + z(y_i)^\top\beta  + \beta_0 
	&\approx \sum_{i=1}^n \log(\kappa(y_{b(i)})) + z(y_{b(i)})^\top\beta  + \beta_0 \\
	&= \sum_{b=1}^B n_b \left(\log(\kappa(y_{b})) + z(y_b)^\top\beta  + \beta_0\right),
\end{align*}
where $b(i)$ denotes the bin that the $i$-th response falls in, and $n_b$ represents the number of samples in bin $b$. Combining the above two parts, the Lagrangian function with response discretization takes the form
\begin{align}\label{eq:loglikelihoodPenalty3}
	\frac{1}{n}\sum_{b=1}^B n_b \left(\log(\kappa(y_{b})) + z(y_b)^\top\beta  + \beta_0\right) - \sum_{b=1}^B  \kappa(y_b) e^{z(y_b)^\top\beta  + \beta_0 } \Delta.
\end{align}
The objective function \eqref{eq:loglikelihoodPenalty3} is equivalent to that of a Poisson regression 
with $B$ observations $\{n_b\}_{1 \le b \le B}$ and mean parameters $\mu_b \propto \kappa(y_b) e^{z(y_b)^\top\beta}$. 
Therefore, Lindsey's method estimates the coefficient $\beta$ by fitting the Poisson regression with predictors $z(y)$ and offset $\log\left(\kappa(y_b)\right)$.  The normalizing constant $\beta_0$ in \eqref{eq:loglikelihoodPenalty3} is absorbed in the Poisson regression's intercept. Despite the discretization error, Lindsey's estimates are consistent, asymptotically normal, and remarkably efficient \citep{moschopoulos1994estimation, efron2004large}. We will demonstrate the efficacy of Lindsey's method in two examples at the end of this section.

The number of bins $B$ balances the statistical performance and the
computational complexity of Lindsey's method: as $B$ increases,  the
discretized objective \eqref{eq:loglikelihoodPenalty3} approaches the
original target \eqref{eq:loglikelihoodPenalty2}, and the resulting
estimator converges to the true likelihood maximizer; on the other
hand, the computations increase linearly in $B$. This can become a
factor later when we fit many of these Poisson models repeatedly.

The relationship with a histogram becomes clear now, as well. We could
use the counts in the $B$ bins to form a density estimate, but this
would be very jumpy. Typically we would control this by reducing the number
of bins. Lindsey's method finesses this by having $B$ large, but
controlling the smoothness of the bin means via the $k\ll B$ basis
functions and associated coefficients.

To control the model complexity further and avoid numeric instability,
we add a regularization term to~\eqref{eq:loglikelihoodPenalty2}. 
For example, we can penalize deviations from Gaussian distributions via the regularizer \citep{wahba1977optimal, silverman1982estimation, silverman1986density}\footnote{We regard exponential distribution as a special case of Gaussian distribution with $\sigma^2 = \infty$.} 
\begin{align}\label{eq:loglikelihoodPenalty}
	\int \left(\frac{d^3}{dy^3}\left(z(y)^\top \beta + 	\beta_0\right)\right)^2 dy.
\end{align} 
The penalty measures the roughness of the tilting function and is zero if and only if the tilting function's exponent is a quadratic polynomial, i.e., a Gaussian distribution. 
We also attach a hyper-parameter $\lambda$ to trade-off the objective \eqref{eq:loglikelihoodPenalty2} and the penalty \eqref{eq:loglikelihoodPenalty}, and tune $\lambda$ to achieve the best performance on validation data sets.\footnote{The hyper-parameter is a function of the penalized Poisson regression's degrees of freedom (see Appendix~\ref{appn:sec:hyperParameter} for more details), and we tune the degrees of freedom to achieve the best performance on validation data sets.}

It is convenient to tailor the spline basis functions to the penalty~\eqref{eq:loglikelihoodPenalty}. Note that, for arbitrary bases $z(y)$, the penalty \eqref{eq:loglikelihoodPenalty} is a quadratic form in $\beta$
\begin{align}\label{eq:loglikelihoodPenalty05}
\int \left(\frac{d^3}{dy^3}\left(z(y)^\top\beta + \beta_0\right)\right)^2 dy
= \sum_{j,l=1}^k \beta_j \beta_l \int z_{j}'''(y) z_{l}'''(y) dy
=: \beta^\top \Omega \beta,
\end{align}
where $\Omega_{jl} = \int z_{j}'''(y) z_{l}'''(y) dy$. We 
transform our splines so that the associated ${\Omega} = \text{\rm
  diag}(\omega_1, \ldots, \omega_k)$ is diagonal and the penalty
reduces to a weighted ridge penalty
\begin{align}\label{eq:loglikelihoodPenalty075}
	\sum_{j=1}^k \omega_j \beta_j^2.
\end{align}
(details in Appendix~\ref{appn:sec:linearTransformation}). Figure~\ref{fig:NSplineTransform} depicts an example of the
transformed spline bases (in increasing order of $\omega_j$) and the corresponding smoothed versions. Among
the transformed bases, the linear and quadratic components (the first
and the second bases in
Figure~\ref{fig:NSplineTransform}) are not shrunk by the roughness
penalty (Claim~\ref{claim:nullSpace}), and higher-complexity splines
are more heavily penalized \citep[chap.~5, for example]{hastie09:_elemen_statis_learn_II}.

\begin{claim}\label{claim:nullSpace}
	Assume $u\in \RR^k$ and $\Omega u = 0$. Then $z(y)^\top u$ is a linear or quadratic function of $y$.
\end{claim}

\begin{figure}[bt]
	\centering
	\begin{minipage}{12cm}
		\centering  
		\includegraphics[clip,  trim = 0cm 0cm 0cm 0.2cm, scale = 0.57]{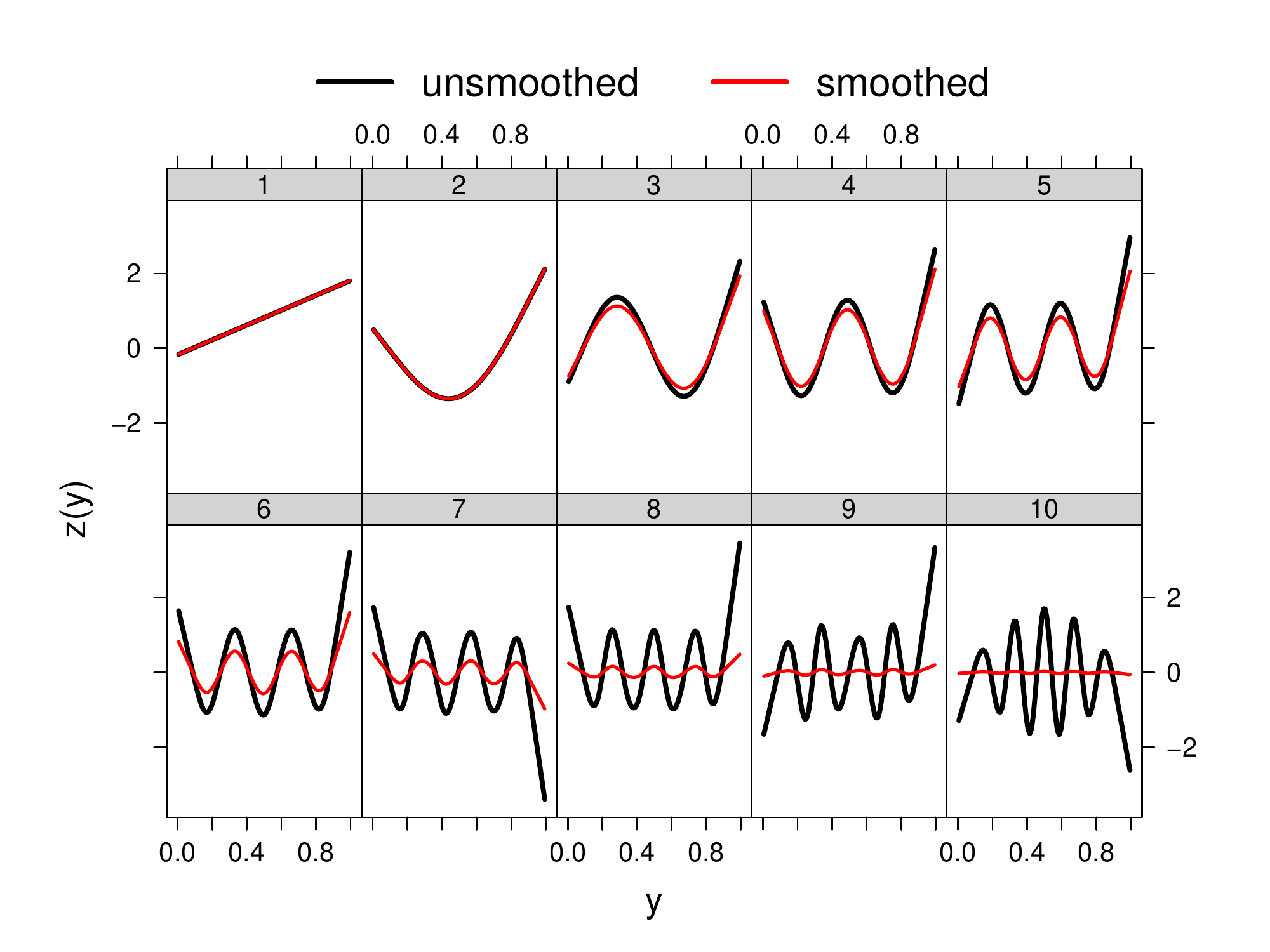}  
	\end{minipage}
	\caption{{Transformed natural cubic spline bases. The transformed basis functions before smoothing (in black) are normalized such that $\int (z_j(y))^2 dy = 1$ and ordered by increasing penalty factors. The damped functions (in red) represent smoothed basis functions (with $6$ degrees of freedom). 
	}}
	\label{fig:NSplineTransform}
\end{figure}

We provide examples of Lindsey's performance in estimating bimodal and skewed distributions in Appendix~\ref{appn:sec:figure}.

\section{LinCDE Trees}\label{sec:LinCDETree}
In this section, we extend the density estimation problem to the conditional density estimation problem. We introduce LinCDE trees combining Lindsey's method and recursive partitioning: LinCDE trees partition the covariate space and estimate a local unconditional density via Lindsey's method in each subregion.

To begin, we restate the target density family \eqref{eq:g(y)} in the
language of exponential families. We call $z(y)$ the sufficient
statistics and
$g(y)$
the natural parameter. We replace the normalizing constant $\beta_0$ by the negative cumulant generating function $\psi(\beta)$ defined as 
\begin{align*}
	e^{\psi(\beta)} = \int \kappa(y) e^{z(y)^\top \beta} dy.
\end{align*}
As a result, the density normalizing constraint of $\kappa(y) e^{z(y)^\top \beta - \psi(\beta)}$ is automatically satisfied.

In the conditional density estimation problem, we consider the target family generalized from \eqref{eq:g(y)}
\begin{align}\label{eq:exponentialFamily2}
	f_{y|x}(y\mid x)  = \kappa(y) e^{z(y)^\top \beta(x) - \psi(\beta(x))}.
\end{align}
The dependence of the response on the covariates is encoded in the parameter function $\beta(x)$. 
\begin{itemize}
	\item If there is only $k=1$ basis function $z(y)$ and the parameter function satisfies $\beta(x) = x^\top \theta$, then the conditional density family \eqref{eq:exponentialFamily2} reduces to a generalized linear model.
	\item If we consider a standard Gaussian prior, basis function $z(y) = y$, and tree-structured $\beta(x)$, then the conditional density family \eqref{eq:exponentialFamily2} reduces to a regression tree model. In fact, the optimization problem based on \eqref{eq:exponentialFamily2} is equivalent to finding the tree $\beta(x)$ that minimizes the sum of squares $\sum_{i=1}^n (y_i - \beta(x_i))^2$.
\end{itemize}

Similar to the density estimation problem, we aim to find the member in the family \eqref{eq:exponentialFamily2} that maximizes the conditional log-likelihood with the ridge penalty \eqref{eq:loglikelihoodPenalty075}
\begin{align}\label{eq:loglikelihoodPenalty4}
\begin{split}
	\ell(\calR_0; \beta) 
	:=& \sum_{i=1}^n \left(\log(\kappa(y_i)) + z(y_i)^\top\beta(x_i) - \psi(\beta(x_i)) - \lambda \sum_{j=1}^k \omega_j \beta_j^2(x_i)\right) \\
	=& \sum_{i=1}^n \left(\log(\kappa(y_i)) + z(y_i)^\top\beta(x_i) - \psi(\beta(x_i))\right) - \lambda\sum_{i=1}^n \sum_{j=1}^k \omega_j \beta_j^2(x_i),
\end{split}
\end{align} 
where $\calR_0$ denotes the full covariate space.
If $\beta(x)$ is constant,  the problem \eqref{eq:loglikelihoodPenalty4} simplifies to the unconditional problem \eqref{eq:loglikelihood}.

The conditional density estimation problem
\eqref{eq:loglikelihoodPenalty4} is more complicated than the
unconditional version due to the covariates $x$:
\begin{enumerate}
\item 
Given a covariate configuration $x$, there is often at most one
observation whose covariates take the value $x$, and it is infeasible
to estimate the multi-dimensional natural parameter $\beta(x)$ based
on a single observation; 
\item There may be a multitude of covariates, and only a few are influential. Proper variable selection or shrinkage is necessary to avoid serious overfitting. 
\end{enumerate}

One way to finesse these difficulties is to use trees
\citep{breiman1984classification}. We divide the covariate space into subregions with approximately homogeneous conditional distributions, and in each subregion, we estimate a density independent of the covariate values. We name the method ``LinCDE trees". In response to the first difficulty, by conditioning on a subregion instead of a specific covariate value, we have more samples for local density estimation. In response to the second difficulty, trees perform internal feature selection, and are thus resistant to the inclusion of many irrelevant covariates. Moreover, the advantages of tree-based methods are automatically inherited, such as being tolerant of all types of covariates, computationally efficient, and easy to interpret.

Before we delve into the details, we again draw a connection between LinCDE
trees and a naive binning approach---fitting a multinomial model using
trees. The naive approach discretizes the response into multiple bins
and predicts conditional cell probabilities through recursive
partitioning. The normalized conditional cell probabilities serve as
an approximation of the conditional densities, and the more bins 
used, the higher resolution the approximation is. The naive approach
is able to detect subregions with homogeneous multinomial
distributions. However, the estimates are bumpy, especially with a
large number of bins. To stabilize the method, restrictions enforcing
smoothness are required, and LinCDE trees realize the goal by modeling
the density exponent using splines.

We now explain how LinCDE trees work. In standard tree algorithms, there are two major steps: 
\begin{itemize}
	\item \textit{Splitting}: partitioning the covariate space into subregions;
	\item \textit{Fitting}: performing estimation in each subregion. The estimator is usually obtained by maximizing a specific objective function. For example, in a regression tree with $\ell_2$ loss, the estimator is the sample average; in a classification tree with misclassification error, the estimator is the majority's label.
\end{itemize} 
The fitting step is a direct application of Lindsey's method in Section~\ref{sec:Lindsey}. In a subregion $\calR$, we treat the natural parameter functions as a constant vector and solve the density estimation problem via Lindsey's method. We denote the objective function value in region $\calR$ with parameter $\beta$ by $\ell(\calR; \beta)$, and let $\hat{\beta}_{\calR} := \argmax_{\beta} \ell(\calR; \beta)$.

Now for the splitting step. Similar to standard regression and
classification trees, we proceed with a greedy algorithm and select
the candidate split that improves the objective the
most. Mathematically, starting from a region $\calR$, we maximize the
improvement statistic
\begin{align}
\label{eq:split1}
	\Delta \ell(\calR, s) 
	:= \ell(\calR_{s,L}; \hat{\beta}_{\calR_{s,L}}) + \ell(\calR_{s,R}; \hat{\beta}_{\calR_{s,R}}) - \ell(\calR; \hat{\beta}_{\calR}),
\end{align}
where $\calR_{s,L}$ and $\calR_{s,R}$ are the regions on the left and right of the candidate split, respectively. 
Direct computation of the difference \eqref{eq:split1} requires
running Lindsey's method twice for \emph{each} candidate split $s$ to
obtain $\hat{\beta}_{\calR_{s,L}}$, $\hat{\beta}_{\calR_{s,R}}$, and
the total computation time is prohibitive. Instead, we approximate the difference \eqref{eq:split1} by a simple quadratic term in Proposition~\ref{prop:approx}, which can be computed much faster.

\begin{proposition}
	 [Improvement approximation for LinCDE trees]
	\label{prop:approx}
	Let $n_{\calR}$, $\bar{z}_{\calR}$ be the sample size and average sufficient statistics in a region $\calR$. Assume that $\nabla^2\psi(\hat{\beta}_{\calR}) + 2 \lambda \Omega$ is invertible, then for a candidate split $s$, 
	\begin{align*}
	\frac{1}{n_{\calR}} \Delta \ell(\calR, s) 
	= \frac{n_{\calR_{s,L}} n_{\calR_{s,R}}}{2 n_{\calR}^2} (\bar{z}_{\calR_{s,L}} - 
	\bar{z}_{\calR_{s,R}})^\top \left(\nabla^2\psi(\hat{\beta}_{\calR}) + 2 \lambda \Omega \right)^{-1} (\bar{z}_{\calR_{s,L}} - \bar{z}_{\calR_{s,R}}) + r_s,
	\end{align*} 	
	where the remainder term satisfies $r_s = O\left(\|\bar{z}_{\calR_{s,L}} - \bar{z}_{\calR}\|_2^3 + \|\bar{z}_{\calR_{s,R}}- \bar{z}_{\calR}\|_2^3\right)$.
\end{proposition}

Proposition~\ref{prop:approx} writes the difference \eqref{eq:split1}
as a quadratic form plus a higher-order residual term.  If $z(y) = y$,
the model amounts to a regression tree, and the residual term is
zero. For general $z(y)$, when the average sufficient statistics
$\bar{z}_{\calR_{s,L}}$, $\bar{z}_{\calR_{s,R}}$ are similar, the
residual term is of smaller order than the quadratic form and can thus
be dropped; when $\bar{z}_{\calR_{s,L}}$, $\bar{z}_{\calR_{s,R}}$ are
considerably different, the residual term is not guaranteed to be
small theoretically. However, we empirically demonstrate in Appendix~\ref{appn:sec:approximation} that at such splits, the quadratic form is still
sufficiently close to the true log-likelihood difference. Based on this
empirical evidence, we use the quadratic approximation to determine the optimal splits.

The quadratic approximation suggested by Proposition~\ref{prop:approx} is the product of the squared difference between the average sufficient statistics in $\calR_L$ and $\calR_R$ normalized by $\nabla^2\psi(\hat{\beta}_{\calR}) + 2 \lambda \Omega$, further multiplied by the sample proportions in $\calR_L$ and $\calR_R$. By selecting the candidate split that maximizes the quadratic term, we will end up with two subregions different in the sufficient statistics means and reasonably balanced in sample sizes.

To compute the quadratic approximation, we need subsample proportions $n_{\calR_{s,L}}/n_{\calR}$, $n_{\calR_{s,R}}/n_{\calR}$, average sufficient statistics $\bar{z}_{\calR_{s,L}}$, $\bar{z}_{\calR_{s,R}}$, and the inverse matrix of $\nabla^2\psi(\hat{\beta}_{\calR}) + 2\lambda \Omega$. 
For the candidate splits based on the same covariate,
$\{n_{\calR_{s,L}}, n_{\calR_{s,R}}, \bar{z}_{\calR_{s,L}},
\bar{z}_{\calR_{s,R}}\}$ can be computed efficiently for all split points by scanning through the samples in $\calR$ once. For all candidate splits, this
takes $O(dn_{\calR}k)$ operations in total. The matrix $\nabla^2\psi(\hat{\beta}_{\calR})$ is shared by all candidate splits and needs to be computed only once. The difficulty is that $\nabla^2\psi(\beta)$ is often unavailable in closed form. However, since $\nabla^2\psi(\beta_{\calR})$ is the covariance matrix of the sufficient statistics $z(y)$ if the responses $y$ are generated from the model parameterized by $\beta_{\calR}$, we apply Lindsey's method to estimate ${\beta}_{\calR}$ and compute the covariance matrix of the sufficient statistics based on the multinomial cell probabilities, which takes $O(k^2B)$. Appendix~\ref{appn:sec:proof} shows the resulting covariance matrix approximates $\nabla^2\psi(\hat{\beta}_{\calR})$ with a fine discretization. The total time complexity of the above splitting procedure is summarized in the following Proposition~\ref{prop:computation}.

\begin{proposition}\label{prop:computation}	
	Assume that there are $S$ candidate splits, $k$ basis
        functions, $d$ covariates,  $B$ discretization bins, and
        $n_{\calR}$ observations in the current region. Then the splitting step for LinCDE trees is of time complexity ${O}(dn_{\calR}k + k^2B + k^3 + Sk^2)$.
\end{proposition}

According to Proposition~\ref{prop:computation}, 
the computation time based on the quadratic approximation is significantly reduced compared to running Lindsey's methods in $\calR_{s,L}$ and $\calR_{s,R}$ for all candidate splits, which takes $\tilde{O}(S(n_{\calR}k + k^2B + k^3))$. 
 
Having found the best split $s_{\max}$, we partition $\calR$ into two subregions $\calR_{s_{\max}, L}$ and $\calR_{s_{\max}, R}$, and repeat the splitting procedure in the two subregions. Along the recursively partitioning, the response distribution's heterogeneity is reduced. The fitting and the splitting steps of LinCDE trees are summarized below, the complete algorithm is given in Algorithm~\ref{algo:LinCDETree}, and implementation details are displayed in Section~\ref{sec:software}. Stopping criteria for LinCDE trees are discussed in Appendix~\ref{appn:sec:stopping}.

\begin{itemize}
	\item \textit{Fitting (LinCDE tree)}. At a region $\calR$:	
	\begin{enumerate}
		\item Count the number of observations $\{n_{\calR,b}\}$ in each bin.
		\item Fit a Poisson regression model with the response variable $\{n_{\calR,b}\}$, regressors $z(y_b)$, the offset $\log(\kappa(y_b))$, and a weighted ridge penalty.\footnotemark~
		Denote the estimated coefficients by $\hat{\beta}_{\calR}$. 
	\end{enumerate}
	
	\item \textit{Splitting (LinCDE tree)}.  At a region $\calR$:
	\begin{enumerate}
		\item Compute $\{n_{\calR_{s,L}}, n_{\calR_{s,R}}, \bar{z}_{\calR_{s,L}}, \bar{z}_{\calR_{s,R}}\}$ for each candidate split, and approximate $\nabla^2{\psi}(\hat{\beta}_{\calR})$ by $z(y)$'s  covariance matrix in $\calR$ using  $\hat{\beta}_{\calR}$.
		\item For each candidate split $s\in \calS$, compute the quadratic approximation $\widehat{\Delta} \ell(\calR, s)$ by Proposition~\ref{prop:approx}, and choose the split $s_{\max} = \argmax_{s \in \calS} \hat{\Delta} \ell(\calR, s)$.  
	\end{enumerate}
\end{itemize}
\footnotetext{Weights (penalty factors) of the ridge penalty 
	are $\boldsymbol{\omega} = [\omega_1, \ldots, \omega_k]^\top$.}

\begin{algorithm}
	\DontPrintSemicolon  
	\SetAlgoLined
	\BlankLine
	\caption{LinCDE tree}\label{algo:LinCDETree}
	Start at the full covariate space.\\
	1. Apply \textit{Fitting (LinCDE tree)} and obtain the natural parameter estimator $\hat{\beta}$.\\
	2. Apply \textit{Splitting (LinCDE tree)} and obtain the optimal split $s_{\max}$.\\
	3. Repeat steps 1 and 2 to the left and right children of $s_{\max}$ until the stopping rule is satisfied, e.g., the maximal tree depth is reached. Output the natural parameter estimator $\hat{\beta}$ in each subregion.
\end{algorithm}

\begin{figure}[bt]
	\centering
	\begin{minipage}{16cm}
	\centering
		\includegraphics[scale = 0.57]{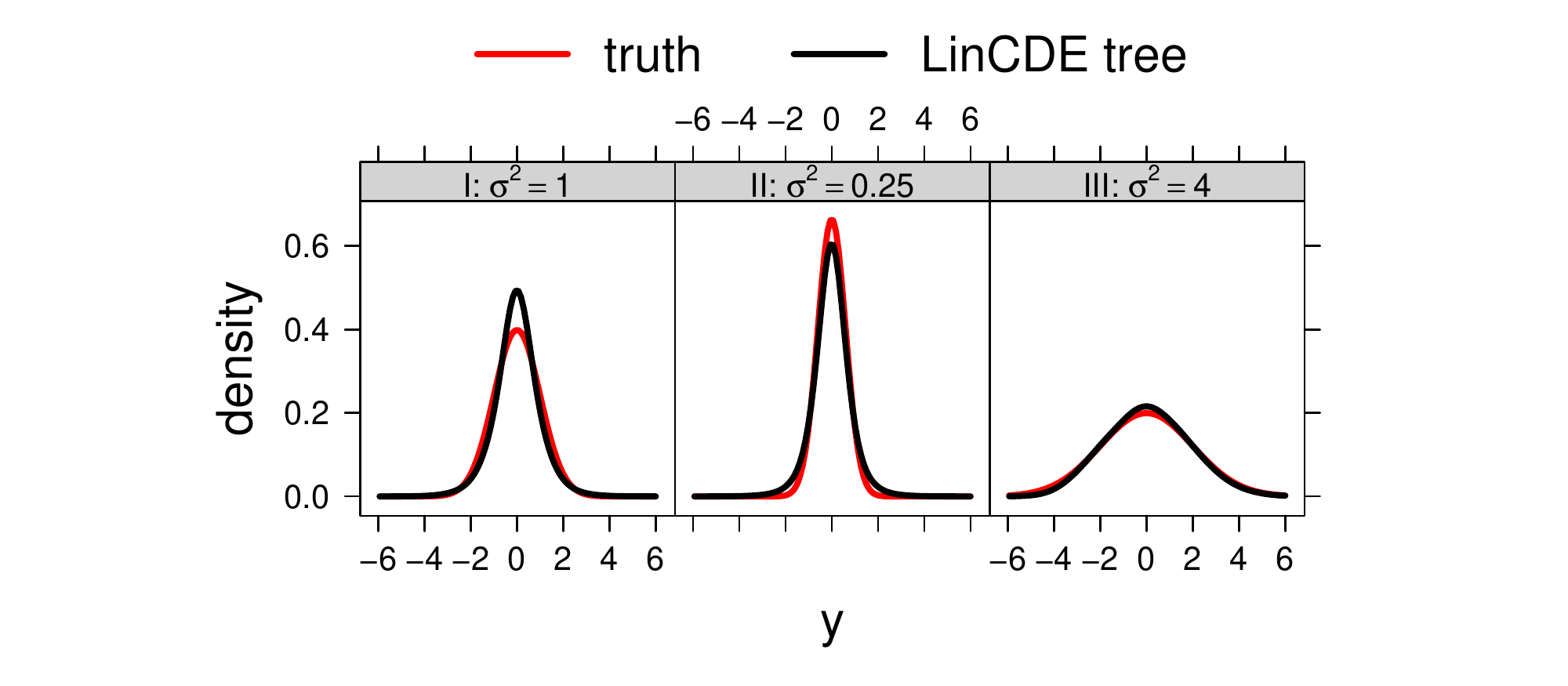} 
	\end{minipage}
	\begin{minipage}{16cm}
	\centering
	\includegraphics[scale = 0.57]{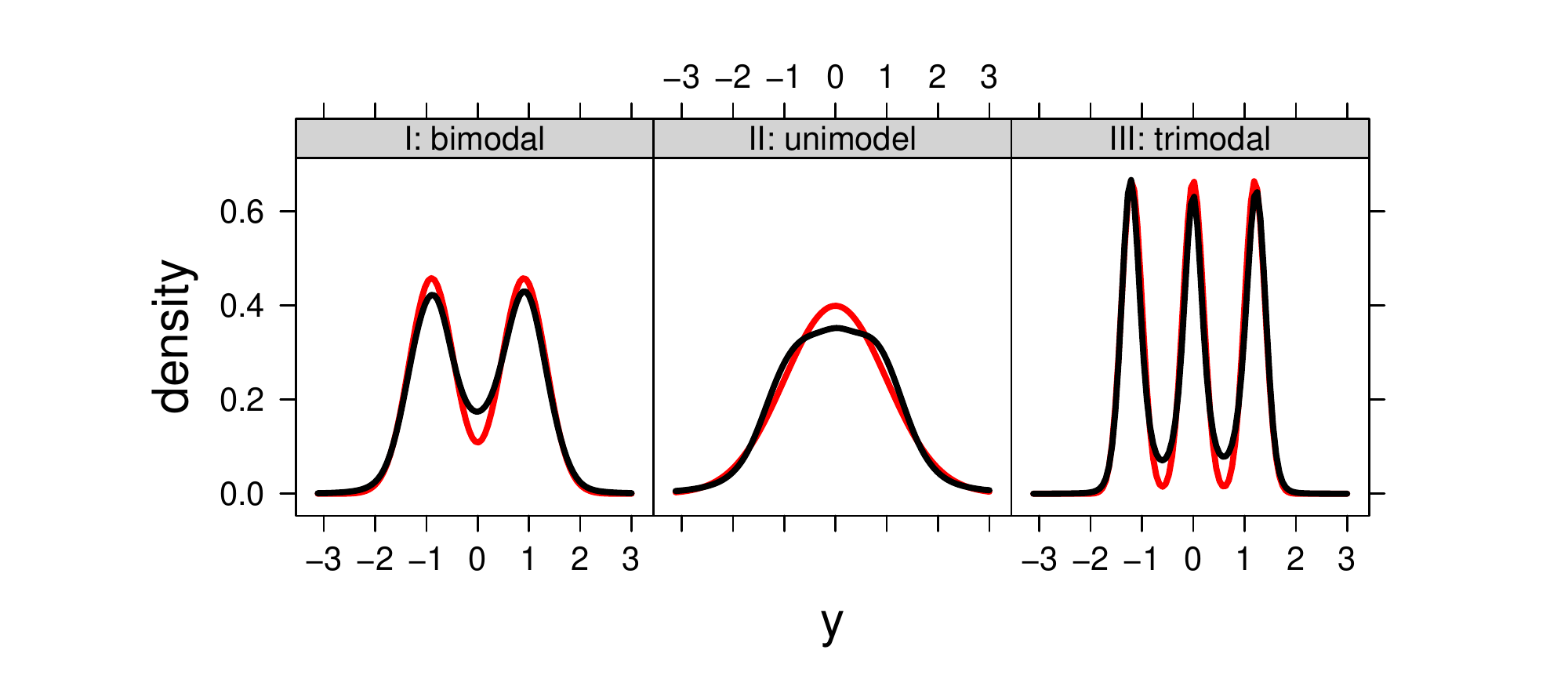} 
	\end{minipage}
	\begin{minipage}{16cm}
		\centering
		\includegraphics[scale = 0.57]{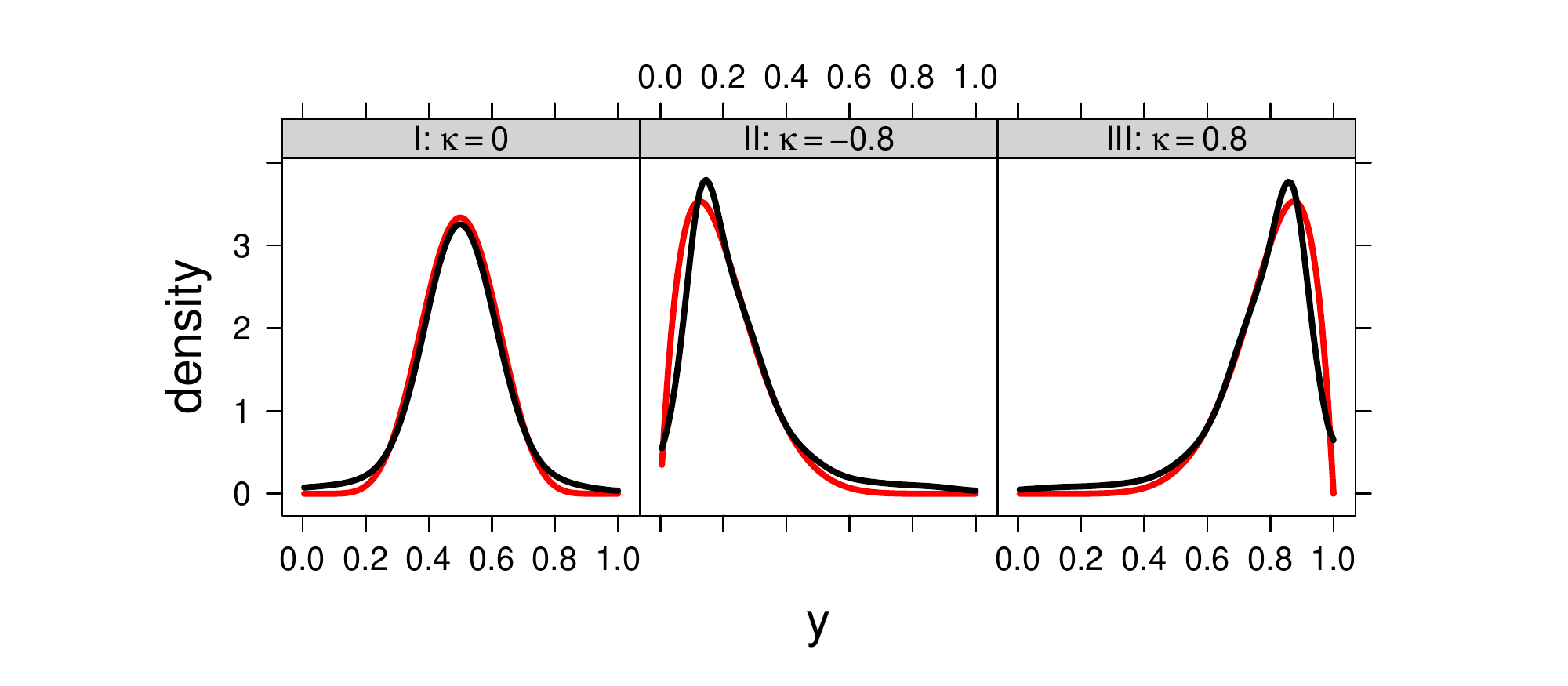} 
	\end{minipage}
	\caption{{LinCDE trees' conditional density estimates of heteroscedastic, 
			multimodal and skewed distributions. The responses are generated from the model \eqref{eq:LinCDEExample}. The local densities are Gaussian with variances $\sigma^2 \in \{0.25, 1, 4\}$ (first row), Gaussian mixtures with $1$, $2$, $3$ components (second row), and Beta distributions with skewness $\kappa \in \{-0.8, 0, 0.8\}$ (third row). In each trial, we sample $400$ observations. We use $10$ natural cubic splines , $5$ degrees of freedom, and a maximal tree depth at $2$. 
			We repeat $100$ times and plot the average fits against the true densities.
}}
	\label{fig:LinCDETreeExample}
\end{figure}

To conclude the section, we demonstrate the effectiveness of LinCDE trees in three toy examples. We generate $10$ covariates randomly uniformly on $[-1,1]$. The response follows
\begin{align}\label{eq:LinCDEExample}
	f_{y|x}(y \mid x) =
	\begin{cases}
		f_1(y), \quad & x^{(1)} <  -0.2, \\
		f_2(y), \quad & x^{(1)} \ge  -0.2, x^{(2)} \ge 0, \\
		f_3(y), \quad & x^{(1)} \ge -0.2, x^{(2)} < 0, \\		
	\end{cases}
\end{align}
with three different local densities $f_l(x)$, $1 \le l \le 3$, varying in variance, number of modes, and skewness. The response distribution is determined by the first two covariates and independent of the rest. In Figure~\ref{fig:LinCDETreeExample}, we plot the average conditional density estimates in the three subregions. LinCDE trees are able to distinguish the densities differing in the above characteristic properties and produce good fits. We also compute the normalized importance score---the proportion of overall improvement in the split-criterion attributed to each splitting variable. In all settings, the first two covariates contribute over $99\%$ importance. In other words, LinCDE trees focus on the first two influential covariates and avoid splitting at nuisance covariates. 

\section{LinCDE Boosting}\label{sec:LinCDEBoosting}

Although LinCDE trees are useful as stand-alone tools, our ultimate
goal is to use them as \emph{weak learners} in a boosting paradigm.
Standard tree boosting \citep{friedman2001greedy} builds an additive
model of shallow trees in a forward stagewise manner. Though a single
shallow tree is high in bias, tree boosting manages to reduce the bias
by successively making small modifications to the current
estimate.\footnote{We remark that another successful ensemble method
 ---random forests \citep{breiman2001random}---are not appropriate for LinCDE trees. Random forests construct a large number of trees with low correlation and average the predictions. Deep trees are grown to ensure low-bias estimates, which is, however, unsatisfactory here because deep LinCDE trees will have leaves with too few observations for density estimation.}

We proceed with the boosting idea and propose LinCDE
boosting. Starting from a null estimate, we iteratively modify the
current estimate by modifying the natural parameter functions via a LinCDE tree. In particular, at the $t$-th iteration,  
\begin{align}
\label{eq:subroutine}
\begin{split}
\gamma^t(x) &= \argmax_{\text{LinCDE tree }\gamma(x)} \ell(\calR_0; \beta^{t}(x) + \gamma(x)), \\
\beta^{t+1}(x) &\leftarrow \beta^{t}(x) + \gamma^t(x).
\end{split}
\end{align}
Section~\ref{subsec:additiveModel} gives details.
We remark that the LinCDE tree modifier $\gamma^t(x)$ for boosting is
an expanded version of that in Section~\ref{sec:LinCDETree}: in
previous LinCDE trees, all samples share the same carrying density
$\kappa(y)$, while in LinCDE trees for boosting, the carrying densities $\kappa(y) e^{z(y)^\top\beta^{t}(x) - \psi(\beta^{t}(x))}$ differ across units. We elaborate on LinCDE trees with heterogeneous carrying densities in Sections~\ref{subsec:fitting} and \ref{subsec:splitting}.

\begin{figure}[bt]
	\centering
	\begin{minipage}{16cm}
		\centering  
		\includegraphics[scale = 0.6]{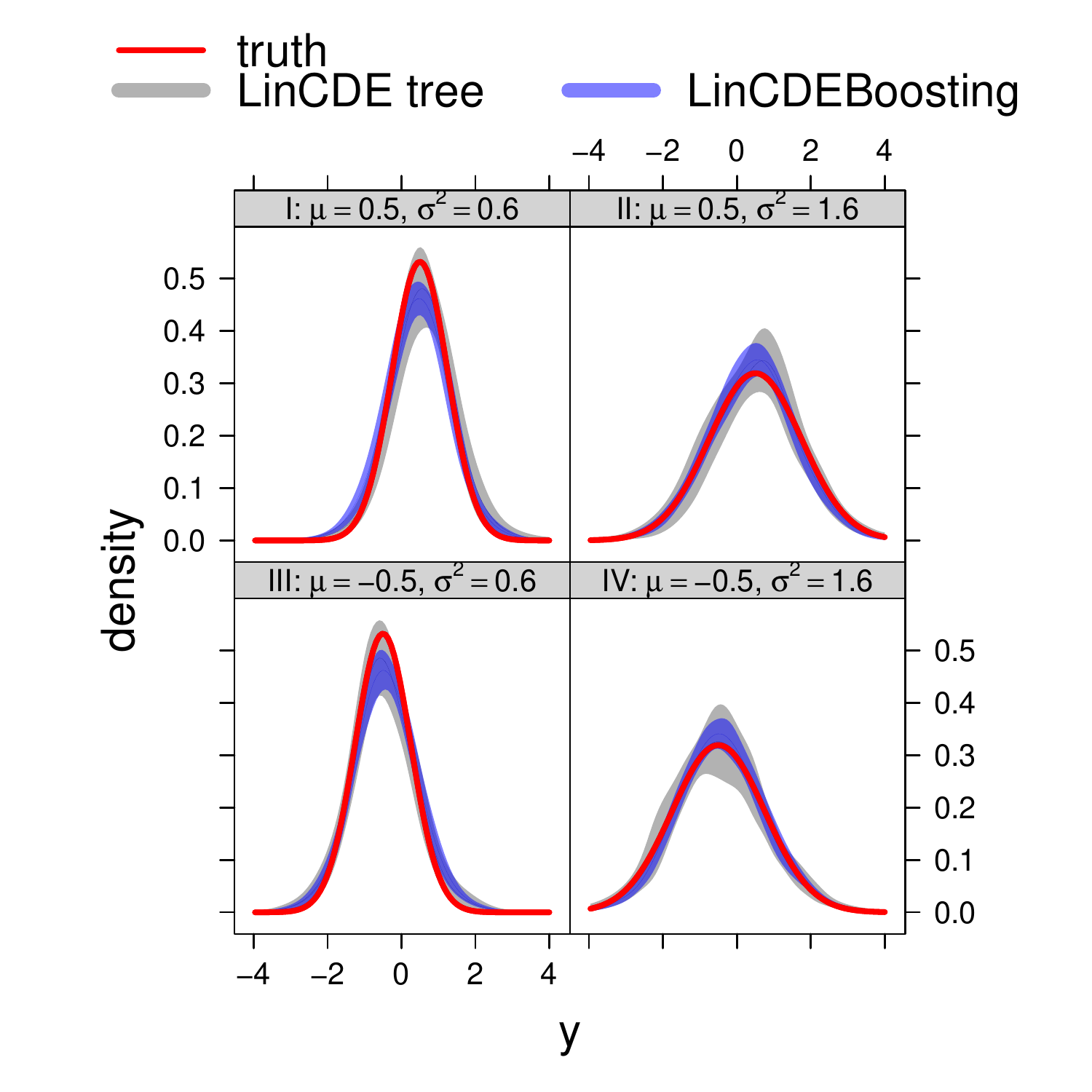}  
	\end{minipage}
	\caption{{Comparison of LinCDE trees and LinCDE boosting. The responses are generated from \eqref{eq:LinCDEBoostingExample}. We pick $4$ landmarks corresponding to different conditional means and variances. From top to bottom, the conditional means decrease from $0.5$ to $-0.5$. From left to right, the conditional variances increase from $0.6$ to $1.6$. In each trial, we sample $400$ observations from the target distribution. We repeat each setting $100$ times, and plot the average estimated conditional densities plus and minus one standard deviation against the true densities.}}
	\label{fig:LinCDEBoostingExample}
\end{figure}

Before discussing the details of LinCDE boosting, we compare LinCDE boosting and LinCDE trees on a toy example in Figure~\ref{fig:LinCDEBoostingExample}. We consider a locally Gaussian distribution with heterogeneous mean and variance
\begin{align}\label{eq:LinCDEBoostingExample}
y = x^{(1)} + 0.5 x^{(2)} \varepsilon, \quad \varepsilon \sim \calN(0,1).
\end{align}
The covariate generating mechanism is the same as in Figure~\ref{fig:LinCDETreeExample}. We plot the average estimated conditional densities plus and minus one standard deviation.
Though both LinCDE trees and LinCDE boosting produce good fits in all settings, the estimation bands of LinCDE boosting are always narrower by around a half. The observation implies LinCDE boosting is more stable than LinCDE trees.

\subsection{Additive Model in the Natural Parameter Scale}\label{subsec:additiveModel}
In LinCDE boosting, we build an additive model in the natural parameter scale of the density \eqref{eq:exponentialFamily2}. We find a sequence of LinCDE tree-based learners with parameter functions $\{\gamma^t(x)\}_{0 \le t \le T-1}$, and aggregate those ``basis" functions to obtain the final estimate\footnote{To stabilize the performance, we may shrink $\gamma^t(x)$ by some learning rate $\eta \in (0,1]$, and let $\beta^T(x) = \sum_{t=0}^{T-1} \eta \gamma^t(x)$.}
\begin{align}\label{eq:tilt0}
	\beta^T(x) = \sum_{t=0}^{T-1} \gamma^t(x).
\end{align}
In other words, at the $t$-th iteration, we tilt the current
conditional density estimate 
\begin{align}\label{eq:tilt}
	\begin{split}
	f_{y|x}^{t+1}(y \mid x) 
	= f_{y|x}^{t}(y \mid x) \cdot e^{z(y)^\top \gamma^{t}(x) - \phi_{\beta^t(x)}(\gamma^{t}(x))}
	\end{split} 
\end{align}
based on knowledge $\gamma^t(x)$ learned by the new tree.
Here $\phi_{\beta^t(x)}(\gamma^{t}(x)) = \psi(\beta^{t}(x) +
\gamma^{t}(x)) - \psi(\beta^{t}(x))$ is the updated normalizing function (depending on $x$).  Appendix~\ref{appn:sec:approxError} shows that if the true conditional density is smooth, the approximation error of LinCDE boosting's function class~\eqref{eq:tilt} with splines $z(y)$ will vanish as the number of  splines $k$ increases.

We determine the LinCDE tree modifiers in \eqref{eq:tilt0} by log-likelihood maximization. We aim to find the modifier that produces the largest improvement in the objective $\ell(\calR_0; \beta(x) + \gamma(x))$ defined as
\begin{align}\label{eq:loglikelihoodPenalty7}
\begin{split}
\sum_{i=1}^n \left(\log(f_{y|x}^{t}(y_i \mid x_i)) + z(y_i)^\top \gamma(x_i) - \phi_{\beta^t(x_i)}(\gamma(x_i))\right) + \lambda \sum_{i=1}^n \sum_{j=1}^k  \omega_j \gamma_j^2(x_i).
\end{split}
\end{align}
Compared to the objective \eqref{eq:loglikelihoodPenalty4} of LinCDE trees, the only difference in \eqref{eq:loglikelihoodPenalty7} is the normalizing function $\phi_{\beta^t(x)}(\gamma^{t}(x))$. When $\beta^t(x)$ is a constant function, the normalizing functions of LinCDE trees and LinCDE boosting coincide. In Section~\ref{subsec:fitting} and \ref{subsec:splitting}, we demonstrate how LinCDE boosting's heterogeneous normalizing function complicates the fitting and splitting steps and propose corresponding solutions.

\subsection{Fitting Step}\label{subsec:fitting}
 
For fitting, given a subregion $\calR$,
the problem \eqref{eq:loglikelihoodPenalty7} can not be solved by Lindsey's method as in LinCDE trees, because $\beta^{t}(x)$ could be non-constant in the subregion $\calR$. Explicitly, instead of the single constraint in \eqref{eq:loglikelihood}, we could have up to $n_\calR$ constraints
\begin{align}\label{eq:constratint2}
	\int \kappa(y) e^{z(y)^\top (\beta^t(x_i) + \gamma^t(x_i)) + \beta_0(x_i)} dy = 1, \quad x_i \in \calR.
\end{align}
As a result, the Lagrangian function \eqref{eq:loglikelihoodPenalty2} as well as subsequent discrete approximations for Lindsey's method are invalid.

Fortunately, we can solve the fitting problem iteratively (\textit{Fitting (LinCDE boosting)} below). Define bin probabilities
\begin{align}\label{eq:cellProb}
\begin{split}
{p}_b(\beta^t(x)) &:= \frac{\kappa(y_b) e^{z^\top(y_b) \beta^t(x) }}{\sum_{b'=1}^B \kappa(y_{b'}) e^{z^\top(y_{b'}) \beta^t(x) }}, \\
\bar{p}_b(\calR; \beta^t(x)) &:=  \frac{1}{n_{\calR}} \sum_{x_i \in \calR}  \bar{p}_b(\beta^t(x_i)).
\end{split}
\end{align} 
We feed the marginal cell probabilities $\bar{p}_b(\calR; \beta^t(x))$
to the fitting step as the baseline for modification. In Step $1$,
Lindsey's method produces a natural parameter modifier and a universal
intercept for all samples in $\calR$. The intercept produced by
Lindsey's method guarantees that the marginal cell probabilities to
sum to unity, but not for every individual $x_i$. In Step $2$, we update the individual normalizing constants to ensure all constraints \eqref{eq:constratint2} are satisfied. In Proposition~\ref{prop:iterative}, we show that the fitting step of LinCDE boosting converges to the maximizer of the objective \eqref{eq:loglikelihoodPenalty7}.

\begin{itemize}
	\item \textit{Fitting (LinCDE boosting)}. In a region $\calR$, initialize $\gamma = 0 \in \RR^k$, $\gamma_0 = 0 \in \RR^{n_{\calR}}$. Count the number of observations $\{n_{\calR,b}\}$ in each bin. 
	\begin{enumerate}
		\item \textit{Updating $\gamma$}. Compute $\bar{p}_b(\calR; \beta^t(x)+\gamma)$ in \eqref{eq:cellProb},
		fit a Poisson regression model with the response variable $\{n_{\calR,b}\}$, regressors $z(y_b)$, the offset $\log(\bar{p}_b(\calR; \beta^t(x) + \gamma))$, and a weighted ridge penalty. 	
		Denote the estimated coefficients by $\Delta \gamma$. Update $\gamma \leftarrow \gamma + \Delta \gamma$. 
		\item \textit{Updating $\gamma_0$ (normalization)}. Compute the normalizing constants for all samples in $\calR$
		\begin{align*}
			\gamma_{0,i} = -\log\left(\sum_{b} p_b(\beta^t(x_i)) e^{z(y_b)^\top \gamma }\right).
		\end{align*}
		\item Repeat steps 1 and 2 until $\|\Delta \gamma\|_2 \le \varepsilon$. Output $\gamma$, $\gamma_0$.
	\end{enumerate}
\end{itemize}

\begin{proposition}\label{prop:iterative}
	Assume that $\lambda = 0$ and $Y$ is supported on the midpoints $\{y_b\}$, then the fitting step of LinCDE boosting converges, and the output $\gamma_{\calR}^t$ satisfies
	\begin{align*}
	\gamma_{\calR}^t = \argmax_{{\gamma}} \ell(\calR; \beta^t(x) + {\gamma}).
	\end{align*} 
\end{proposition}
We offer some intuition behind Proposition~\ref{prop:iterative}; i.e., why the fitting step of LinCDE boosting will stop at the likelihood maximizer. If $\beta^t(x) + \gamma$ is already optimal, then the average sufficient statistics $z(y)$ under marginal probabilities \eqref{eq:cellProb} should match the observations, which yields the KKT condition of the Poisson regression in Lindsey's method. As a result, Lindsey's method will produce zero updates, and the algorithm converges.

\subsection{Splitting Step}\label{subsec:splitting}

Reminiscent of the splitting step for  LinCDE trees, we seek the split that produces the largest improvement in the objective \eqref{eq:loglikelihoodPenalty7}. Proposition~\ref{prop:approx} is not valid due to the heterogeneity in $\beta^t(x)$, and we propose an expanded version.
\begin{proposition}\label{prop:approx2}
	In a region $\calR$, let $n_{\calR}$ be the sample size and $\gamma^t_{\calR}$ be the optimal update. Define the average sufficient statistics residuals as
	\begin{align*}
	\bar{r}^t_{\calR} := \frac{1}{n_\calR} \sum_{x_i \in \calR} \left(z_i - \nabla\psi(\beta^t(x_i) + \gamma_{\calR}^t)\right).
	\end{align*}
	Given a candidate split $s$, define
	\begin{align*}
	\Psi_{s}^t(\gamma^t_\calR) :=& \frac{n_{\calR_{s,R}}}{n_{\calR}} \left(\frac{1}{n_{\calR_{s,L}}} \sum_{x_i \in \calR_{s,L}} \nabla^2\psi\left(\beta^t(x_i) + \gamma^t_\calR\right) \right)^{-1} \\
	&+\frac{n_{\calR_{s,L}}}{n_{\calR}} \left(\frac{1}{n_{\calR_{s,R}}} \sum_{x_i \in \calR_{s,R}} \nabla^2\psi\left(\beta^t(x_i) + \gamma^t_\calR\right) \right)^{-1}.
	\end{align*}	
	Then the improvement of the unpenalized conditional log-likelihood satisfies
	\begin{align*}
	\frac{1}{n_{\calR}}\Delta \ell^t(\calR, s)
	= \frac{n_{\calR_{s,L}} n_{\calR_{s,R}}}{2 n_{\calR}^2} \left(\bar{r}^t_{\calR_{s,L}} - 
	\bar{r}^t_{\calR_{s,R}}\right)^\top \Psi_s^t(\gamma^t_\calR) \left(\bar{r}^t_{\calR_{s,L}} - \bar{r}^t_{\calR_{s,R}}\right) + r_s,
	\end{align*} 	
	where $r_s = O(\|\bar{r}^t_{\calR_{s,L}} - \bar{r}^t_{\calR}\|_2^3 + \|\bar{r}^t_{\calR_{s,R}}- \bar{r}^t_{\calR}\|_2^3)$.
\end{proposition}

The average sufficient-statistic residuals $\bar{r}^t_{\calR}$
measures the deviation of the current estimator from the
observations. By maximizing the quadratic approximation in
Proposition~\ref{prop:approx2}, we find the candidate split $s$ such
that $\bar{r}^t_{\calR_{s,L}}$ and $\bar{r}^t_{\calR_{s,R}}$ are far
apart, and modify the current estimator differently in the left and
right children determined by the selected split. The updated splitting procedure is summarized in \textit{Splitting (LinCDE boosting)}.

\begin{itemize}	
	\item \textit{Splitting (LinCDE boosting)}.  In a region $\calR$:
	\begin{enumerate}
		\item Compute $\{n_{\calR_{s,L}}, n_{\calR_{s,R}},
                  \bar{z}_{\calR_{s,L}}, \bar{z}_{\calR_{s,R}}\}$ for
                  each candidate split $s$, and approximate
                  $\tilde{\Psi}^t(\gamma^t_\calR)$ in \eqref{eq:Psi}
                  by the average covariance matrix of $z(y)$ in $\calR$.
		\item For each candidate split $s\in \calS$, compute the quadratic approximation $\widehat{\Delta} \ell(\calR, s)$ by Proposition~\ref{prop:approx2}, and choose the split $s_{\max} = \argmax_{s \in \calS} \hat{\Delta} \ell(\calR, s)$.  
	\end{enumerate}	
\end{itemize}

The computation of the quadratic approximation is largely the same, except that the normalization matrix $\Psi_{s}^t(\gamma^t_\calR)$ varies across candidate splits and requires separate computation. To relieve the computational burden, we propose the following surrogate independent of candidate splits\footnote{In practice, we add a universal diagonal matrix to $\frac{1}{n_{\calR}} \sum_{x_i \in \calR} \nabla^2\psi(\beta(x_i) + \gamma^t_\calR)$ to stabilize the matrix inversion.}
\begin{align}\label{eq:Psi}
\tilde{\Psi}^t(\gamma^t_\calR) = \left(\frac{1}{n_{\calR}} \sum_{x_i \in \calR} \nabla^2\psi\left(\beta(x_i) + \gamma^t_\calR\right) \right)^{-1}.
\end{align}
The surrogate $\tilde{\Psi}^t(\gamma^t_\calR)$ coincides with $\Psi_s^t(\gamma^t_\calR)$ if $\beta^t(x)$ is a constant vector, or the normalizing function $\psi(\beta)$ is quadratic.

Proposition~\ref{prop:computation2} gives the computational time complexity of the splitting procedure (\textit{Splitting (LinCDE boosting)}). The computation time scales linearly with regard to the sample size multiplied by dimension and the number of candidate splits. The extra computation compared to LinCDE trees comes from residual calculations and individual normalizations.
\begin{proposition}\label{prop:computation2}	
	Assume that there are $S$ candidate splits, then the splitting
        step for LinCDE boosting is of computational time
        complexity $\tilde{O}(dn_{\calR}kB + n_{\calR}k^2B + k^3 + Sk^2)$.
\end{proposition}

\begin{algorithm}
	\DontPrintSemicolon  
	\SetAlgoLined
	\BlankLine
	\caption{LinCDE boosting}\label{algo:LinCDESub}
	Initialize the natural parameter function $\beta^0(x)$.\footnotemark \\
	\For{t = 1:T}{
		1. Apply Algorithm~\ref{algo:LinCDETree} with \textit{Fitting (LinCDE boosting)}, \textit{Splitting (LinCDE boosting)}, and obtain the optimal LinCDE tree modifier $\hat{\gamma}^{t-1}(x)$.\\
		2. Update 
		\begin{align*}
			\hat{\beta}^{t}(x) \leftarrow \hat{\beta}^{t-1}(x) + \hat{\gamma}^{t-1}(x).
		\end{align*}
	}
	Output $\hat{\beta}^T(x)$.
\end{algorithm}
\footnotetext{The initialization $\beta^0(x)$ is usually a constant vector, e.g., zero vector,  independent of the covariates.}

\section{Pretreatment}\label{sec:pretreatment}
In this section, we discuss two pretreatments: response transformation and centering. The pretreatments are helpful when the response is heavy-tailed and when the conditional distributions $f_{y|x}(y \mid x)$ vary wildly in location.

\subsection{Response Transformation}
Heavy-tailed response distributions are common in practice, such as income and waiting time. If the response is heavy-tailed, then in Lindsey's method, 
most bins will be approximately empty. As a result the model tends to
be over-parameterized and the estimates tend to overfit.

\begin{figure}[bt]
	\centering
	\begin{minipage}{5cm}
		\centering  
		\includegraphics[clip, trim = 0.1cm 0cm 0cm 0cm, scale = 0.62]{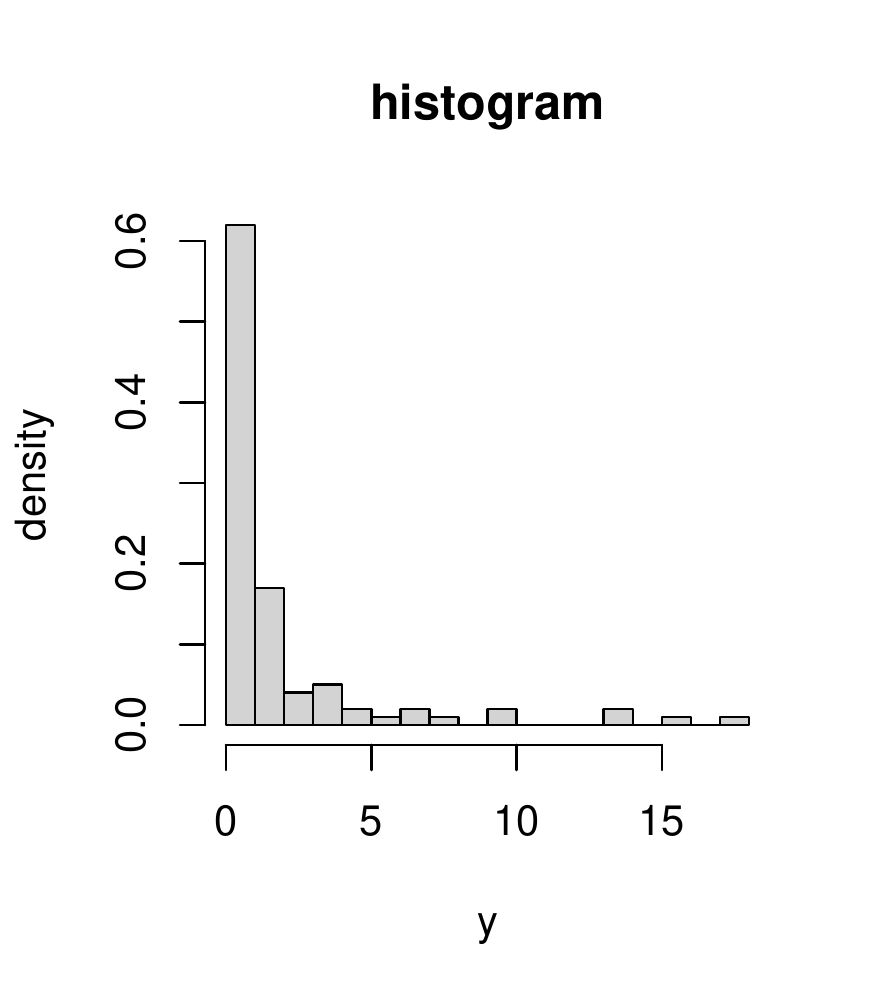}  
	\end{minipage}
	\begin{minipage}{5cm}
		\centering  
		\includegraphics[clip, trim = 0.1cm 0cm 0cm 0cm,scale = 0.62]{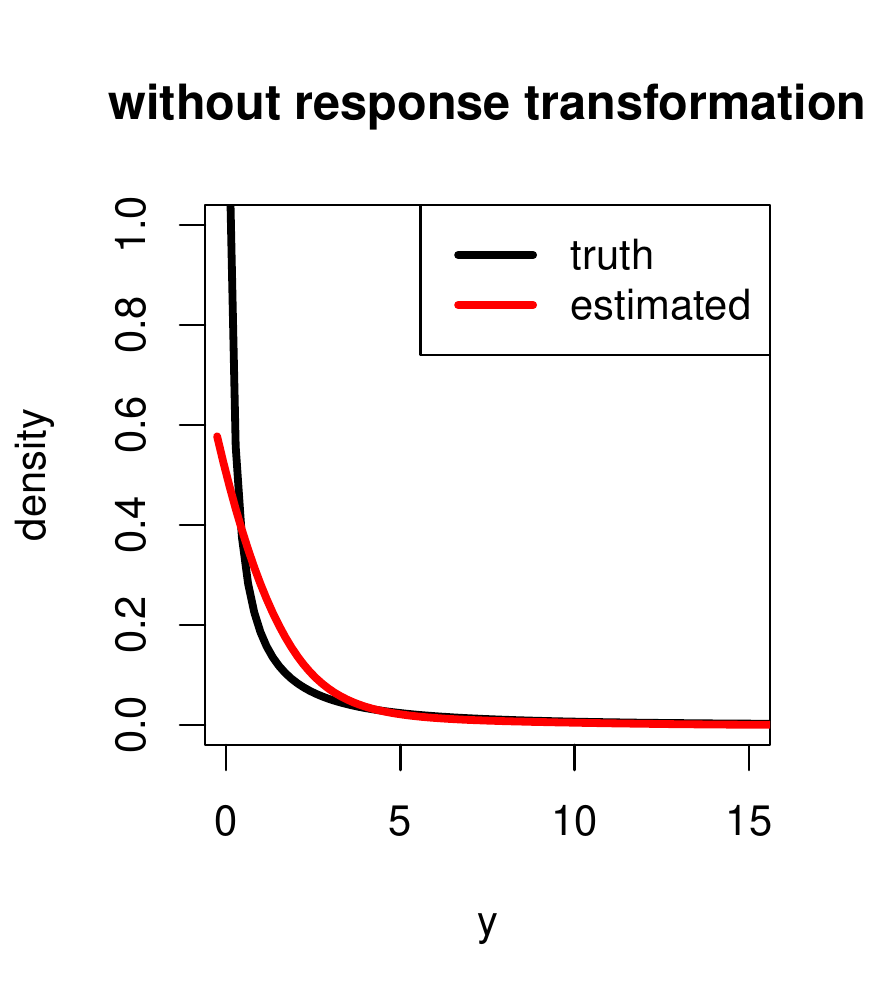}  
	\end{minipage}
	\begin{minipage}{5cm}
		\centering  
		\includegraphics[clip, trim = 0.1cm 0cm 0cm 0cm,scale = 0.62]{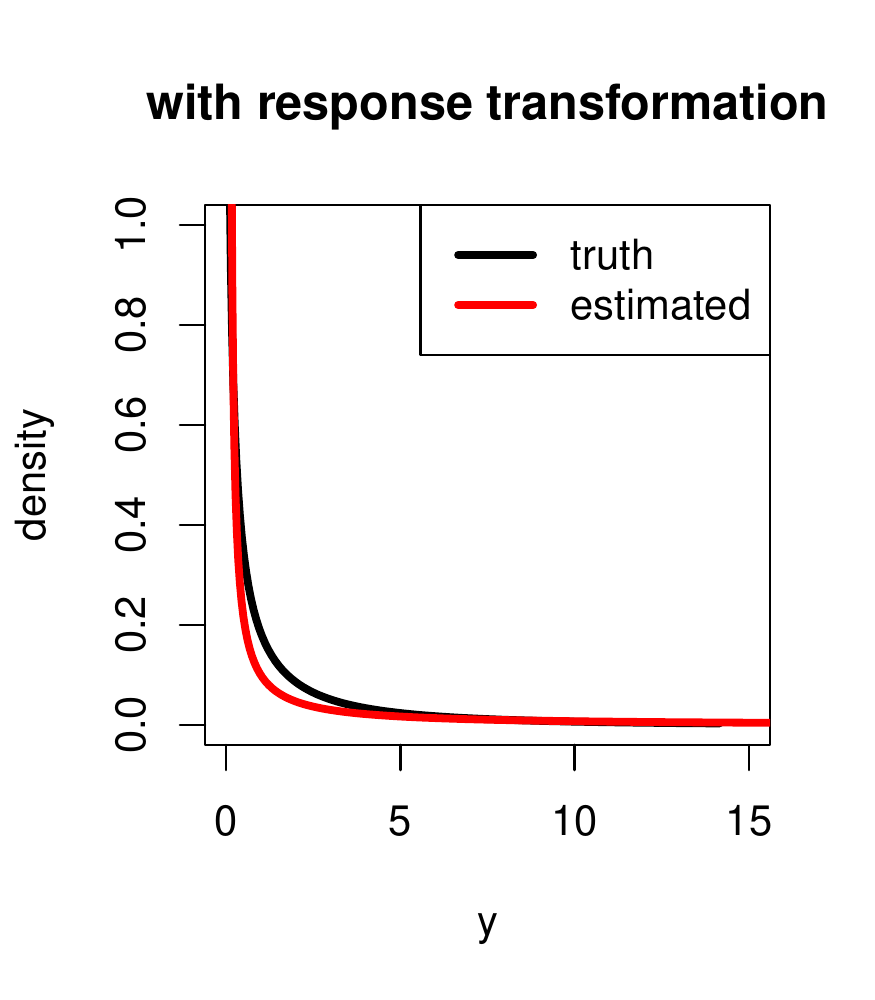}  
	\end{minipage}
	\caption{{Response transformation and Lindsey's method. We generate $100$ responses $y = w^2$, $w \sim$ exp(1), and apply Lindsey's method. The left panel plots the histogram of the responses, and we observe several extreme values ($\ge 10$). The middle panel plots the true density and Lindsey's estimate. Due to a limited number ($20$) of bins and a wide response range caused by outliers, Lindsey's estimate is inaccurate (not sharp enough) around zero. The right panel demonstrates Lindsey's estimate using log-transformed responses. The log-transformation helps the estimation around zero and does not sacrifice the fit at the tail.}}
	\label{fig:outlier}
\end{figure}

In response to the heavy-tailed responses, we recommend transforming
the response first to be marginally normally distributed. Often log and cube-root transformations are useful. In a more principled way, we can apply the Box-Cox transformation to the responses and choose the optimal power parameter to produce the best approximation of a Gaussian distribution curve. Once the model is fit to the transformed data, we map the estimated conditional densities of the transformed responses back to those of the original observations. See Figure~\ref{fig:outlier} for an example.

\subsection{Centering}

\begin{figure}[bt]
	\centering
	\begin{minipage}{16cm}
		\centering  
		\includegraphics[scale = 0.6]{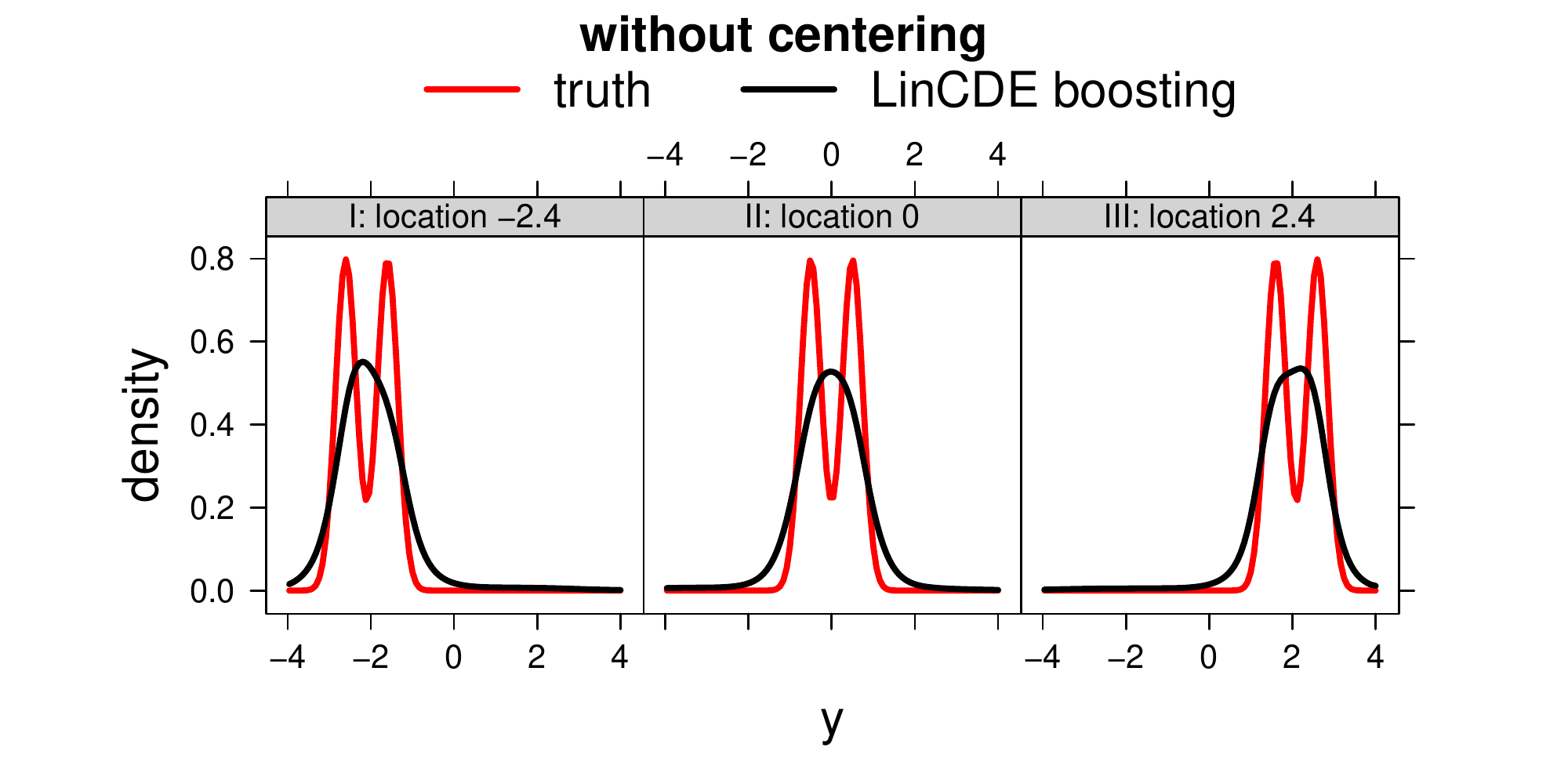}  
	\end{minipage}
	\begin{minipage}{16cm}
		\centering  
		\includegraphics[scale = 0.6]{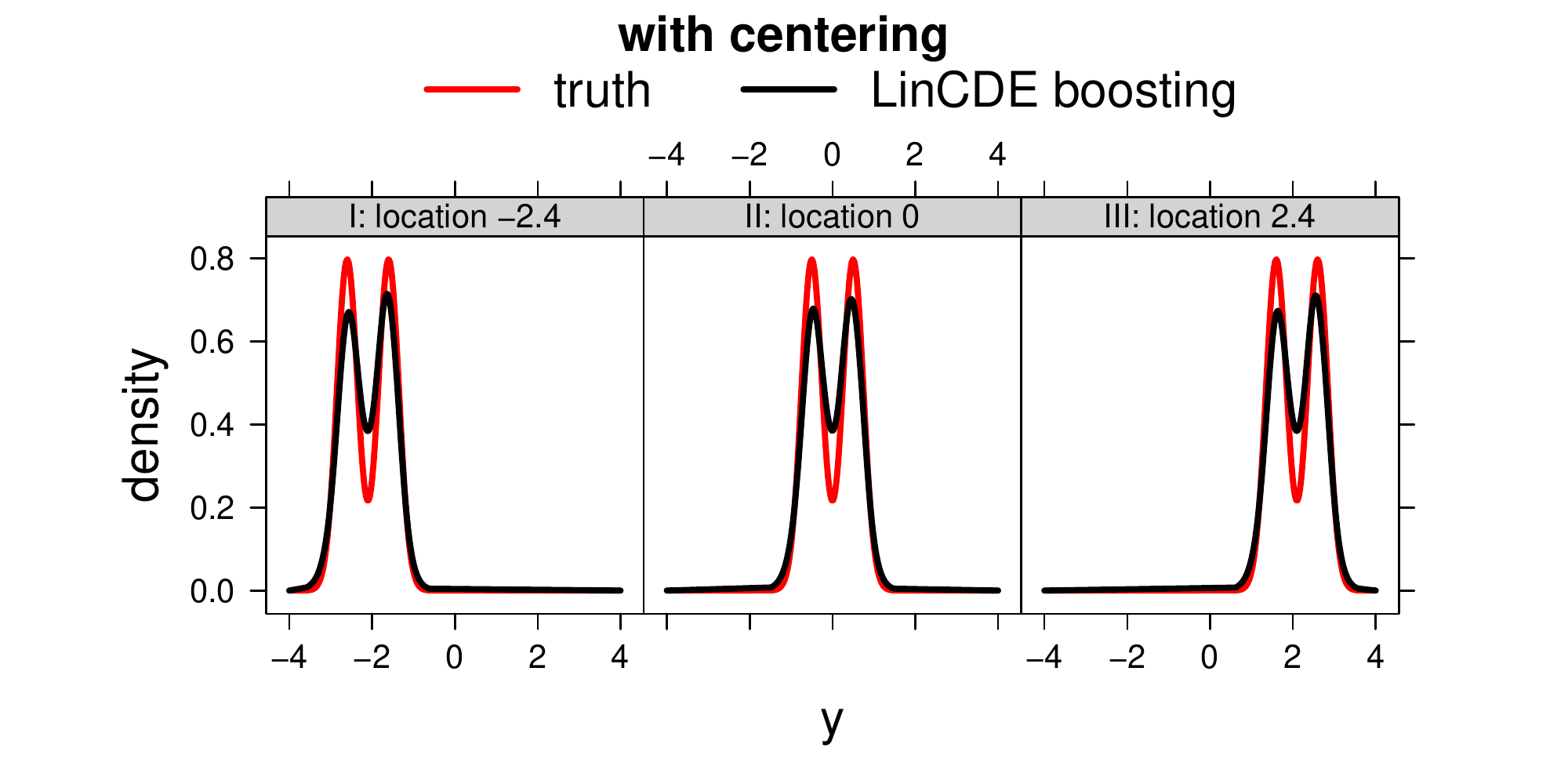}  
	\end{minipage}
	\caption{{Conditional density estimation with and without centering. We consider the conditional density \eqref{eq:meanAugmentationExample} and pick $3$ landmarks corresponding to different locations. The first row plots LinCDE boosting's estimates without centering, and the second row plots the estimates augmented with true means. In each trial, we sample $1000$ observations from the target distribution. We repeat each setting $100$ times, and plot the average estimated conditional densities. In both settings, LinCDE boosting uses $k=10$ sufficient statistics and $20$ response bins. }}
	\label{fig:meanAugmentation}
\end{figure}

For a distribution whose conditional components differ wildly in location, LinCDE needs a large number of sufficient statistics to capture local distributional characteristics. For instance, Figure~\ref{fig:meanAugmentation} displays a conditional Gaussian mixture with location shift
\begin{align}\label{eq:meanAugmentationExample}
	\begin{split}
		&y = 3 x^{(1)} + w z^{(1)} + (1-w) z^{(2)}, \\ 
		\qquad w \sim \text{Ber}(0.5), \quad & z^{(1)} \sim \calN(-0.5, 0.06), \quad z^{(2)} \sim \calN(0.5, 0.06), \quad z^{(1)} \independent z^{(2)}.
	\end{split}
\end{align}
When we apply LinCDE boosting with $k=10$ sufficient statistics, the estimates do not reproduce the bimodalities due to a lack of flexibility. We call this the ``disjoint support'' problem. 

A straightforward solution to the disjoint support problem is to
increase the number $k$ so that the sufficient statistics $z(y)$ are
adequately expressive. As a consequence, the number of components in
the parameter function $\beta(x)$ goes up. This approach is prone to
overfitting, especially when there are a small number ($\sim 20$)
samples in a terminal node. In addition, this approach will
significantly slow down the splitting procedure, which scales $O(k^3)$
by Proposition~\ref{prop:computation2}.

Our solution is to \emph{center} the response prior to fitting the LinCDE
model. Since the difference in location causes the disjoint support
problem, we suggest aligning the centers of the conditional densities
in advance. Explicitly, we first estimate the locations via some
conditional mean estimator and then subtract the estimates from the
responses. The support of the residuals are less heterogeneous, and we
apply LinCDE boosting to these residuals to capture additional
distributional structures. Finally, we transform the resulting density
estimates back to those of the responses. The procedure is summarized
in Algorithm~\ref{algo:meanAugmentation}.

\begin{algorithm}[h!]
	\DontPrintSemicolon  
	\SetAlgoLined
	\BlankLine
	\caption{Centering}\label{algo:meanAugmentation}
	1. Estimate the conditional mean $\hat{h}(x)$ using the
        training data $\{(x_i, y_i)\}$. Compute the residuals $r_i = y_i - \hat{h}(x_i)$.\\
	2. Apply LinCDE boosting to $\{(x_i, r_i)\}$, and obtain $\hat{f}_{R|X}(r \mid x)$.\\
	3. Define $\hat{f}_{Y|X}(y \mid x) = \hat{f}_{R|X}(y - \hat{h}(x) \mid x)$ and output $\hat{f}_{Y|X}$.
\end{algorithm}

Centering splits the task of conditional distribution estimation into
conditional mean estimation and distributional property
estimation. For centering we have available a variety of popular
conditional mean estimators, such as the standard random forest,
boosting, and neural networks. Once the data are centered, LinCDE
boosting has a more manageable task. Figure~\ref{fig:meanAugmentation} shows that with centering, LinCDE boosting is able to reproduce the bimodal structure in the above example with the same set of sufficient statistics.

\section{Simulation}\label{sec:simulation} 

In this section, we demonstrate the efficacy of LinCDE boosting on simulated examples. 

\subsection{Data and Methods}
Consider $d=20$ covariates randomly generated from uniform $[-1,1]$. The responses given the covariates are sampled from the following distributions:
\begin{itemize}
	\item \textit{Locally Gaussian distribution (LGD)}: 
	\begin{align*}
		Y \mid X = x ~\sim~ \calN\left(0.5 x^{(1)} + x^{(1)} x^{(2)}, ~\left(0.5 + 0.25 x^{(2)}\right)^2\right).
	\end{align*}
	At a covariate configuration, the response is Gaussian with the mean determined by $x^{(1)}$ and $x^{(2)}$, and the variance determined by $x^{(2)}$. Covariates $x^{(3)}$ to $x^{(20)}$ are nuisances variables; 
	\item \textit{Locally Gaussian or Gaussian mixture distribution (LGGMD)}: 
	\begin{align*}
		Y \mid X = x ~\sim ~
		\begin{cases}	
			\!\begin{aligned}
				& 0.5 \calN \left(\mu(x^{(1)}) - 0.5, ~\sigma_{+}^2(x^{(3)})\right) \\
				& + 0.5 \calN \left(\mu(x^{(1)}) + 0.5, ~\sigma_{-}^2(x^{(3)})\right),
			\end{aligned} 
		&x^{(2)} \le 0.2, \\
		\calN\left(\mu(x^{(1)}), ~\sigma^2\right),	&x^{(2)} > 0.2,
		\end{cases}
	\end{align*}
	where the means and variances are
	\begin{align*}
		&\mu\left(x^{(1)}\right) = 0.25x^{(1)}, ~ \sigma^2 = 0.3,~\\
		&\sigma_{+}^2\left(x^{(3)}\right) = 0.25\left(0.25 x^{(3)} + 0.5\right)^2,~\\
		&\sigma_{-}^2\left(x^{(3)}\right) = 0.25\left(0.25 x^{(3)} - 0.5\right)^2.		
	\end{align*}
	The mean is determined by $x^{(1)}$. The modality depends on
        $x^{(2)}$: in the subregion $x^{(2)} \ge 0.2$, the response
        follows a bimodal Gaussian mixture distribution, while in the
        complementary subregion, the response follows a unimodal
        Gaussian distribution. The skewness or symmetry is controlled
        by $x^{(3)}$ in the Gaussian mixture subregion: larger
        absolute values of $x^{(3)}$ imply higher asymmetry. Overall,
        the conditional distribution has location, shape, and symmetry
        dependent on the first three covariates. Covariates $x^{(4)}$
        to $x^{(20)}$ are nuisance variables.
\end{itemize}
The training data set consists of $1000$ i.i.d. samples. The performance is evaluated on an independent test data set of size $1000$.

We compare LinCDE boosting with quantile regression forest and distribution boosting.\footnote{LinCDE boosting: \url{https://github.com/ZijunGao/LinCDE}. Quantile regression forest: R package \textit{quantregForest} \citep{quantregForestPackage}. Distribution boosting: R package \textit{conTree} \citep{conTreePackage}.}
There are a number of tuning parameters in LinCDE boosting. The
primary parameter is the number of trees (iteration number). Secondary
tuning parameters include the tree size, the learning rate, and the ridge penalty parameter. On a separate validation data set, we experimented with a grid of secondary parameters, each associated with a sequence of iteration numbers, and select the best-performing configuration. 
By default, we use $k=10$ transformed natural cubic splines and a
Gaussian carrying density We use a small learning rate $\eta=0.01$ to avoid overfitting. We use $40$ discretization bins for training, and $20$ or $50$ for testing. The simulation examples do not have heavy-tail or disjoint support issues, and thus no pretreatments are needed.

\subsection{Results of Conditional Density Estimation}
Let the oracle be provided with the true density, and the null method estimates a marginal Gaussian distribution.
We consider the following metric
\begin{align}\label{eq:R2}
	\frac{\ell_{\text{method}} - \ell_{\text{null}}}{\ell_{\text{oracle}} - \ell_{\text{null}}},
\end{align}
where $\ell_{\cdot}$ denotes the test conditional log-likelihood of a specific method.
The criterion is analogous to the goodness-of-fit measure $R^2$ of
linear regression. It measures the performance of the method relative
to the oracle; larger values indicate better fits, and the ideal value is one.

Quantile regression forests and distribution boosting estimate conditional quantiles instead of densities. To convert the quantile estimates to density estimates, we define a grid of bins with endpoints $y_{b,L}$ and $y_{b,R}$, and approximate the density in bin $b$ by
\begin{align}\label{eq:quantileToPdf2} 
	\hat{f}_b = \frac{\hat{q}^{-1}(y_{b,R}) - \hat{q}^{-1}(y_{b,L})}{y_{b,R} - y_{b,L}},
\end{align}
where $\hat{q}^{-1}(y)$ represents the inverse function of the quantile estimates. As the bin width shrinks, $\hat{f}_b$ is less biased but of larger variance.
In simulations, we display the results with $20$ bins and $50$ bins (Appendix~\ref{appn:sec:figure}). We observe that LinCDE boosting is robust to the bin size, while distribution boosting and quantile regression forests prefer $20$ bins due to the smaller variances.

\begin{figure}[bt]
    \centering
    \begin{minipage}{7cm}
    \centering  
\includegraphics[width  = 7cm, height = 7cm]{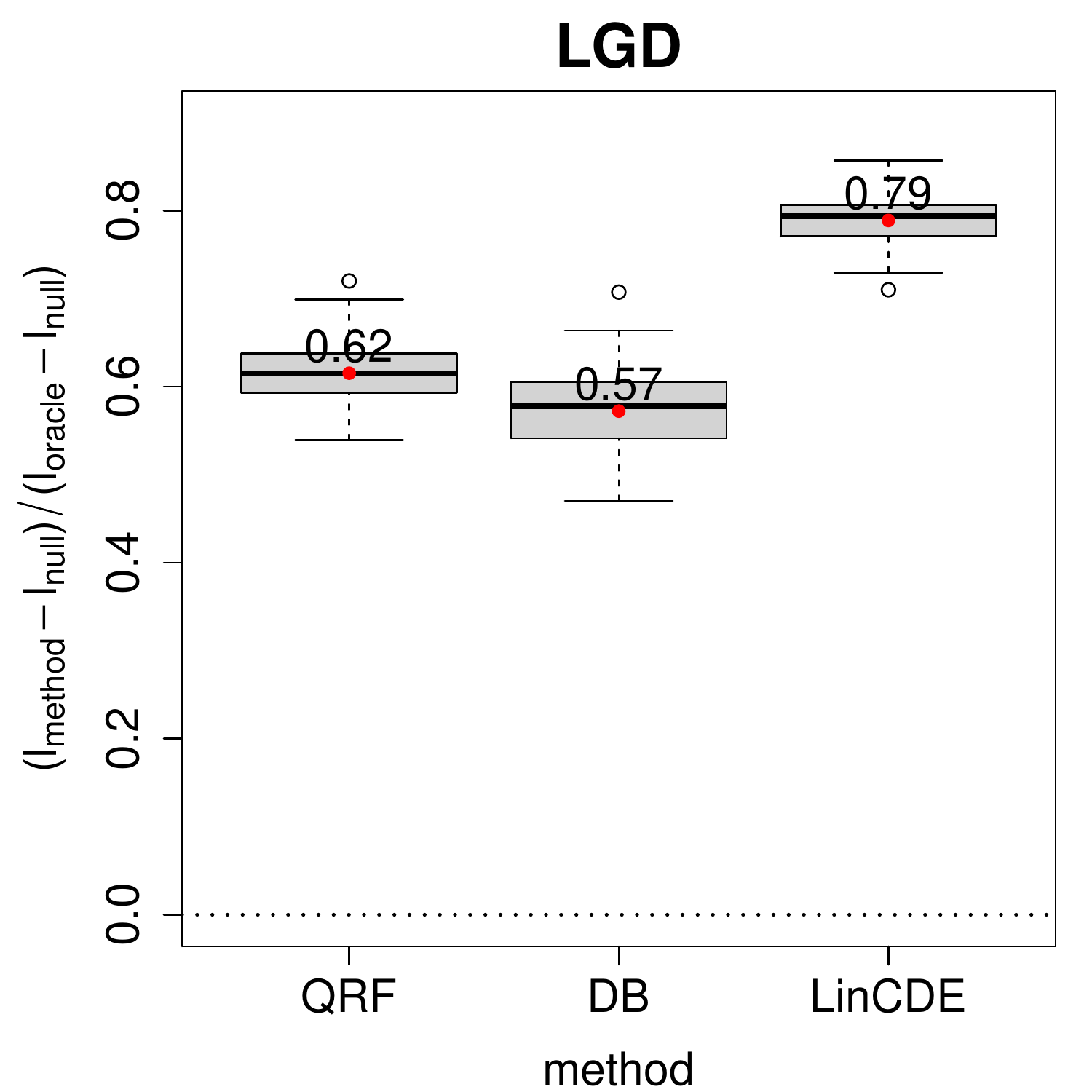} 
    \end{minipage}
    \begin{minipage}{7cm}
    \centering  
\includegraphics[width  = 7cm, height = 7cm]{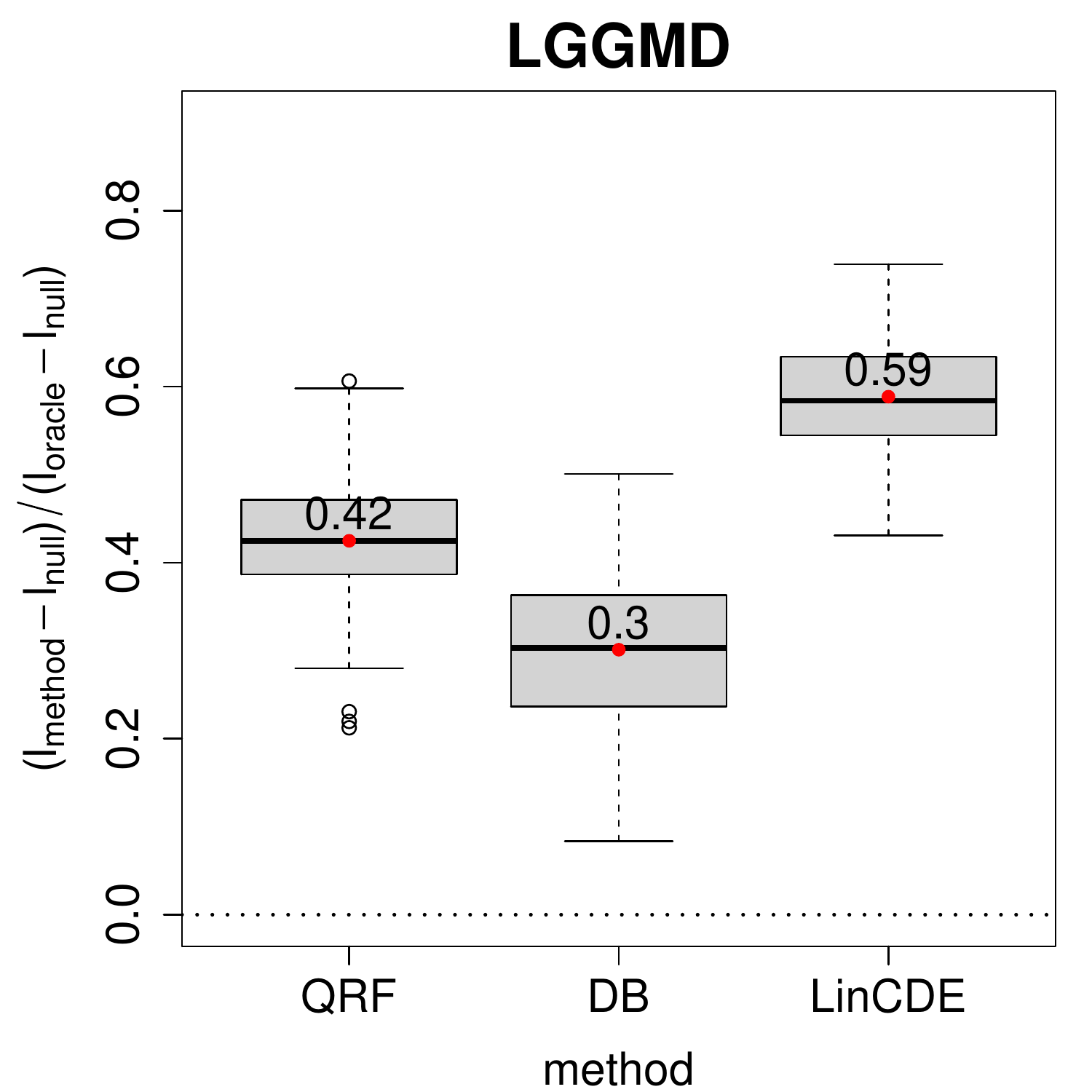}  
    \end{minipage}
    \caption{{Box plots of goodness-of-fit measures
        \eqref{eq:R2} in the setting \textit{LGD} (left panel) and the
        setting \textit{LGGMD} (right panel). The goodness-of-fit measure is based on log-likelihoods, and a larger value indicates a better estimate.
        We compare quantile
        regression forests (QRF), distribution boosting (DB), and
        LinCDE boosting. Densities of quantile regression forests and
        distribution boosting are computed according to \eqref{eq:quantileToPdf2} with $20$ bins. }}
\label{fig:R2} 
\end{figure}

Figure~\ref{fig:R2} presents the goodness-of-fit measure \eqref{eq:R2} of the three methods under the \textit{LGD} and \textit{LGGMD} settings. In both settings, LinCDE boosting leads in performance, improving the null method by $60\%$ to $80\%$ of the oracle's improvements.

\begin{figure}[bt]
	\centering
	\begin{minipage}{12cm}
		\centering  
		\includegraphics[scale = 0.6]{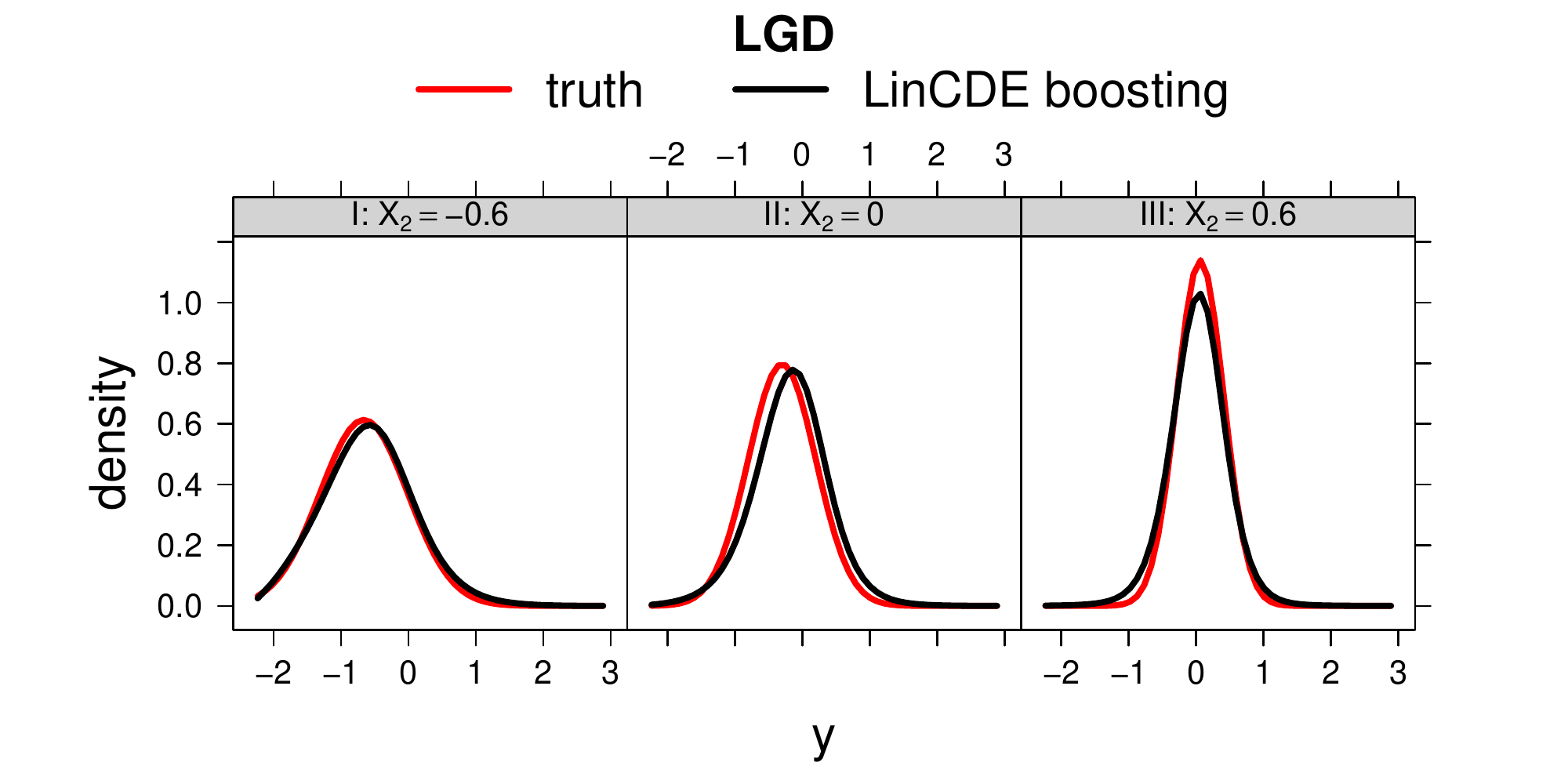} 
	\end{minipage}
	\begin{minipage}{12cm}
		\centering  
		\includegraphics[scale = 0.6]{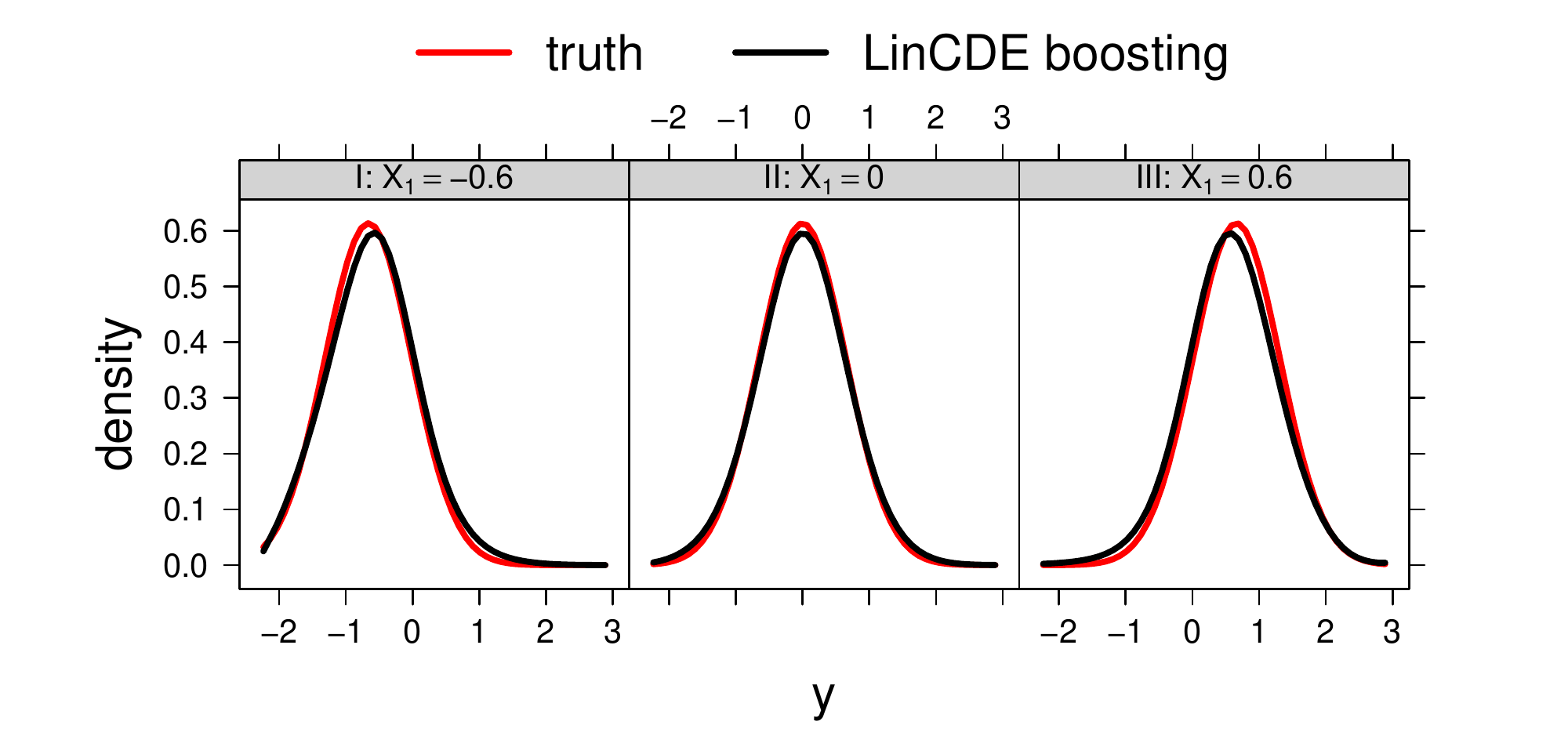} 
	\end{minipage}
	\caption{{Conditional densities estimated by LinCDE
            boosting in the \textit{LGD} setting. We take $x^{(2)} \in \{-0.6, 0, 0.6\}$ (upper panel) and $x^{(1)} \in \{-0.6, 0, 0.6\}$ (lower panel) and fix other covariates. 
    		The estimated conditional densities are close to the truth.
    }}
	\label{fig:densityGLM}
\end{figure}

\begin{figure}[bt]
	\centering
	\begin{minipage}{15cm}
		\centering  
		\includegraphics[scale=0.53]{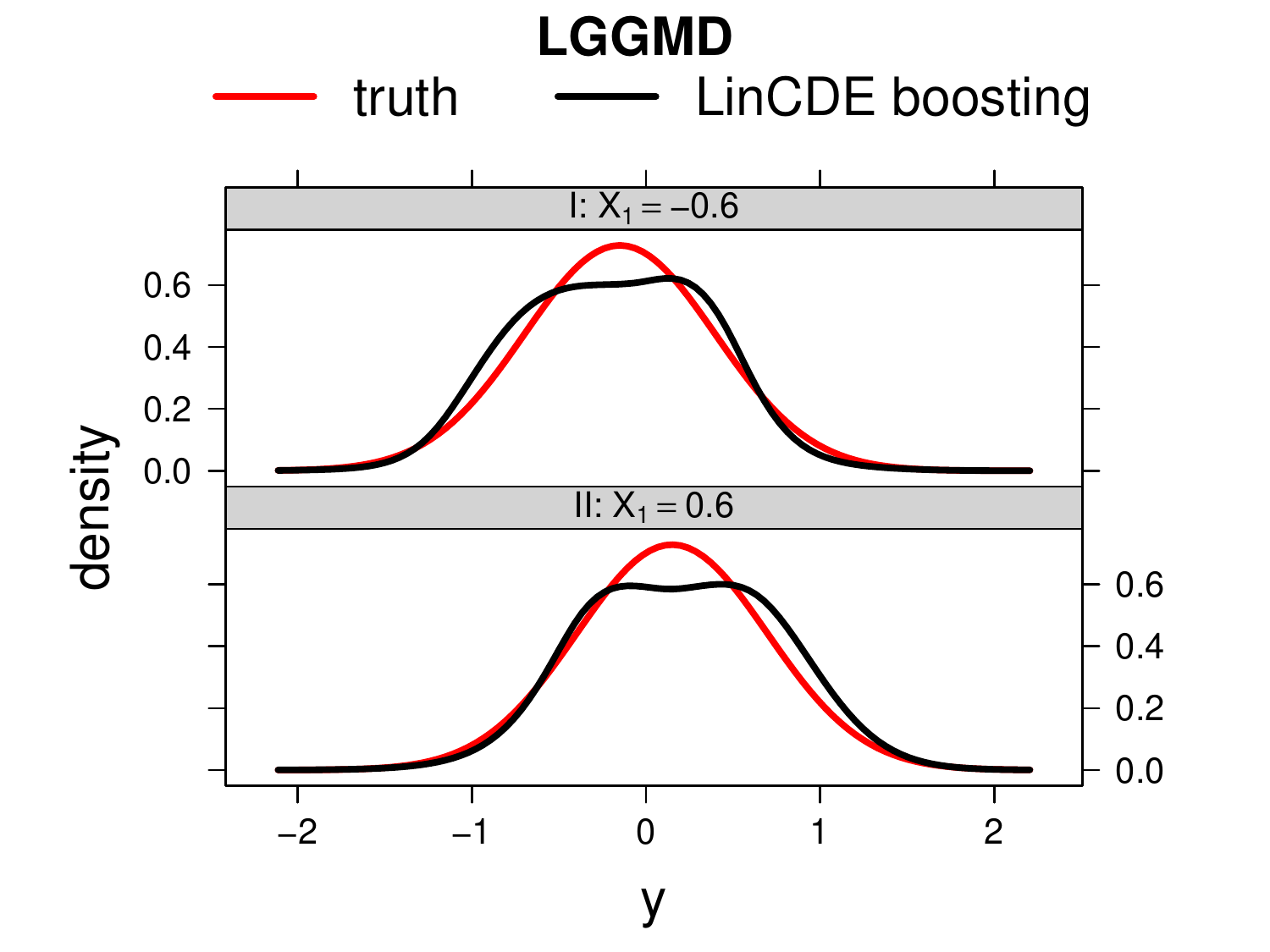} 
	\end{minipage}\\
	\begin{minipage}{7.5cm}
		\centering  
		\includegraphics[scale=0.53]{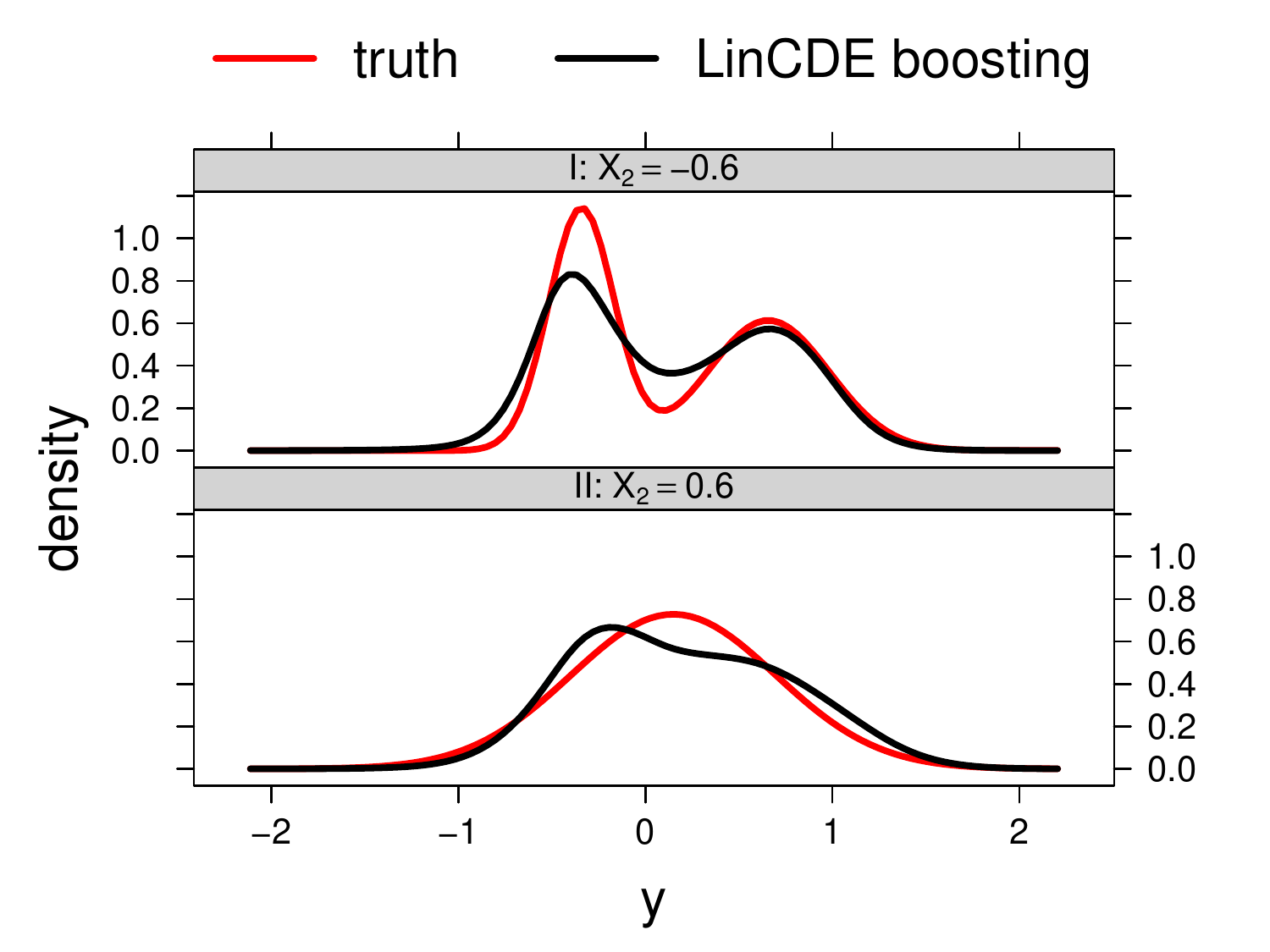} 
	\end{minipage}
		\begin{minipage}{7.5cm}
		\centering  
		\includegraphics[scale=0.53]{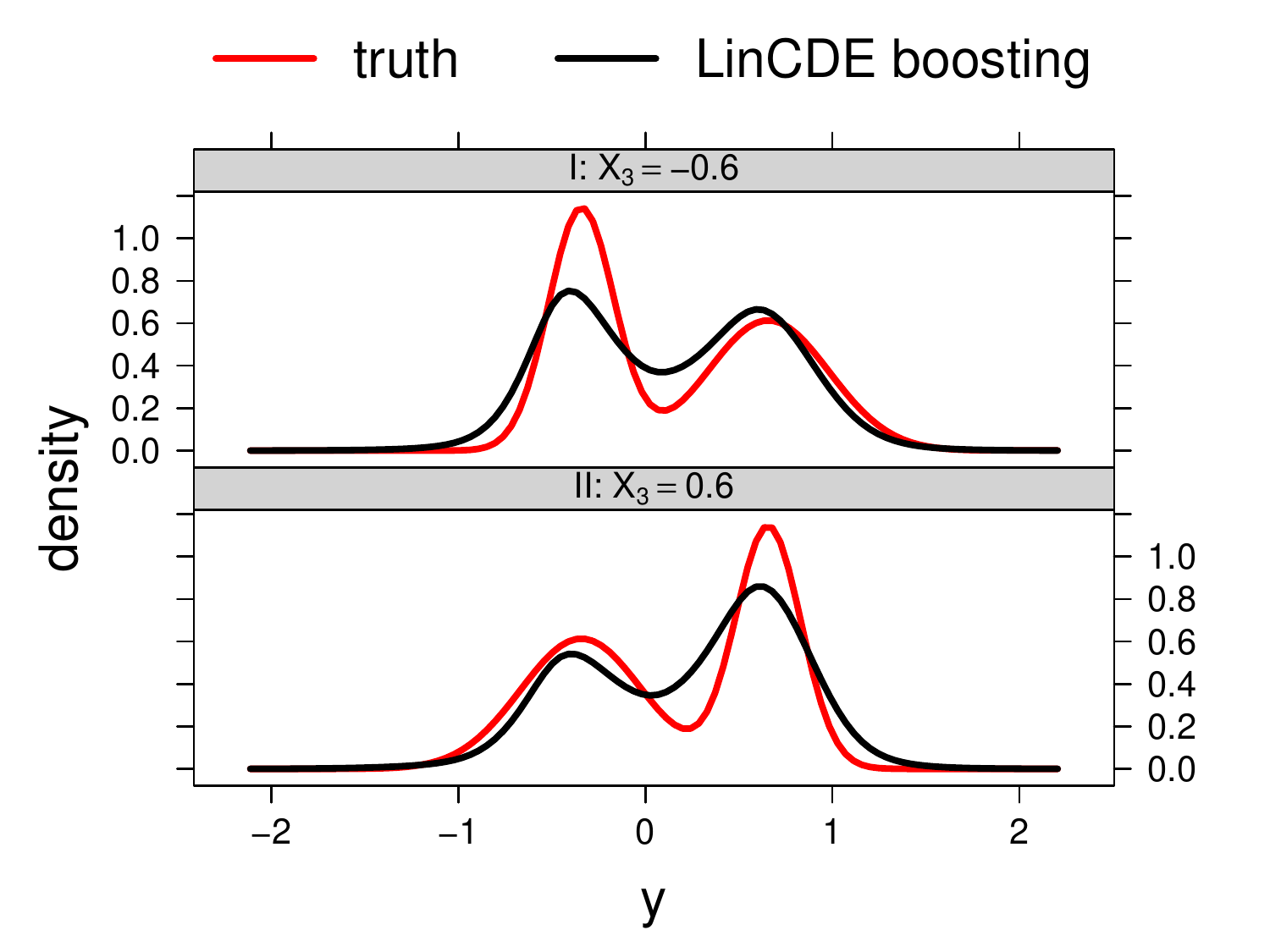} 
	\end{minipage}
	\caption{{Conditional densities estimated by LinCDE
            boosting in the \textit{LGGMD} setting. We take $x^{(1)}$, $x^{(2)}$, $x^{(3)} \in \{-0.6,0.6\}$ respectively. The estimated conditional densities vary in location as $x^{(1)}$ changes, in shape as $x^{(2)}$ changes, and in symmetry as $x^{(3)}$ changes. The estimated conditional densities are close to the truth.
    }}
	\label{fig:densityGM}
\end{figure}

\begin{figure}[bt]
	\centering
	\begin{minipage}{7cm}
		\centering  
		\includegraphics[scale = 0.5]{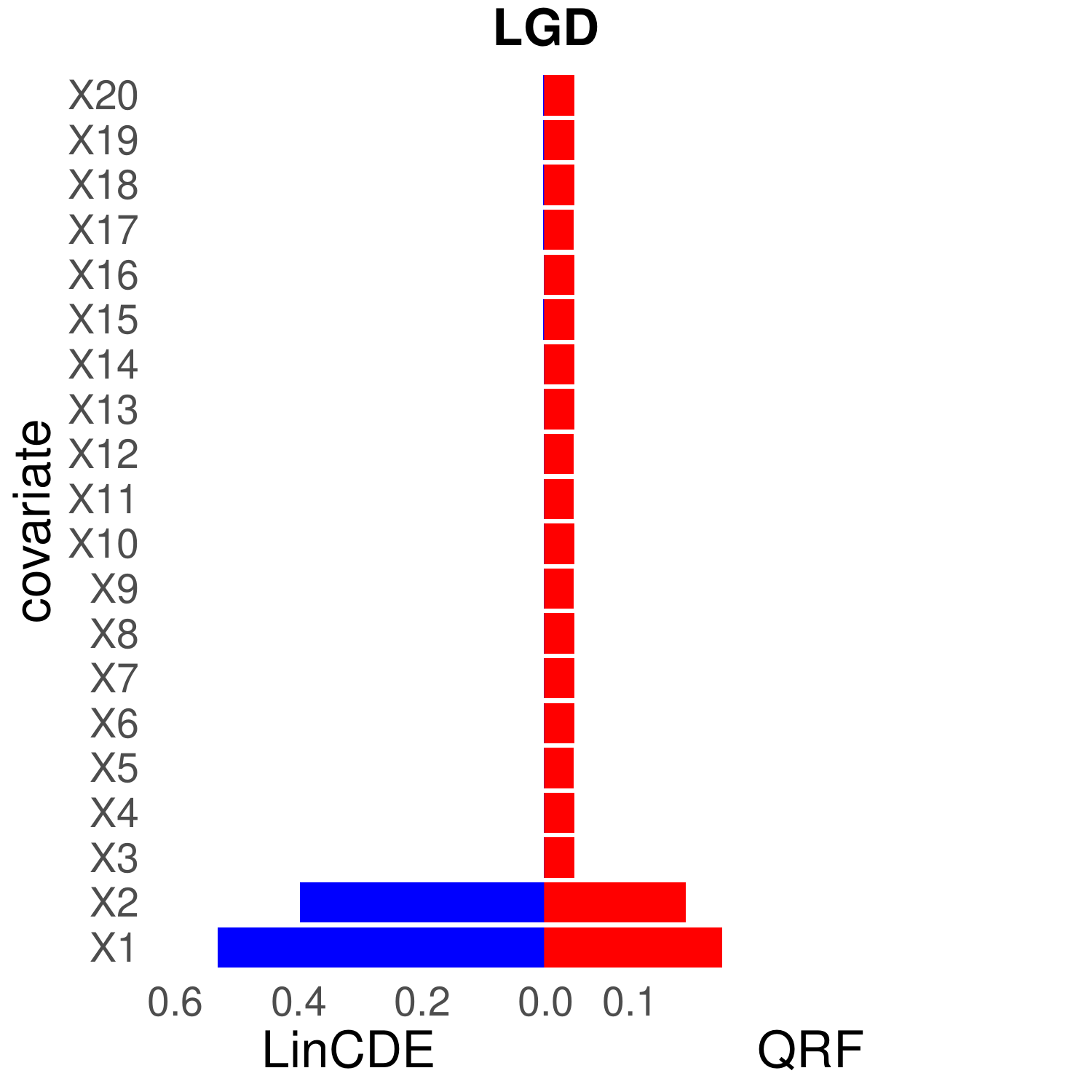} 
	\end{minipage} 
	\begin{minipage}{7cm}
		\centering  
		\includegraphics[scale=0.5]{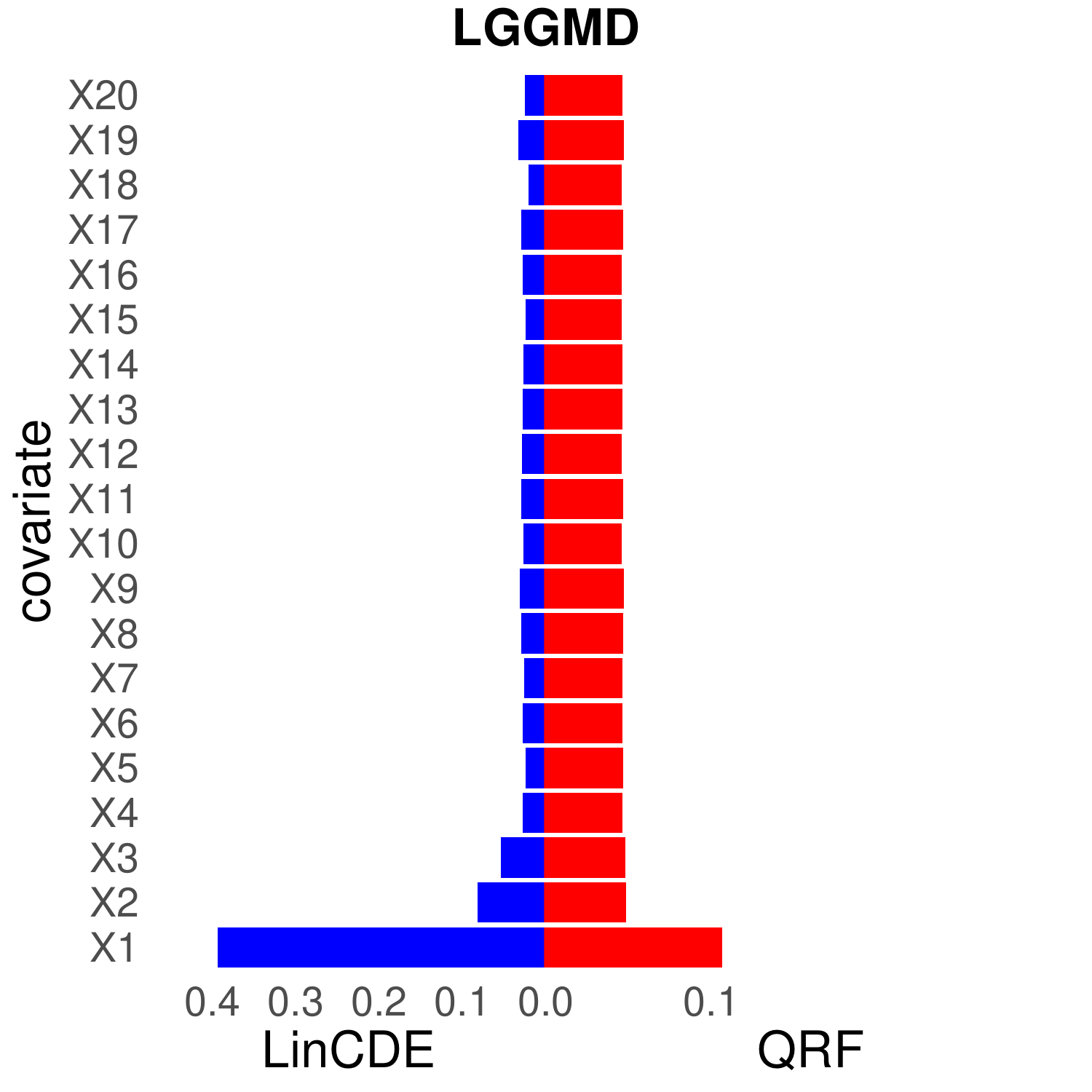}  
	\end{minipage}
	\caption{{Importance scores of LinCDE boosting and quantile regression forest in the \textit{LGD} (left panel) and \textit{LGGMD} (right panel) settings. We normalize the importance scores to sum to one. In the \textit{LGD} setting, both methods detect all the influential covariates $x^{(1)}$ and $x^{(2)}$. In the \textit{LGGMD} setting, LinCDE boosting identifies all the influential covariates $x^{(1)}$, $x^{(2))}$, and $x^{(3)}$, while quantile regression forest only identifies $x^{(1)}$.  
	 }}
	\label{fig:IS}
\end{figure}

Figures~\ref{fig:densityGLM} and \ref{fig:densityGM} depict the estimated conditional densities of LinCDE boosting in different subregions.
In both settings, LinCDE boosting identifies the roles of important
covariates: in the \textit{LGD} setting, the estimated conditional
densities vary in location as $x^{(1)}$ changes, and in scale as
$x^{(2)}$ changes; in the \textit{LGGMD} setting, the estimated
conditional densities vary in location as $x^{(1)}$ changes, in shape
as $x^{(2)}$ changes, and in symmetry as $x^{(3)}$ changes. To further
illustrate the ability of LinCDE boosting to detect influential covariates, we present the importance scores in Figure~\ref{fig:IS}. In the \textit{LGD} setting, LinCDE boosting puts around $87\%$ of the importance on $x^{(1)}$ and $x^{(2)}$, while quantile regression forest distributes more importance on the nuisances ($x^{(1)}$ and $x^{(2)}$ accounting for $40\%$). In the \textit{LGGMD} setting, LinCDE boosting is able to detect all influential covariates $x^{(1)}$, $x^{(2)}$, $x^{(3)}$, while the quantile regression forest only recognizes $x^{(1)}$.

\subsection{Results of Conditional CDF Estimation}
Here we evaluate the conditional CDF estimates of the three methods,
bringing the comparisons closer to the home court of distribution
boosting. 
We consider the average absolute error (AAE) used by \citet{friedman2019contrast}
\begin{align}\label{eq:AAE}
	\text{AAE} = \frac{1}{n} \sum_{i=1}^n \frac{1}{m} \sum_{j=1}^m 
	\left|\hat{F}(q(u_j \mid x_i) \mid x_i) - F(q(u_j \mid x_i) \mid x_i)\right|,
\end{align}
where $\{u_j\}$ is an evenly spaced grid on $[0,1]$, and $q(u \mid x)$ denotes the $u$ quantile at the covariate value $x$.
To compute the CDF estimates, for distribution boosting and quantile regression forest, we directly invert the estimated quantiles to CDFs. For LinCDE boosting, we compute the multinomial cell probabilities with a fine grid ($50$ bins) and obtain the CDFs based on the cell probabilities.

\begin{figure}[bt]
	\centering
	\begin{minipage}{7cm}
		\centering  
		\includegraphics[width  = 7cm, height = 7cm]{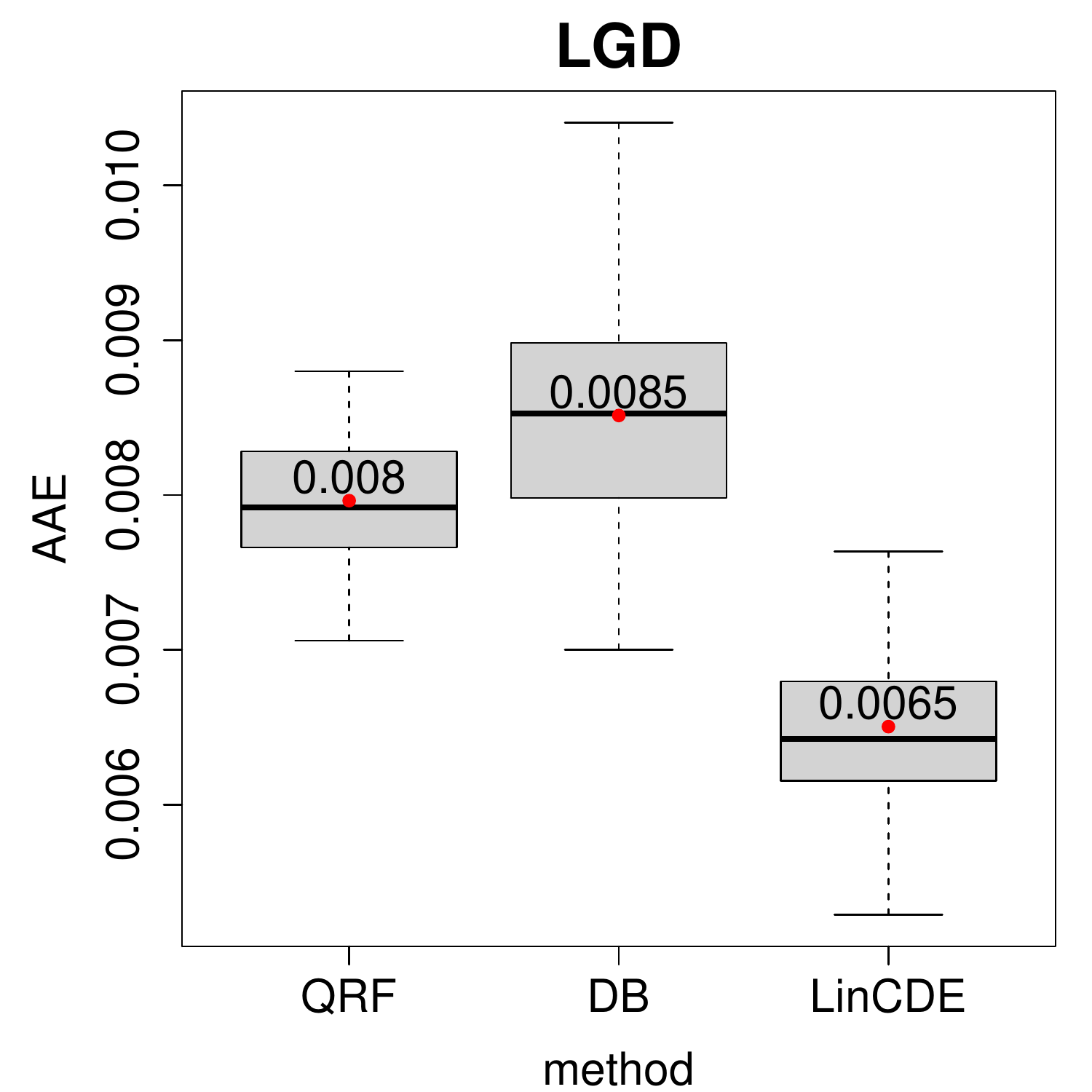} 
	\end{minipage} 
	\begin{minipage}{7cm}
		\centering  
		\includegraphics[width  = 7cm, height = 7cm]{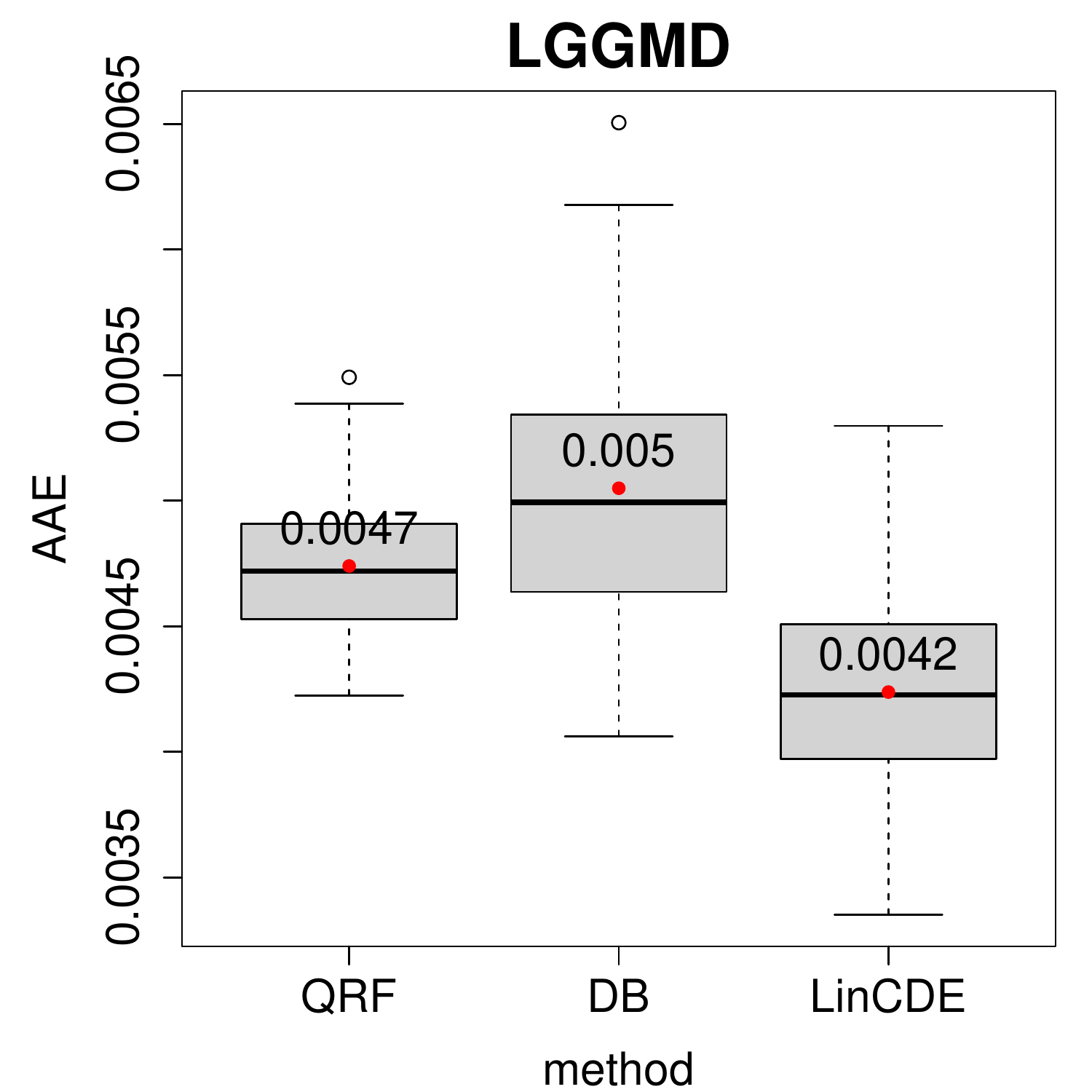}  
	\end{minipage}
	\caption{{Box plots of AAE \eqref{eq:AAE} in the LGD (left panel) and \textit{LGGMD} (right panel) settings. AAE is a metric naturally defined for conditional CDF estimates \citep{friedman2019contrast}, and a smaller value indicates a better estimate.}}
	\label{fig:AAE}
\end{figure}

Figure~\ref{fig:AAE} depicts the AAE metrics. In both settings, LinCDE boosting produces the smallest AAE. Notice that
\begin{align*}
	\hat{F}(q(u_j) \mid x_i) - u_j
	&= \hat{F}(q(u_j \mid x_i) \mid x_i) - \hat{F}(\hat{q}(u_j \mid x_i) \mid x_i) \\
	&\approx \hat{f}(q(u_j \mid x_i) \mid x_i) \cdot (q(u_j \mid x_i) - \hat{q}(u_j \mid x_i)).
\end{align*}
Though distribution boosting and quantile regression forest estimate
the quantiles well, the CDF estimates can be harmed by the implicit
density estimator multiplied. In Appendix~\ref{appn:sec:figure}, we also compare the CDF
estimates using  Cram\'er-von Mises distance and observe consistent
patterns to what we see here. 

\subsection{Results of Conditional Quantile Estimation}
Here the comparisons are in the home court of quantile random forests.
We evaluate the conditional quantile estimates of the three
methods. We compute the quantile losses at $\{5\%, 25\%, 50\%, 75\%,
95\%\}$ levels (Table~\ref{tab:pinball}).  For LinCDE boosting, we
compute the multinomial cell probabilities ($50$ bins) and obtain the
quantiles based on the cell probabilities. Despite the fact that
quantile-based metrics should favor quantile-based methods, we observe
that the performance of LinCDE boosting is similar.

\begin{table}[h!]
	\centering
	\begin{tabular}{c|c|ccccc|cc}
		\hline
		\hline
		\multirow{2}{*}{data}  & \multirow{2}{*}{method} & \multicolumn{5}{c|}{quantile loss}  & coverage &          width\\ \cline{3-7} 
		&                         & 5  \%     & 25  \%    & 50 \% & 75 \%  & 95 \%  &$90\%$ PI& $90\%$ PI \\ \hline
		\multirow{6}{*}{\textit{LGD}}   & \multirow{2}{*}{QRF}    & 0.058   & 0.174   & 0.218   & 0.174   & 0.056   &93.3\%& 2.02\\
		&                         & (0.001) & (0.02)  & (0.002) & (0.02)  & (0.001) &(0.9\%)& (0.048)\\ \cline{2-9} 
		& \multirow{2}{*}{DB}     & 0.058   & 0.176   & 0.218   & 0.174   & 0.057  & 92.5\%&1.95 \\
		&                         & (0.001) & (0.03)  & (0.003) & (0.03)  & (0.002) &(1.2\%)&(0.051)\\ \cline{2-9} 
		& \multirow{2}{*}{LinCDE} & 0.055   & 0.168   & 0.212   & 0.169   & 0.054 &91.9\%& 1.84 \\
		&                         & (0.001) & (0.01)  & (0.001) & (0.02)  & (0.001) &(1.0\%)&(0.045)\\ \hline
		\multirow{6}{*}{\textit{LGGMD}} & \multirow{2}{*}{QRF}    & 0.054   & 0.180   & 0.246   & 0.181   & 0.055   &89.5\%&1.76\\
		&                         & (0.001) & (0.001) & (0.002) & (0.001) & (0.001) &(0.7\%)&(0.034)\\ \cline{2-9} 
		& \multirow{2}{*}{DB}     & 0.054   & 0.182   & 0.246   & 0.182   & 0.055   &91.2\%&1.88\\
		&                         & (0.001) & (0.002) & (0.001) & (0.002) & (0.001) &(0.9\%)&(0.038)\\ \cline{2-9} 
		& \multirow{2}{*}{LinCDE} & 0.053   & 0.181   & 0.246   & 0.181   & 0.055   &90.5\%& 1.78\\
		&                         & (0.001) & (0.001) & (0.001) & (0.001) & (0.001) &(0.7\%)& (0.032)\\ \hline
	\end{tabular}
	\caption{{Table of quantile losses (the smaller the better) at $\{5\%, 25\%, 50\%, 75\%, 95\%\}$ levels and $90\%$ prediction interval coverages (ideally $90\%$), interval widths (the narrower the better) in the \textit{LGD} and \textit{LGGMD} settings. Standard deviations are in the parentheses. PI stands for prediction interval.}}
	\label{tab:pinball}
\end{table}

We also construct $50\%$ and $90\%$ prediction intervals based on the quantiles. In Table \ref{tab:pinball}, we display the coverages and widths of the $90\%$ prediction intervals. All methods produce fairly good coverages. Results of $50\%$ prediction intervals are consistent and can be found in Appendix~\ref{appn:sec:figure}.

\subsection{Computation Time}
We compare the computation time of the three methods.\footnote{The experiments are run on a personal computer with a dual-core CPU and 8GB memory.} We normalize the training time by the number of trees. In Table~\ref{tab:time}, quantile random forest is the fastest, followed by LinCDE boosting. LinCDE boosting takes about $5$ seconds for $100$ iterations.

\begin{table}[h!]
	\centering
	\begin{tabular}{c|ccc}
		\hline\hline
		time(s) & QRF & DB  & LinCDE\\ \hline
		\textit{LGD}     & 1.3 (0.018)                                                           & 13 (0.21)              & 4.8 (0.36)      \\
		\textit{LGGMD}   & 1.4 (0.15)                                                            & 14 (1.0)             & 5.1 (0.93)      \\ \hline
	\end{tabular}
\caption{{Table of computation time in seconds. We present the
    training time for $n=1000$ samples and  $d=20$ features per $100$ trees. Standard deviations are in parentheses.}}
\label{tab:time}
\end{table}

\section{Real Data Analysis}\label{sec:realData}
In this section, we analyze real data sets with LinCDE boosting. The pipeline is as follows:
first, we split the samples into training and test data sets; next, we perform $5$-fold cross-validation on the training data set to select the hyper-parameters; finally, we apply the estimators with the selected hyper-parameters and evaluate multiple criteria on the test data set. We repeat the procedure $20$ times and average the results. As for the centering, we use random forests as the conditional mean learner.

\subsection{Old Faithful Geyser Data}
The Old Faithful Geyser data records the eruptions from the ``Old
Faithful'' geyser in the Yellowstone National Park \citep{azzalini1990look} and represents continuous measurement from August $1$ to August $15$, $1985$. The data consists of $299$ observations and $2$ variables: eruption time and waiting time for the eruption. We estimate the conditional distribution of the eruption time given the waiting time. 

\begin{figure}[bt]
	\centering
	\begin{minipage}{5.8cm}
		\centering  
		\includegraphics[scale = 0.53]{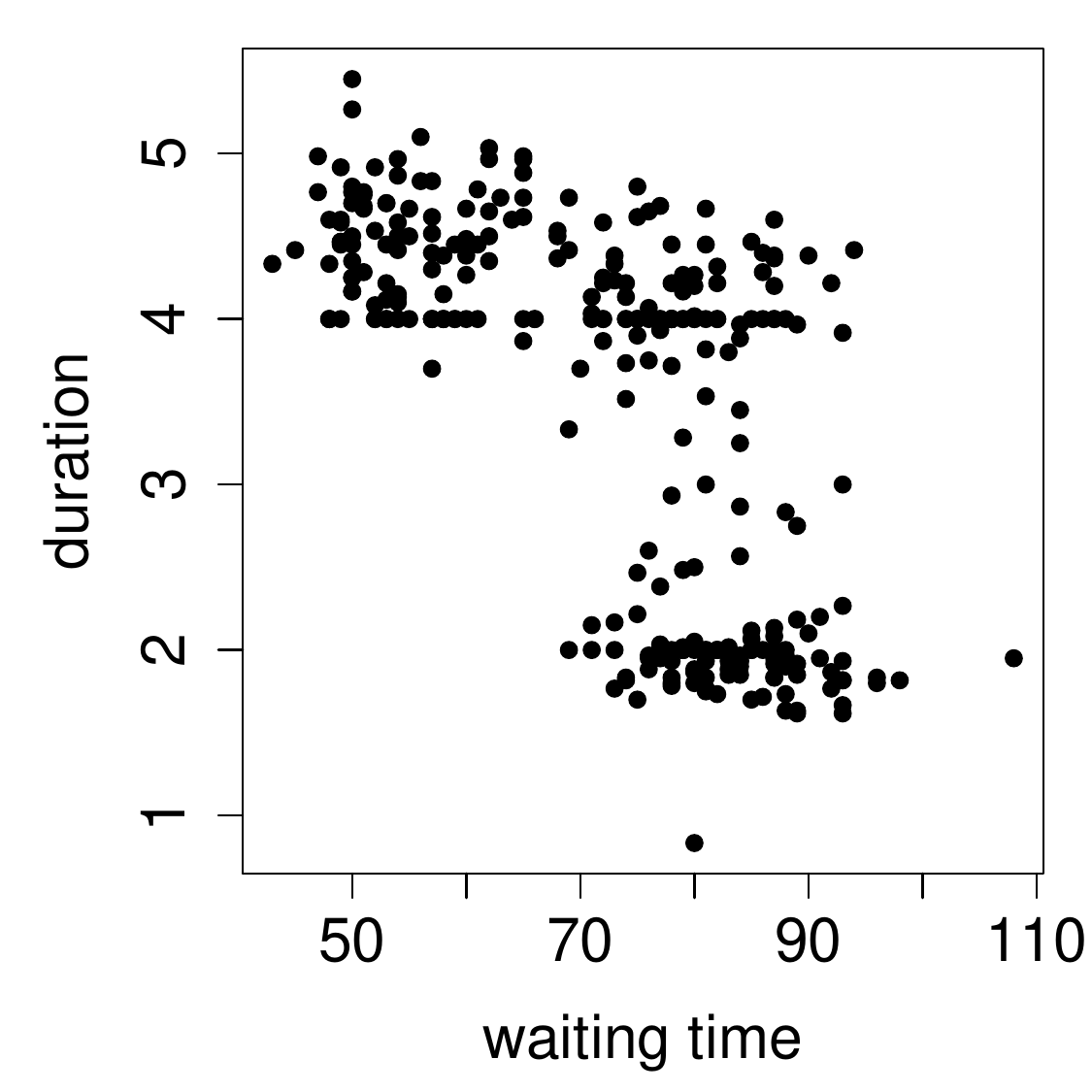} 
	\end{minipage}
	\begin{minipage}{9.2cm}
		\centering  
		\includegraphics[scale = 0.55]{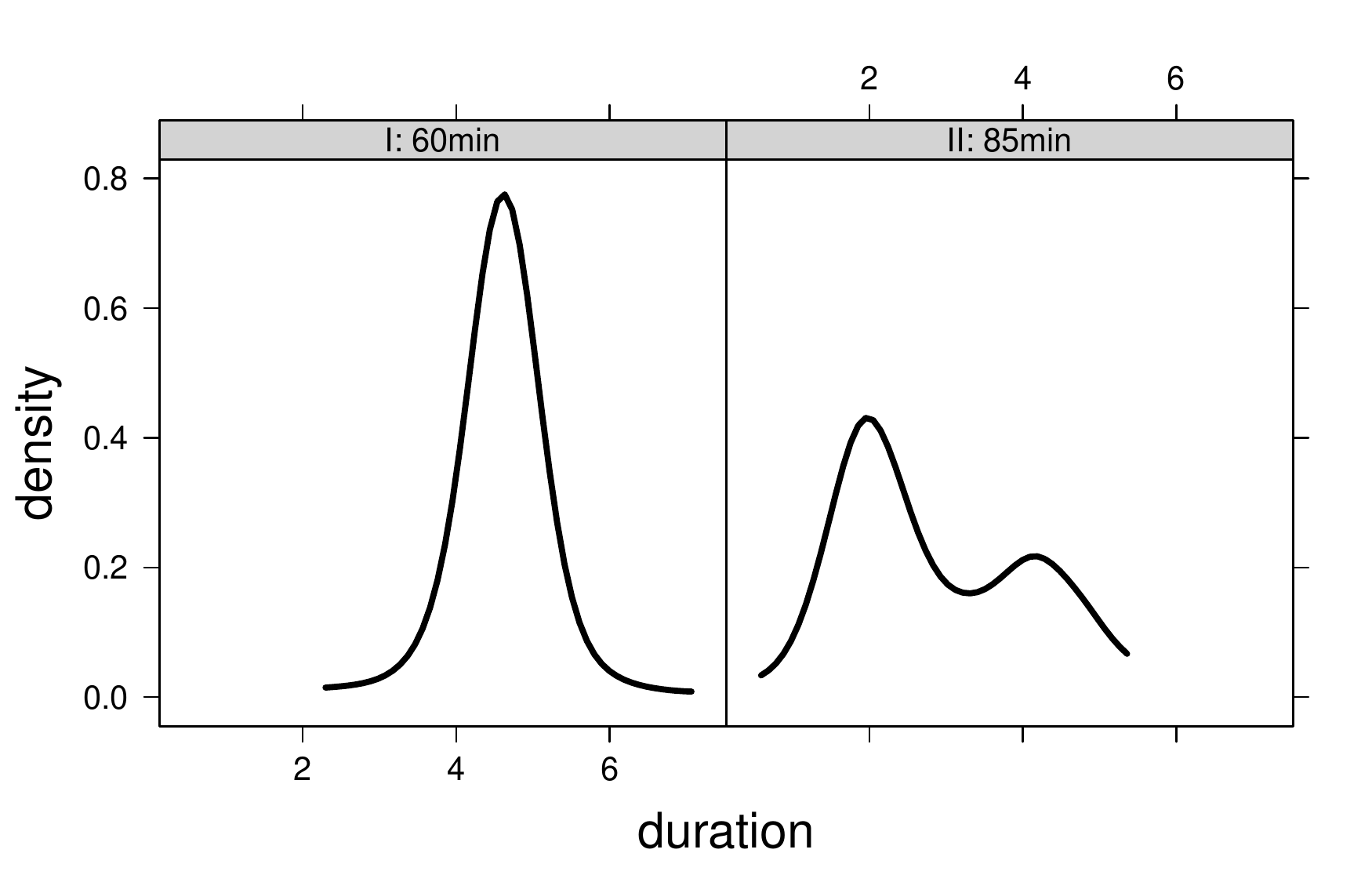}  
	\end{minipage}
	\caption{{LinCDE boosting applied to the Old Faithful Geyser data. On the left, we plot the eruption time versus the waiting time. On the right, we display the estimated conditional densities by LinCDE boosting at waiting times $60$min and $85$min.}}
	\label{fig:oldFaithfulGeyser}
\end{figure}

In Figure~\ref{fig:oldFaithfulGeyser}, we plot the eruption time
versus the waiting time. There is a clear cutoff at $70$min: for any
waiting time over $70$min, the distribution of eruption time is
bimodal, while for any waiting time below $70$min, the distribution is
unimodal. In Figure~\ref{fig:oldFaithfulGeyser}, we display the
estimated conditional densities of LinCDE boosting at waiting time
$85$min and $60$min. LinCDE boosting is capable of detecting the
difference in modality. In Table~\ref{tab:realData}, we summarize the
comparison between LinCDE boosting, distribution boosting, and
quantile regression forest regarding the negative log-likelihoods and
quantile losses. LinCDE boosting outperforms in log-likelihood, and is
competitive in quantile losses. We remark the AAE and Cram\'er-von
Mises distance can not be computed since we do not have the true conditional distributions.

\begin{table}[h!]
	\centering
	\begin{tabular}{c|c|c|ccccc}
		\hline\hline
		\multirow{2}{*}{data}                & \multirow{2}{*}{method} & \multirow{2}{*}{-log-like} & \multicolumn{5}{c}{quantile loss}                     \\ \cline{4-8} 
		&                         &                            & 5\%       & 25 \%     & 50\%      & 75 \%     & 95 \%     \\ \hline
		\multirow{6}{*}{Geyser} & \multirow{2}{*}{QRF}    & 1.55                       & 0.09    & 0.30    & 0.42    & 0.33    & 0.10    \\
		&                         & (0.12)                     & (0.01)  & (0.02)  & (0.03)  & (0.02)  & (0.01)  \\ \cline{2-8} 
		& \multirow{2}{*}{DB}     & 1.28                       & 0.09    & 0.27    & 0.37    & 0.30    & 0.09    \\
		&                         & (0.10)                     & (0.01)  & (0.02)  & (0.03)  & (0.02)  & (0.01)  \\ \cline{2-8} 
		& \multirow{2}{*}{LinCDE} & \textbf{1.16}                       & 0.09    & 0.28    & 0.37    & 0.30    & 0.09    \\
		&                         & (0.07)                     & (0.01)  & (0.02)  & (0.03)  & (0.02)  & (0.01)  \\ \hline
		\multirow{6}{*}{Height (age 14 - 40)} & \multirow{2}{*}{QRF}    & 3.30                       & 0.63    & 1.72    & 2.19    & 1.77    & 0.69   \\
		&                         & (0.03)                     & (0.03)  & (0.06)  & (0.07)  & (0.07)  & (0.04)  \\ \cline{2-8} 
		& \multirow{2}{*}{DB}     & 3.29                       & 0.65    & 1.84    & 2.24    & 1.90   & 0.72    \\
		&                         & (0.04)                     & (0.03)  & (0.05)  & (0.06)  & (0.07)  & (0.05)  \\ \cline{2-8} 
		& \multirow{2}{*}{LinCDE} & \textbf{3.19}                       & 0.64    & 1.82    & 2.20    & 1.88    & 0.71    \\
		&                         & (0.03)                     & (0.04)  & (0.05)  & (0.06)  & (0.07)  & (0.04)  \\ \hline
		\multirow{6}{*}{Height (age 14 - 17)}              & \multirow{2}{*}{QRF}    & 3.93                       & 0.95    & 2.99    & 3.87    & 3.19    & 1.12    \\
		&                         & (0.17)                       & (0.12)  & (0.24)  & (0.25)  & (0.23)  & (0.20)  \\ \cline{2-8} 
		& \multirow{2}{*}{DB}     & 4.21                       & 1.04    & 3.62    & 4.90    & 4.39    & 1.77    \\
		&                         & (0.16)                       & (0.04)  & (0.19)  & (0.28)  & (0.38)  & (0.32)  \\ \cline{2-8} 
		& \multirow{2}{*}{LinCDE} & \textbf{3.61}                       & 0.84    & 2.92    & 3.79    & 3.05    & 0.96    \\
		&                         & (0.06)                       & (0.05)  & (0.16)  & (0.23)  & (0.19)  & (0.10)  \\ \hline
		\multirow{6}{*}{Bone density}        & \multirow{2}{*}{QRF}    & -1.67                      & 0.004   & 0.012   & 0.015  & 0.013   & 0.004   \\
		&                         & (0.11)                     & (0.001) & (0.001) & (0.001) & (0.001) & (0.001) \\ \cline{2-8} 
		& \multirow{2}{*}{DB}     & -1.49                      & 0.007   & 0.015   & 0.014   & 0.016   & 0.007   \\
		&                         & (0.03)                     & (0.000) & (0.001) & (0.001) & (0.001) & (0.000) \\ \cline{2-8} 
		& \multirow{2}{*}{LinCDE} & \textbf{-1.89}                      & 0.004   & 0.011   & 0.014   & 0.013   & 0.005   \\
		&                         & (0.06)                     & (0.000) & (0.001) & (0.001) & (0.001) & (0.001) \\ \hline
	\end{tabular}
	\caption{{Comparison of LinCDE boosting, QRF, and DB on real data sets. We display negative log-likelihoods and quantile losses at $\{5\%, 25\%,50\%,75\%,95\%\}$ levels. (For all metrics, the smaller the better.) Standard deviations are in the parentheses. }}
\label{tab:realData}
\end{table}

\subsection{Human Height Data}
The human height data is taken from the NHANES data set: a series of health and nutrition surveys collected by the US National Center for Health Statistics (NCHS). We estimate the conditional distribution of the standing height. We consider two subsets: $542$ samples in the age range $14$ to $17$, and $1956$ samples in the age range $14$ to $40$.
In the smaller subset, we only consider two covariates: age and poverty;
in the larger subset, we consider $9$ covariates, including age, poverty, race, gender, etc. In the smaller subset, we tune by cross-validation; in the larger subset, we split the data set for validation, training, and test (proportion $2:1:1$), and tune on the hold-out validation data.

\begin{figure}[bt]
	\centering
	\begin{minipage}{7cm}
		\centering  
		\includegraphics[scale = 0.45]{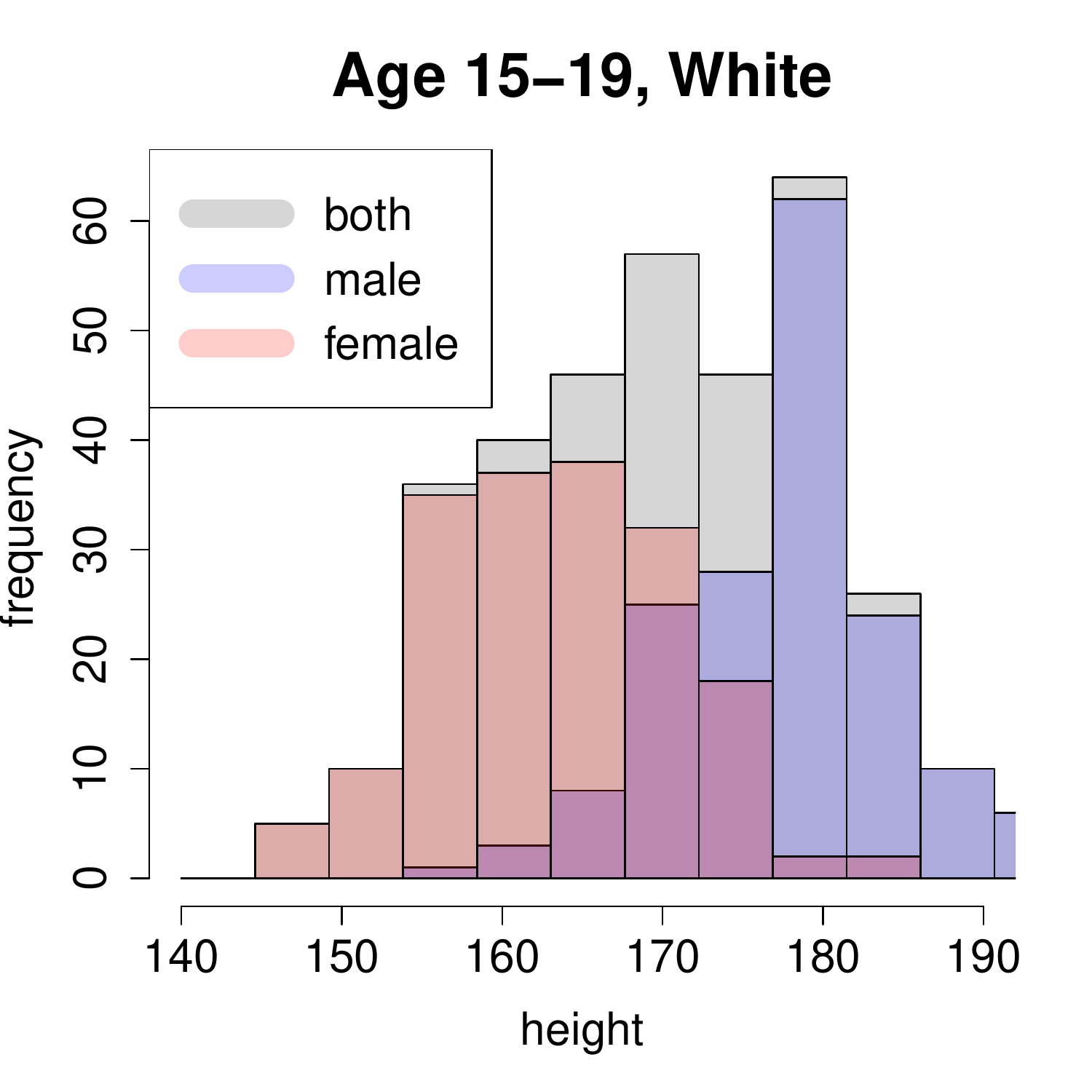}  
	\end{minipage}
	\begin{minipage}{7cm}
	\centering  
	\includegraphics[scale = 0.45]{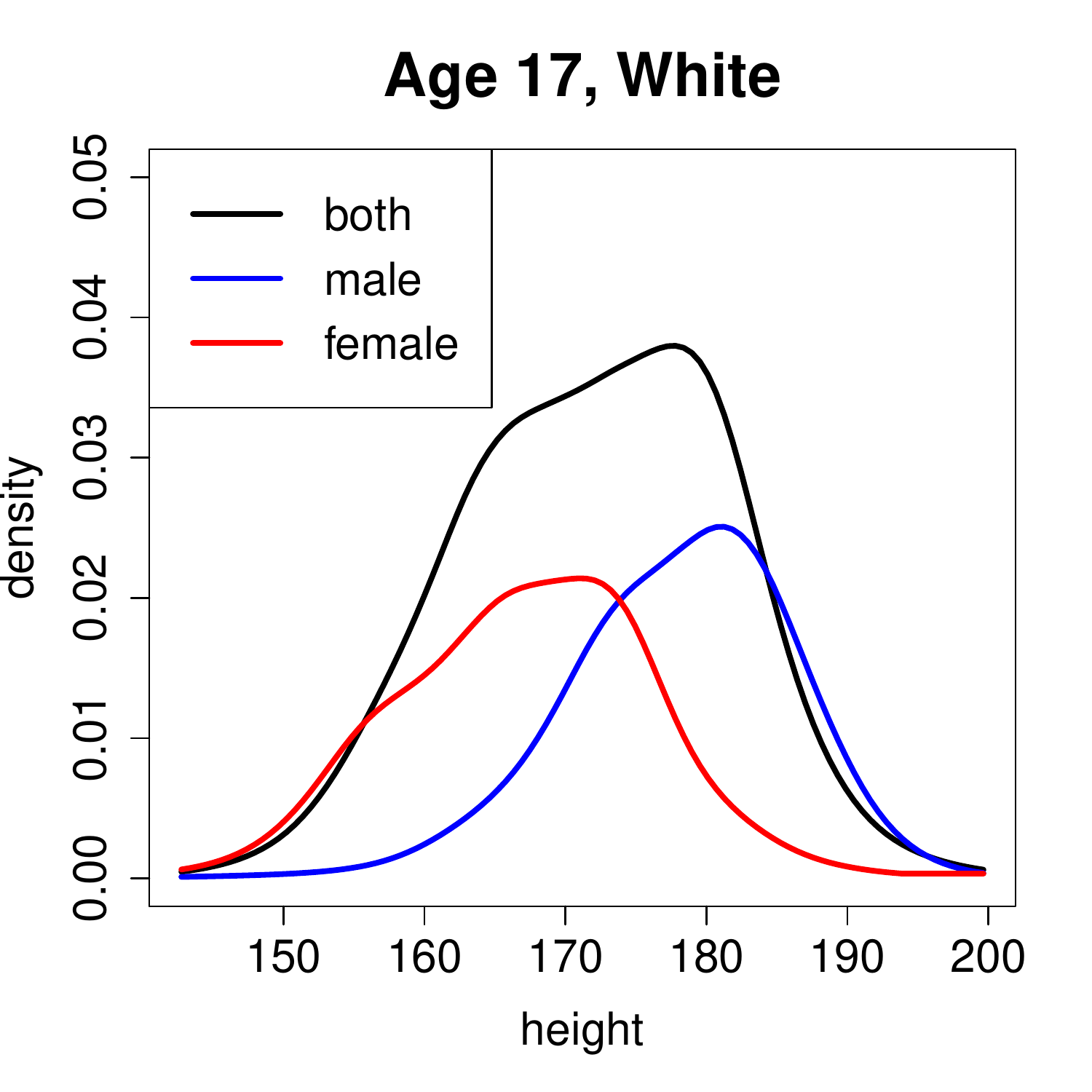}  
\end{minipage}
	\caption{{LinCDE boosting applied to the height data. On the left, we plot the histogram of heights of white teenagers in the age range $15$-$19$. On the right, we plot the average estimated density by LinCDE boosting from the larger data set at age $17$, race white, and other covariates fixed at the corresponding medians. The estimated conditional density without the gender is on the borderline between unimodal and bimodal. Furthermore, we contrast the histograms of male and female heights against that of the combined data in the left plot, explaining the formation of the quasi-bimodality. We also depict the estimated densities of females, males (right panel), which accord with the histograms (left panel).}} 
	\label{fig:height}
\end{figure}

The distribution of heights combining male and female is used as a
typical illustration of bimodality
\citep{peck2011statistics}. However, \citet{schilling2002human} point
out the separation between the heights of men and women is not large enough to produce the bimodality. In Figure~\ref{fig:height}, we demonstrate the histogram of heights of white teenagers in the age range $15$-$19$. The distribution of the combined data is sightly bimodal. Boys' heights are larger and more concentrated. We also provide LinCDE boosting's conditional density estimates obtained from the larger data set. Overall, the estimates accord with the histograms. The estimates without the covariate gender is on the borderline of unimodal and bimodal. The estimates with the gender explains the formation of the quasi-bimodality.

The comparisons of log-likelihood and quantile losses are summarized in Table~\ref{tab:realData}. In both the larger and the smaller data sets, LinCDE boosting performs the best concerning the log-likelihood. The advantage is more significant in the smaller data set. 
The reason is that in the larger data set, there are more covariates, and the conditional mean explains a larger proportion of the variation in response, while in the smaller data set, the conditional distribution after the centering contains more information, such as the bimodality, which can be learnt by LinCDE boosting.

\subsection{Relative Spinal Bone Mineral Density Data}
The relative spinal bone mineral density (spnbmd) data contains $485$ observations on $261$ North American adolescents. The response is the difference in spnbmd taken on two consecutive visits divided by the average. There are three covariates: sex, race, and age (the average age over the two visits). We estimate the conditional distributions of the spnbmd. The comparisons of log-likelihood and quantile losses are summarized in Table~\ref{tab:realData}. LinCDE boosting performs the best concerning the log-likelihood.

\begin{figure}[bt]
	\centering
	\begin{minipage}{5cm}
		\centering  
		\includegraphics[scale = 0.46]{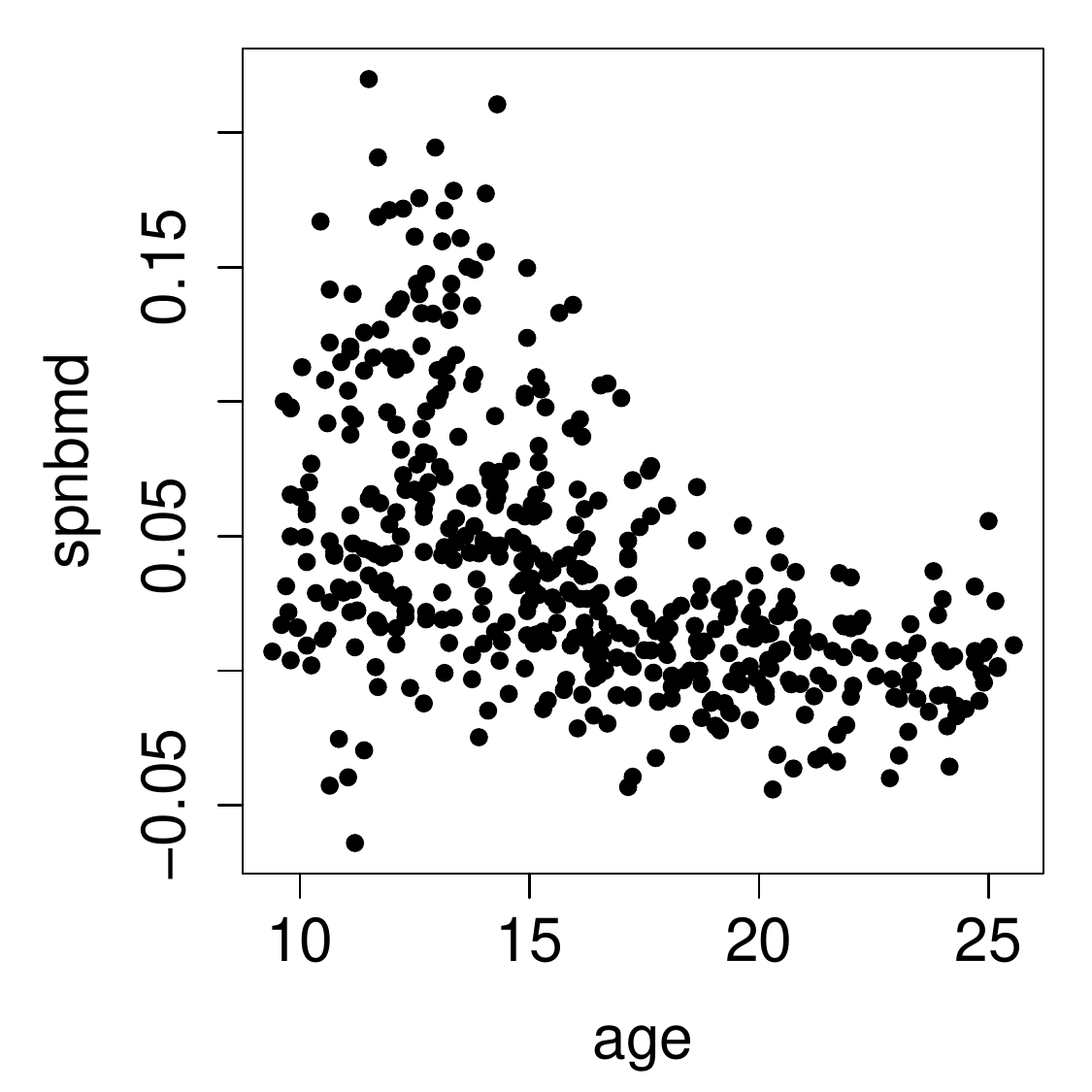} 
	\end{minipage}
	\begin{minipage}{10cm}
		\centering  
		\includegraphics[scale = 0.52]{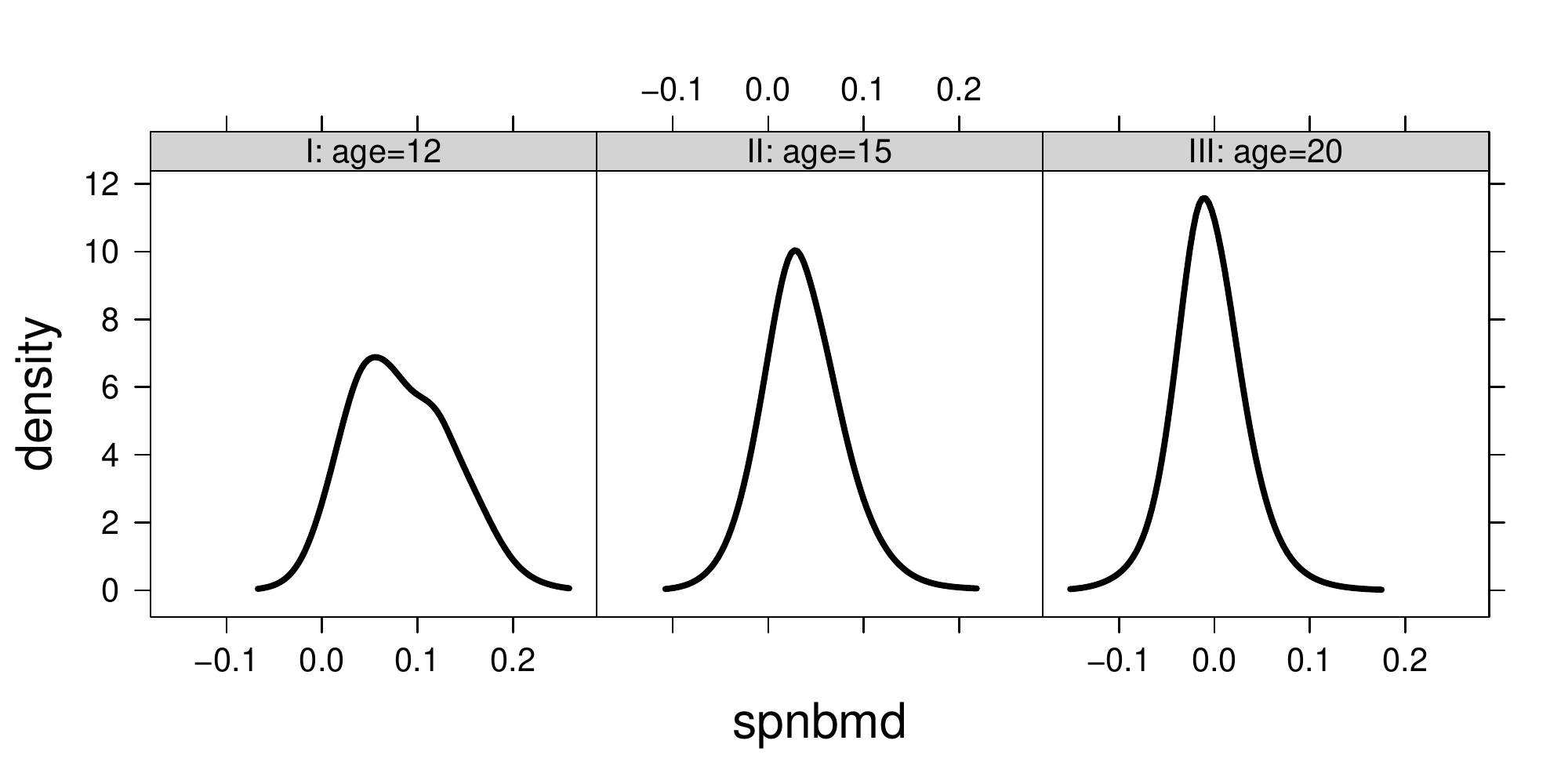}  
	\end{minipage}
	\caption{{LinCDE boosting applied to the relative spinal bone mineral density data. On the left, we display the scatterplot of spnbmd versus age. There is serious heteroscedasticity. On the right, we plot LinCDE boosting's estimates at age $12$, $15$ and $20$ of white females. The spreads of the conditional densities decrease as the age grows. At age $12$, the spnbmd distribution is right-skewed, while those at age $15$ and $20$ are approximately symmetric.}}
	\label{fig:bone}
\end{figure}

The scatterplot of spnbmd versus age demonstrates serious heteroscedasticity: the variances reach the climax at age $12$ and decrease afterward. In Figure~\ref{fig:bone}, we plot LinCDE boosting's estimates at age $12$, $15$ and $20$ of white females. The spreads of the conditional densities decrease as the age grows. At age $12$, the spnmd distribution is right-skewed, while those at age $15$ and $20$ are approximately symmetric.

\section{Discussion}\label{sec:discussion}
In this paper, we propose LinCDE boosting for conditional density
estimation. LinCDE boosting poses no specific parametric assumptions
of the density family. The estimates reflect a variety of
distributional characteristics. In the presence of unrelated nuisance covariates, LinCDE boosting is able to focus on the influential ones.

So far, we have discussed only univariate responses. Multivariate
responses emerge in multiple practical scenarios: locations on a $2$D
surface, joint distributions of health indices, to name a
few. Lindsey's method and thus LinCDE boosting can be easily generalized to multivariate responses. 
Assume the responses are $p$-dimensional, multivariate LinCDE boosting considers the density~\eqref{eq:exponentialFamily2} with sufficient statistics involving $y^{(1)}$ to $y^{(p)}$. 
As an illustrative example, if we use sufficient statistics $\{y^{(i)}\}$ and $\{y^{(i)} y^{(j)}\}$, the resulting density will be a multivariate Gaussian \citep[see][chap.~8.3 for the galaxy data example]{efron2016computer}. 
The response discretization now divides the hyper-rectangle response support into equal-sized $p$-dimensional bins and the rest of LinCDE boosting procedures carry over.
The cost of multivariate responses is the exponentially growing number of bins and sufficient statistics, which requires more samples as well as computational power. In contrast, conditional quantiles for multi-dimensional responses are relatively less straightforward but several promising proposals have been proposed \citep{barnett1976ordering, chaudhuri1996geometric, kong2012quantile, carlier2016vector}. We refer readers to \citet{koenker2017quantile} and references therein.

There are several exciting applications of LinCDE boosting.
\begin{itemize}
	\item \textit{Online learning}. Online learning processes the data that become available in a sequential order, such as stock prices and online auctions. As opposed to batch learning techniques which generate the best predictor by learning on the entire training data set once, online learning updates the best predictor for future data at each step. Online updating of LinCDE boosting is simple: we input the previous conditional density estimates as offsets, and modify them to fit new data.
	\item \textit{Conditional density ratio estimation}. A stream of work studies the density ratio model (DRM), particularly the semi-parametric DRM. The density ratio can be used for importance sampling, two-sample testing, outlier detection \citep[see][for an extensive review]{sugiyama2012density}. Boosting tilts the baseline estimate parametrically based on the smaller group. 
\end{itemize}

One future research direction is adding adaptive ridge penalty in LinCDE boosting.
Recall the fits in the \textit{LGGMD} setting (Figure~\ref{fig:densityGM}) where the conditional densities change from a relatively smooth Gaussian density to a curvy bimodal Gaussian mixture, the estimates get stuck in between: in the smooth Gaussian subregions, the estimates produce unnecessary curvatures; in the bumpy Gaussian mixture subregions, the estimates are not sufficiently wavy. The lack-of-fit can be attributed to the universal constraint on the degrees of freedom: the constraint may be too stringent in some subregions while lenient in others.

\section{Software}\label{sec:software}

Software for LinCDE is made available as an R package at \url{https://github.com/ZijunGao/LinCDE}. The package can be installed from GitHub with 
\begin{lstlisting}[language  = R]
	install.packages("devtools")
	devtools::install_github("ZijunGao/LinCDE", build_vignettes = TRUE)
\end{lstlisting}
The package comes with a detailed vignette discussing hyper-parameter tuning and demonstrating a number of simulated and real data examples.

\acks{This research was partially supported by grants DMS-2013736 and IIS
	1837931 from the National Science Foundation, and grant 5R01 EB
	001988-21 from the National Institutes of Health.
}

\newpage

\appendix

\section{Regularization Parameter and Degrees of Freedom}\label{appn:sec:hyperParameter}

We discuss how the hyper-parameter $\lambda$ relates to the degrees of freedom.\footnote{The degrees of freedom are derived under the Poisson approximation \eqref{eq:loglikelihoodPenalty3} of the original likelihood.}
According to Eq. (12.74) in \citet{efron2016computer}, the effective number of parameters in Poisson models takes the form
\begin{align}\label{eq:df1}
\text{df} = \sum_{b=1}^B \text{cov}(\hat{\eta}_b, n_b),
\end{align} 
where $n_b$ are responses and $\hat{\eta}_b$ are estimates of the natural parameter. In the following Proposition~\ref{prop:df}, we obtain an approximation of Eq.~\eqref{eq:df1} in the ridge Poisson regression via a quadratic expansion of the objective function. 

\begin{proposition}\label{prop:df} 
	Assume the responses $n_b$ are generated independently from the Poisson model with conditional means $\mu^*_b = \kappa_b e^{z_b^\top \beta^*}$ ($z$ includes the intercept) and the corresponding natural parameter $\eta^*_b = \log(\mu^*_b)$. Let $\lambda' > 0$ be the hyper-parameter of the ridge penalty.\footnote{$\lambda' = n\lambda$ for the $\lambda$ in \eqref{eq:loglikelihoodPenalty}.}
	\begin{itemize}
		\item For arbitrary $\beta$, the second order Taylor expansion at $\beta^*$ of the Poisson log-likelihood with ridge penalty is 
		\begin{align}\label{eq:prop:approx}
		\begin{split}
		\quad \sum_{b=1}^B n_b &\left(\log(\kappa_b) + z_b^\top\beta \right) - \kappa_b e^{z_b^\top\beta} - log(n_b!) - \lambda' \sum_{j=1}^k \omega_j \beta_j^2 \\
		&= -\frac{1}{2} (Z\beta + K - \zeta)^\top W (Z\beta + K - \zeta) - \lambda' \beta^\top \Omega \beta + C, \\ 
		Z_{b\cdot } &= z_b^\top, \quad K_b = \log(\kappa_b), \quad \zeta_b = \eta_b^* + \frac{n_b - \mu_b^*}{\mu_b^*}, \\
		W &= \text{\rm diag}(\mu_1^*, \ldots, \mu_B^*), \quad \Omega = \text{\rm diag}(\omega_1, \ldots, \omega_k),
		\end{split}
		\end{align} 
		for some constant $C > 0$ independent of $\beta$.
		\item Let $\hat{\beta}$ be the minimizer of the quadratic approximation \eqref{eq:prop:approx}, then 
		\begin{align}\label{eq:prop:df2}
		\text{df} = \sum_{b=1}^B \text{cov}\left(\log(\kappa_b) + z_b^\top \hat{\beta}, y_b\right) = \text{tr}\left(\left(H_{\beta^*}+ 2 \lambda'  \Omega \right)^{-1} H_{\beta^*}\right),
		\end{align} 
		where $H_{\beta^*} = Z^\top W Z$ is the Hessian matrix of the negative Poisson log-likelihood evaluated at $\beta^*$.
	\end{itemize}
\end{proposition}

We prove Proposition \ref{prop:df} in Appendix \ref{appn:sec:proof}. As a corollary, if $\Omega$ is a scalar multiple of the identity matrix, Eq.~\eqref{eq:prop:df2} agrees with the degrees of freedom formula Eq. (7.34) in \citet{hastie09:_elemen_statis_learn_II}. In practice, $\beta^*$ is unknown and we plug $\hat{\beta}$ in Eq.~\eqref{eq:prop:df2} to compute the number of effective parameters. We remark that Eq.~\eqref{eq:prop:df2} includes one degree of freedom for the intercept and in the LinCDE package we will use Eq.~\eqref{eq:prop:df2} minus one as the degrees of freedom. 

\begin{figure}[bt]
	\centering
	\begin{minipage}{15cm}
		\centering  
		\includegraphics[clip,  trim = 0cm 0.5cm 0cm 1.2cm, scale = 0.7]{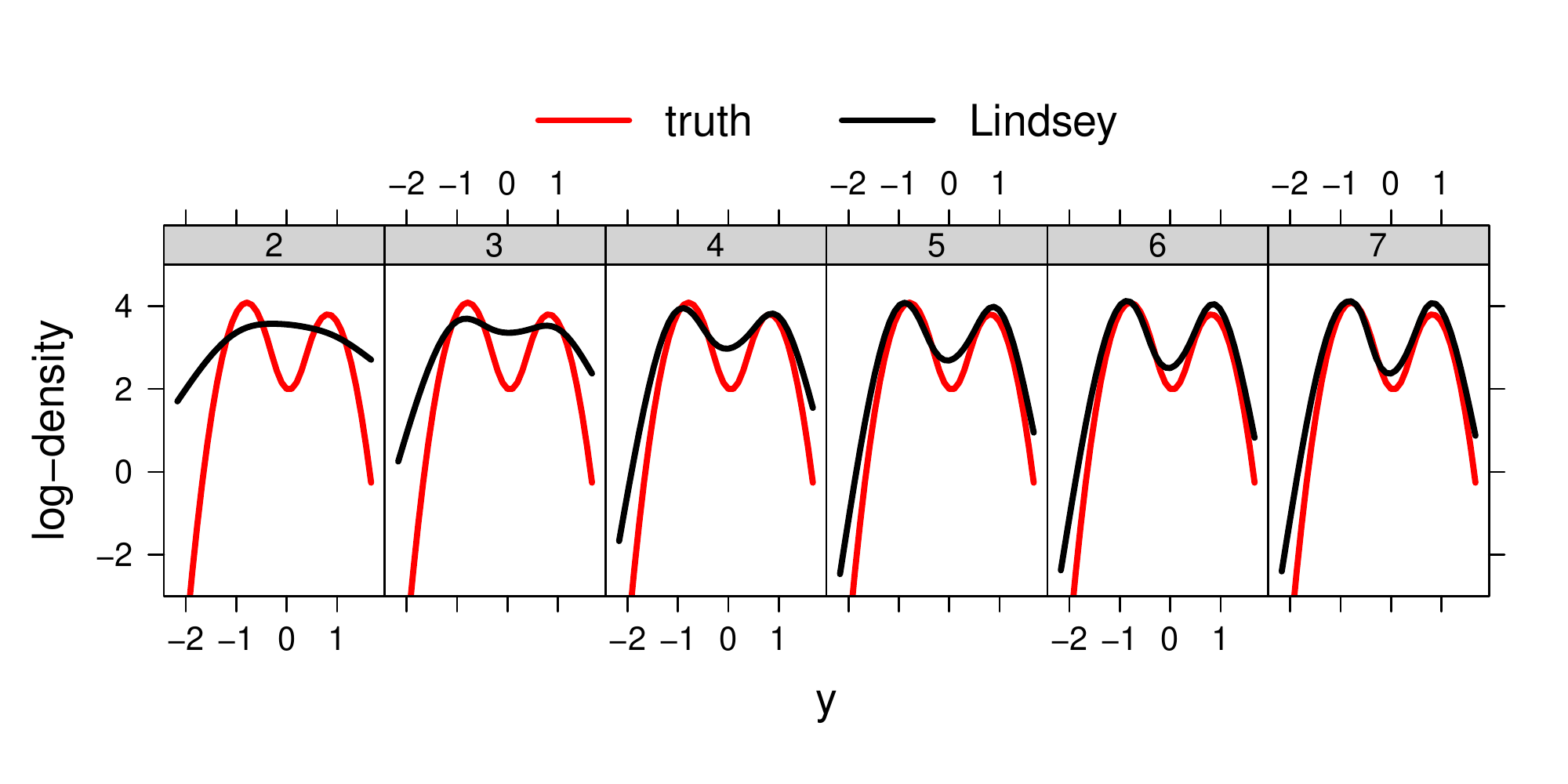}  
	\end{minipage}
	\caption{{Estimated log-densities by Lindsey's method under different degrees of freedom. 
	We use $10$ transformed natural cubic splines as basis functions and increase the degrees of freedom in Lindsey's method from $2$ to $7$ (left to right).}}
	\label{fig:degreeOfFreedom}
\end{figure}

Figure~\ref{fig:degreeOfFreedom} plots the estimated log-densities of a Gaussian mixture distribution from Lindsey's method under different degrees of freedom. As the degrees of freedom (excluding the intercept) increase from $2$ to $7$, the log-density curves evolve from approximately quadratic to significantly bimodal. At $7$ degrees of freedom, the estimated log-density is reasonably close to the underlying truth.

\section{Linear Transformation for Basis Construction}\label{appn:sec:linearTransformation}
We expand on the linear transformation used in the basis construction in Section~\ref{sec:Lindsey}. Let the eigen-decomposition of $\Omega$ be $U D U^\top$, where $U \in \RR^{k \times k}$ is orthonormal, and $D \in \RR^{k \times k}$ is a diagonal matrix with non-negative diagonal values ordered decreasingly. Define the linear-transformed cubic spline bases $\tilde{z}(y) = U^\top z(y)$ and the corresponding coefficients $\tilde{\beta} = U^\top \beta$. Then $\tilde{z}(y)^\top \tilde{\beta} = z(y)^\top\beta$ and $\tilde{\Omega} = U^\top \Omega U = U^\top U D U^\top U = D$. Therefore, $z(y) \mapsto U^\top z(y)$ is the desired linear transformation.

\section{Approximation Performance of Proposition~\ref{prop:approx}}\label{appn:sec:approximation}
\begin{figure}[bt]
	\centering
	\begin{minipage}{16cm}
		\centering  
		\includegraphics[scale = 0.7]{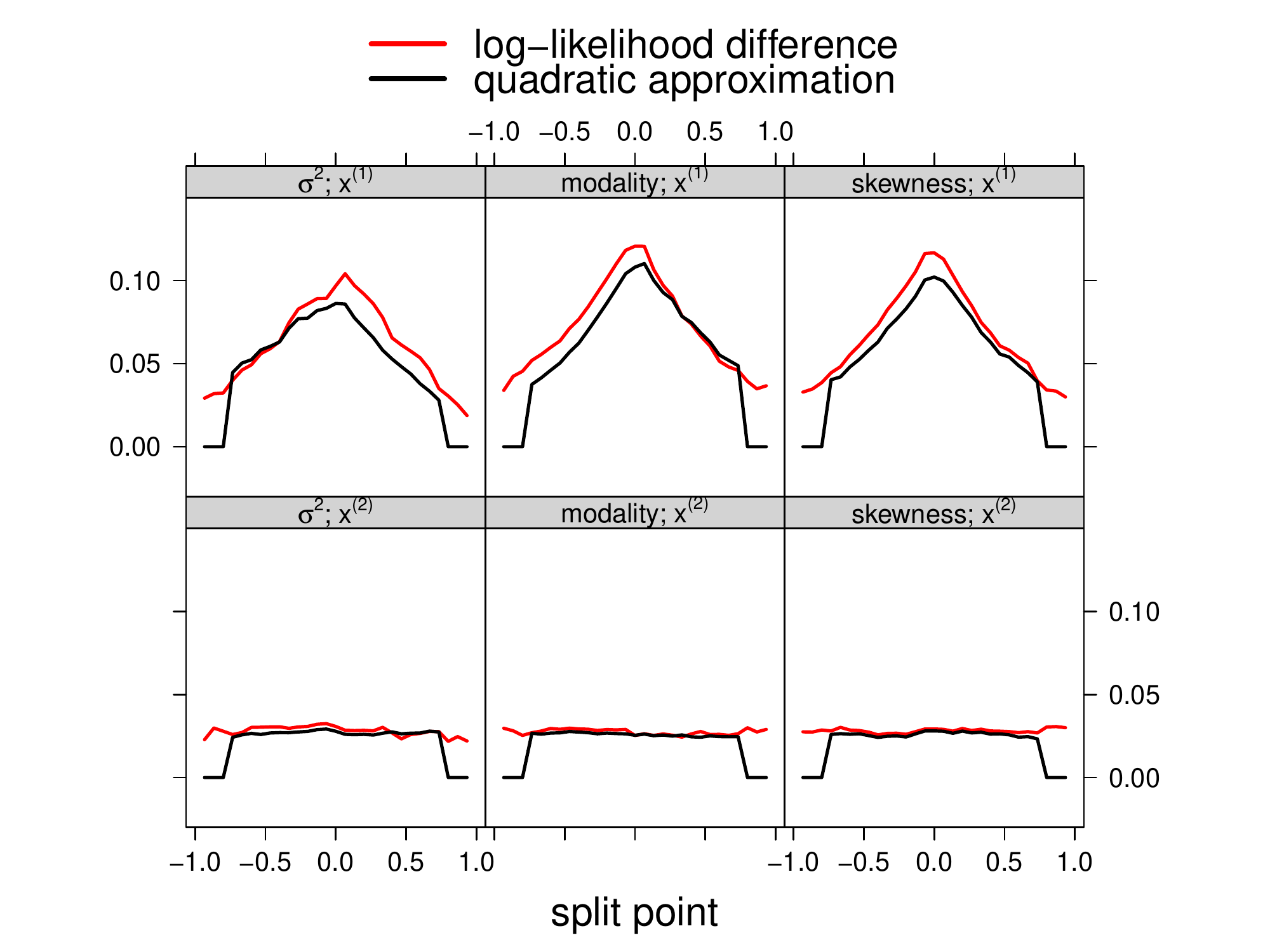}  
		\subcaption*{\textit{Tree-structured conditional densities}}
	\end{minipage}
	\caption{{Comparison of the log-likelihood difference \eqref{eq:split1} and the quadratic approximation in Proposition~\ref{prop:approx}. There are two subregions: $x^{(1)} \le 0$ and $x^{(1)} > 0$. In different subregions, the conditional densities are different in conditional variance $(\sigma^2)$, modality, or skewness (each column corresponds to a type of difference). In the same subregion, the conditional density does not change. In each trial, we sample $100$ observations and $5$ covariates. We use $5$ transformed natural cubic splines as basis functions, and $30$ candidate splits for each covariate equally spread across their ranges. We plot the log-likelihood difference (red) and the quadratic approximation (black) for candidate splits of $x^{(1)}$ (influential) and $x^{(2)}$ (nuisance), respectively. The results are aggregated over $100$ times. }}
	\label{fig:approximationTree}
\end{figure}

\begin{figure}[bt]
	\centering
	\begin{minipage}{16cm}
		\centering  
		\includegraphics[scale = 0.7]{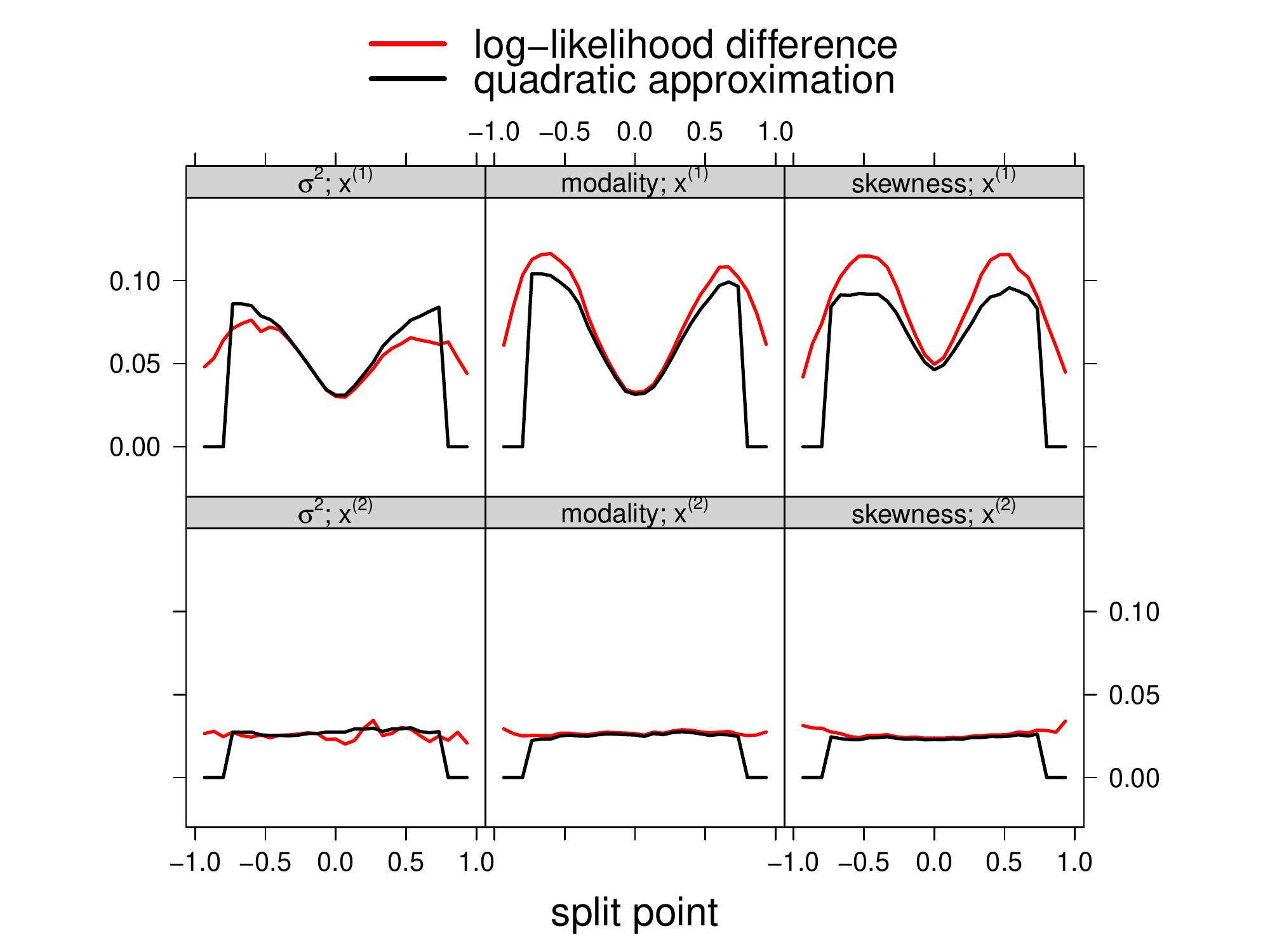}  
		\subcaption*{\textit{Continuously varying conditional densities}}
	\end{minipage}
	\caption{{Comparison of the log-likelihood difference \eqref{eq:split1} and the quadratic approximation in Proposition~\ref{prop:approx}. The conditional densities vary continuously with $|x^{(1)}|$ in variance $(\sigma^2)$, modality, or skewness (each column corresponds to a type of difference). In each trial, we sample $100$ observations and $5$ covariates. 
	We use $5$ transformed natural cubic splines as basis functions, and $30$ candidate splits for each covariate equally spread across their ranges.  
	We plot the log-likelihood difference (red) and the quadratic approximation (black) for candidate splits of $x^{(1)}$ (influential) and $x^{(2)}$ (nuisance), respectively. The results are aggregated over $100$ times. }}
	\label{fig:approximationGLM}
\end{figure}

In Figures~\ref{fig:approximationTree} and \ref{fig:approximationGLM}, we empirically demonstrate the efficacy of the quadratic approximation suggested by Proposition~\ref{prop:approx} . We let the conditional densities depend solely on $x^{(1)}$ and jump (Figure~\ref{fig:approximationTree}) or vary continuously (Figure~\ref{fig:approximationGLM}) in conditional variance, modality, or skewness. We observe that at candidate splits where the left child and the right child are similar, e.g., all the candidate splits based on $x^{(2)}$, the quadratic form and the exact log-likelihood difference are almost the same. At candidate splits where the left child and the right child are different, e.g., $x^{(1)} = 0$ in Figure~\ref{fig:approximationTree}, the exact log-likelihood difference is large, and the quadratic form is sufficiently close to the difference to determine the optimal split. We remark that to gain robustness, we set the quadratic approximation to zero if one of the candidate split's children contain less than $10$ samples, which leads to the imperfect approximation at the boundary candidate splits.

\section{Stopping Criteria for LinCDE Trees}\label{appn:sec:stopping}
We discuss stopping criteria for LinCDE trees. There is no universally optimal choice. If we build a single tree learner, the preferred strategy, at least for regression and classification trees according to \citet{breiman1984classification}, is to grow a large tree, then prune the tree using the cost-complexity pruning. If we train a random forest learner, then \citet{breiman2001random} recommends stopping the splitting process only when some minimum node size, default to be $5$ in the package \textit{randomForest} \citep{randomForestPackage}, is reached. As for tree boosting, \citet{friedman2001greedy} uses trees with the same number of terminal nodes. The number of terminal nodes is treated as a hyper-parameter of the boosting algorithm and tuned to maximize the performance on the data set at hand. The aforementioned stopping criteria generalize to LinCDE trees straightforwardly. Two options are currently available in codes: (1) stop when the tree depth reaches some prefixed level; (2) stop when the decrease in the objective fails to surpass a certain number---a greedy top-down approach. We don't recommend the criterion of stopping until a terminal node is pure, because there will be insufficient samples at terminal nodes for density estimation.

\section{Approximation Error for LinCDE Boosting}\label{appn:sec:approxError}

We characterize the expressiveness of LinCDE boosting's function class class~\eqref{eq:tilt} with splines $z(y)$. 

Without loss of generality, we will assume $f_{Y|X}(y|x)$ is supported on $[0,1] \times [0,1]^d$.
Let $S_2^s$ be the Sobolev space of functions $h(y)$ on  $[0,1]$ for which $h^{(s-1)}$ is absolutely continuous and $\int (h^{(s)}(y))^2 dy$ is finite. 
Denote the space of tree boosting models with arbitrary depth and number of trees by $\calN_1 := \{\sum_{i=1}^N a_i \1_{x^{(j)} \in I_j }, N \ge 1, I_j \subseteq[0,1] \}$. 
Denote the space of splines of order $\zeta$ on $[0,1]$ with equal-spaced knots $\{1,\ldots, k-\zeta+1\} \times \delta$, $\delta = 1/(k-\zeta+2)$, $\zeta \le k/2$, by $\Omega_{k,\zeta}$.
Let $\calF_k$ be the exponential family where $z_j(y) \in \Omega_{k,\zeta}$, $\zeta \ge s \ge 1$, and the parameter functions $\beta_j(x) \in \calN_1$, $1 \le j \le k$.
\begin{proposition}\label{prop:approxError}
	Assume the log-conditional-density function $h(y | x) := \log(f_{Y|X}(y | x) )  \in S_2^s$, $s \le k/2$, and $\|h(\cdot |  x)\|_\infty, \|h^{(s)}(\cdot | x)\|_2 \le B$,  then 
	\begin{align}\label{prop:eq:approxError}
		\min_{\tilde{f}_{Y|X}\in \calF_k}\EE\left[D\left(f_{Y|X}(y | X) ~||~ \tilde{f}_{Y|X}(y | X)\right)\right] \le \frac{C}{k^{2s}},
	\end{align}	
	where $C >0$ is a constant depending on $B$. 
\end{proposition}	
The proof of Proposition \ref{prop:approxError} can be found in Section \ref{appn:sec:proof}.

\section{Proofs}\label{appn:sec:proof}
In this section, we provide proofs of aforementioned propositions and claims.

\subsection{Proof of Claim \ref{claim:nullSpace}}
{\bf Proof}.
For $u \in \RR^k$, $u$ lies in the null space of $\Omega_{z}$ if and only if $u^\top \Omega u = 0$. By the definition of $\Omega$, 
\begin{align*}
0 = u^\top \Omega u 
= \int (z'''(y)^\top u)^2 dy,
\end{align*}
which implies that $(z(y)^\top u)''' = z'''(y)^\top u  = 0$ almost everywhere.  On one hand, if $z(y)^\top u$ is linear or quadratic, then $(z(y)^\top u)'''$ is automatically zero everywhere. On the other hand, since $z(y)$ are cubic spline bases, $z(y)^\top u$ is piece-wise cubic, and $(z(y)^\top u)''' = 0$ implies that $z(y)^\top u$ is piece-wise linear or quadratic. Because $z'(y)$ and  $z''(y)$ are both continuous, $z(y)^\top u$ is also second-order continuous, $z(y)^\top u$ must be linear or quadratic in $y$.
\hfill\BlackBox

\subsection{Proof of Proposition \ref{prop:df}}
{\bf Proof}. 
For arbitrary $\mu_b$, $\eta_b$ such that $\eta_b = \log(\mu_b)$, the second order Taylor expansion at $\mu_b^*$ of the Poisson log-likelihood of the $b$-th sample $\ell(y_b ; \eta_b)$ is
\begin{align}\label{eq:proof:loglike}
\ell(y_b ; \eta_b) 
\approx \ell(y_b ; \eta_b^*)  + \frac{\partial}{\partial \eta_b}\ell(y_b ; \eta_b^*)(\eta_b - \eta_b^*) + \frac{1}{2}\frac{\partial^2}{\partial \eta_b^2}\ell(y_b ; \eta_b^*)(\eta_b - \eta_b^*)^2.
\end{align}
By definition,
\begin{align}\label{eq:proof:derivative}
\begin{split}
\ell(y_b ; \eta_b^*) = y_b \eta_b^* - \kappa_b e^{\eta_b^*} + C, 
\quad \frac{\partial}{\partial \eta_b}\ell(y_b ; \eta_b^*) = y_b - \kappa_b e^{\eta_b^*}, \quad
\frac{\partial^2}{\partial \eta_b^2}\ell(y_b ; \eta_b^*) = - \kappa_b e^{\eta_b^*},
\end{split}
\end{align}
for some constant $C$ independent of $\eta_b^*$.
Plug Eq.~\eqref{eq:proof:derivative} into Eq.~\eqref{eq:proof:loglike},
\begin{align}\label{eq:proof:loglike2}
\begin{split}
\ell(y_b ; \eta_b) 
&\approx y_b \eta_b^* - \kappa_b e^{\eta_b^*}  + (y_b - \kappa_b e^{\eta_b^*})(\eta_b - \eta_b^*) - \frac{1}{2}\kappa_b e^{\eta_b^*} (\eta_b - \eta_b^*)^2 + C\\
&= -\frac{1}{2} \kappa_b e^{\eta_b^*} \left(\eta_b - \eta_b^* - \frac{y_b - \kappa_b e^{\eta_b^*}}{\kappa_b e^{\eta_b^*}} \right)^2 + r(y_b, \eta_b^*),
\end{split}
\end{align}
where the remainder $r(y_b, \eta_b^*)$ is independent of $\eta_b$. Plug $\eta_b = \log(\kappa_b) + z_b^\top \beta$, $\mu_b^* = e^{\eta_b^*}$ in Eq.~\eqref{eq:proof:loglike2}, we sum over all samples and finish the proof of Eq.~\eqref{eq:prop:approx}.

We different the quadratic approximation \eqref{eq:prop:approx} with respect to $\beta$ and obtain the score function
\begin{align*}
- Z^\top W (Z \beta  + K - \zeta) - 2 \lambda' \Omega \beta = 0.
\end{align*}
We solve the score function to obtain
\begin{align*}
\hat{\beta} = \left(Z^\top W Z + 2 \lambda'  \Omega \right)^{-1} Z^\top W (\zeta - K).
\end{align*}
Then plug $\hat{\eta} = K + Z \hat{\beta}$ into Eq.~\eqref{eq:df1} and we get
\begin{align*}
\text{df} = \sum_{b=1}^B \text{cov}(\hat{\eta}_b, y_b)
&= \text{tr}\left(\text{cov}\left(Z\left(Z^\top W Z + 2 \lambda'  \Omega \right)^{-1} Z^\top W (\zeta - K), Y \right)\right) \\
&= \text{tr}\left(Z\left(Z^\top W Z + 2 \lambda'  \Omega \right)^{-1} Z^\top W \text{cov}(\zeta, Y) \right).
\end{align*}
Notice that $\text{cov}(\zeta_b, y_{b'}) = \text{cov}(y_b, y_{b'})/\mu_b^* = \1_{\{b=b'\}}$, then 
\begin{align*}
\text{df} 
= \text{tr}\left(Z\left(Z^\top W Z + 2 \lambda'  \Omega \right)^{-1} Z^\top W \right)
= \text{tr}\left(\left(Z^\top W Z + 2 \lambda'  \Omega \right)^{-1} Z^\top W Z\right).
\end{align*}
This gives Eq.~\eqref{eq:prop:df2}. The Hessian of the negative Poisson log-likelihood at $\beta^*$ is 
\begin{align*}
H_{\beta^*} 
&= -\nabla_{\beta}^2|_{\beta = \beta^*} \sum_{b=1}^B \ell(y_b; \eta_b) 
= \sum_{b=1}^B \nabla_{\beta}^2|_{\beta = \beta^*} \kappa_b e^{z_b^\top \beta} \\
&= \sum_{b=1}^B \kappa_b  e^{z_b^\top \beta^*} z_b z_b^\top   
= Z^\top W Z^\top.
\end{align*}
\hfill\BlackBox

\subsection{Proof of Proposition \ref{prop:approx}}
{\bf Proof}.
We simplify the differences in the log-likelihood,
\begin{align}\label{eq:proof:delta}
\begin{split}
\Delta \ell(\calR, s) 
=& \sum_{x_i \in \calR_L}\left(\log(\kappa(y_i)) + z(y_i)^\top \hat{\beta}_{\calR_{s,L}} - \psi(\hat{\beta}_{\calR_{s,L}})\right) \\
&+ \sum_{x_i \in \calR_R} \left(\log(\kappa(y_i)) + z(y_i)^\top \hat{\beta}_{\calR_{s,R}} - \psi(\hat{\beta}_{\calR_{s,R}}) \right) \\
&-  \sum_{x_i \in \calR}\left(\log(\kappa(y_i)) + z(y_i)^\top \hat{\beta}_{\calR} - \psi(\hat{\beta}_{\calR})\right) \\
&- \lambda n_{\calR_{s,L}} \hat{\beta}_{\calR_{s,L}}^\top \Omega \hat{\beta}_{\calR_{s,L}} - \lambda n_{\calR_{s,R}} \hat{\beta}_{\calR_{s,R}}^\top \Omega \hat{\beta}_{\calR_{s,R}} +  \lambda n_{\calR} \hat{\beta}_{\calR}^\top \Omega \hat{\beta}_{\calR}\\
=& 	n_{\calR_{s,L}} \bar{z}_{\calR_{s,L}}^\top  (\hat{\beta}_{\calR_{s,L}} - \hat{\beta}_{\calR}) + n_{\calR_{s,R}} \bar{z}_{\calR_{s,R}}^\top  (\hat{\beta}_{\calR_{s,R}} - \hat{\beta}_{\calR}) \\
&- n_{\calR_{s,L}}(\psi(\hat{\beta}_{\calR_{s,L}})  - \psi(\hat{\beta}_{\calR}) ) - n_{\calR_{s,R}}(\psi(\hat{\beta}_{\calR_{s,R}})  - \psi(\hat{\beta}_{\calR}) )\\
&- \lambda n_{\calR_{s,L}} \hat{\beta}_{\calR_{s,L}}^\top \Omega \hat{\beta}_{\calR_{s,L}} - \lambda n_{\calR_{s,R}} \hat{\beta}_{\calR_{s,R}}^\top \Omega \hat{\beta}_{\calR_{s,R}} +  \lambda n_{\calR} \hat{\beta}_{\calR}^\top \Omega \hat{\beta}_{\calR},
\end{split}
\end{align} 	
where we use $n_{\calR_{s,L}} \bar{z}_{\calR_{s,L}} + n_{\calR_{s,R}}\bar{z}_{\calR_{s,R}} = n \bar{z}_{\calR}$.
The score equation of $\hat{\beta}_{\calR}$ implies
\begin{align}\label{eq:proof:betaL}
\bar{z}_{\calR}
= \nabla\psi(\hat{\beta}_{\calR}) + 2 \lambda \Omega \hat{\beta}_{\calR}.
\end{align} 
and similarly for $\hat{\beta}_{\calR_{s,L}}$, $\hat{\beta}_{\calR_{s,R}}$.
Plug Eq.~\eqref{eq:proof:betaL} into Eq.~\eqref{eq:proof:delta}, we obtain
\begin{align}\label{eq:proof:betaL3}
\begin{split}
\Delta \ell(\calR, s) 
=& n_{\calR_{s,L}} \nabla\psi(\hat{\beta}_{\calR_{s,L}})^\top (\hat{\beta}_{\calR_{s,L}} - \hat{\beta}_{\calR}) + n_{\calR_{s,R}} \nabla\psi(\hat{\beta}_{\calR_{s,R}})^\top (\hat{\beta}_{\calR_{s,R}} - \hat{\beta}_{\calR}) \\
&- n_{\calR_{s,L}}(\psi(\hat{\beta}_{\calR_{s,L}})  - \psi(\hat{\beta}_{\calR}) ) - n_{\calR_{s,R}}(\psi(\hat{\beta}_{\calR_{s,R}})  - \psi(\hat{\beta}_{\calR}) )\\
&+ \lambda n_{\calR_{s,L}} \left(\hat{\beta}_{\calR_{s,L}} - \hat{\beta}_{\calR}\right)^\top \Omega \left(\hat{\beta}_{\calR_{s,L}} - \hat{\beta}_{\calR}\right) \\
&+ \lambda n_{\calR_{s,R}} \left(\hat{\beta}_{\calR_{s,R}} - \hat{\beta}_{\calR}\right)^\top \Omega \left(\hat{\beta}_{\calR_{s,R}} - \hat{\beta}_{\calR}\right).
\end{split}
\end{align}
By the Taylor expansion of  $\psi(\beta)$ and $\nabla\psi(\beta)$ at $\hat{\beta}_{\calR}$, 
\begin{align}\label{eq:proof:betaL4}
\begin{split}
&\nabla\psi(\hat{\beta}_{\calR_{s,L}})^\top (\hat{\beta}_{\calR_{s,L}} -\hat{\beta}_{\calR}) - (\psi(\hat{\beta}_{\calR_{s,L}}) - \psi(\hat{\beta}_{\calR}))\\
=& \nabla\psi(\hat{\beta}_{\calR_{s,L}})^\top (\hat{\beta}_{\calR_{s,L}} -\hat{\beta}_{\calR}) - \nabla\psi(\hat{\beta}_{\calR})^\top (\hat{\beta}_{\calR_{s,L}} -\hat{\beta}_{\calR}) \\
&- \frac{1}{2} (\hat{\beta}_{\calR_{s,L}} -\hat{\beta}_{\calR})^\top \nabla^2\psi(\hat{\beta}_{\calR})(\hat{\beta}_{\calR_{s,L}} -\hat{\beta}_{\calR}) + O(\|(\hat{\beta}_{\calR_{s,L}} - \hat{\beta}_{\calR})\|_2^3)\\
=& \frac{1}{2} (\hat{\beta}_{\calR_{s,L}} -\hat{\beta}_{\calR})^\top \nabla^2\psi(\hat{\beta}_{\calR})(\hat{\beta}_{\calR_{s,L}} -\hat{\beta}_{\calR}) + O(\|(\hat{\beta}_{\calR_{s,L}} - \hat{\beta}_{\calR})\|_2^3).
\end{split}
\end{align} 
Plug Eq.~\eqref{eq:proof:betaL4} into Eq.~\eqref{eq:proof:betaL3}, we get
\begin{align}\label{eq:proof:betaL5}
\begin{split}
\Delta \ell(\calR, s) 
=& \frac{1}{2} n_{\calR_{s,L}} (\hat{\beta}_{\calR_{s,L}} -\hat{\beta}_{\calR})^\top \nabla^2\psi(\hat{\beta}_{\calR})(\hat{\beta}_{\calR_{s,L}} -\hat{\beta}_{\calR}) \\
&+ \frac{1}{2} n_{\calR_{s,R}}  (\hat{\beta}_{\calR_{s,R}} -\hat{\beta}_{\calR})^\top \nabla^2\psi(\hat{\beta}_{\calR})(\hat{\beta}_{\calR_{s,R}} -\hat{\beta}_{\calR})  \\
&+\lambda n_{\calR_{s,L}} \left(\hat{\beta}_{\calR_{s,L}} - \hat{\beta}_{\calR}\right)^\top \Omega \left(\hat{\beta}_{\calR_{s,L}} - \hat{\beta}_{\calR}\right) \\
&+ \lambda n_{\calR_{s,R}} \left(\hat{\beta}_{\calR_{s,R}} - \hat{\beta}_{\calR}\right)^\top \Omega \left(\hat{\beta}_{\calR_{s,R}} - \hat{\beta}_{\calR}\right) \\
&+ O\left(\|(\hat{\beta}_{\calR_{s,L}} - \hat{\beta}_{\calR})\|_2^3 + \|(\hat{\beta}_{\calR_{s,R}} - \hat{\beta}_{\calR})\|_2^3\right) \\
=& \frac{1}{2} n_{\calR_{s,L}} (\hat{\beta}_{\calR_{s,L}} -\hat{\beta}_{\calR})^\top \left(\nabla^2\psi(\hat{\beta}_{\calR}) + 2 \lambda \Omega \right)(\hat{\beta}_{\calR_{s,L}} -\hat{\beta}_{\calR}) \\
&+ \frac{1}{2} n_{\calR_{s,R}}  (\hat{\beta}_{\calR_{s,R}} -\hat{\beta}_{\calR})^\top \left(\nabla^2\psi(\hat{\beta}_{\calR}) + 2 \lambda \Omega\right) (\hat{\beta}_{\calR_{s,R}} -\hat{\beta}_{\calR}) \\
&+ O\left(\|(\hat{\beta}_{\calR_{s,L}} - \hat{\beta}_{\calR})\|_2^3 + \|(\hat{\beta}_{\calR_{s,R}} - \hat{\beta}_{\calR})\|_2^3\right).
\end{split}
\end{align}
Finally by Eq.~\eqref{eq:proof:betaL} and the assumption that $\nabla^2\psi(\hat{\beta}_{\calR}) + 2 \lambda \Omega$ is invertible,
\begin{align}\label{eq:proof:betaL6}
\begin{split}
\hat{\beta}_{\calR_{s,L}} - \hat{\beta}_{\calR}
&= \left(\nabla^2\psi(\hat{\beta}_{\calR}) + 2 \lambda \Omega\right)^{-1}(\bar{z}_{\calR_{z,L}}  - \bar{z}_{\calR}) + O\left(\|(\hat{\beta}_{\calR_{s,L}} - \hat{\beta}_{\calR})\|_2^2\right).
\end{split}
\end{align} 
Then $\hat{\beta}_{\calR_{s,L}} - \hat{\beta}_{\calR} \asymp \bar{z}_{\calR_{s,L}} -\bar{z}_{\calR}$ and similarly for the right child. Plug Eq.~\eqref{eq:proof:betaL6} into Eq.~\eqref{eq:proof:betaL5},
\begin{align*}
\Delta \ell(\calR, s) 
=& \frac{1}{2} n_{\calR_{s,L}} (\bar{z}_{\calR_{s,L}} -\bar{z}_{\calR})^\top \left(\nabla^2\psi(\hat{\beta}_{\calR}) + 2 \lambda \Omega \right)^{-1}(\bar{z}_{\calR_{s,L}} - \bar{z}_{\calR}) \\
&+ \frac{1}{2} n_{\calR_{s,R}}  (\bar{z}_{\calR_{s,R}} -\bar{z}_{\calR})^\top \left(\nabla^2\psi(\hat{\beta}_{\calR}) + 2 \lambda \Omega\right)^{-1} (\bar{z}_{\calR_{s,R}} -\bar{z}_{\calR}) \\
&+ O\left(\|(\bar{z}_{\calR_{s,L}} - \bar{z}_{\calR})\|_2^3 + \|(\bar{z}_{\calR_{s,R}} - \bar{z}_{\calR})\|_2^3\right) \\
=& \frac{n_{\calR_{s,L}}n_{\calR_{s,L}}}{2n_{\calR}} (\bar{z}_{\calR_{s,L}} -\bar{z}_{\calR_{s,R}})^\top \left(\nabla^2\psi(\hat{\beta}_{\calR}) + 2 \lambda \Omega \right)^{-1}(\bar{z}_{\calR_{s,L}} - \bar{z}_{\calR_{s,R}})  \\
&+ O\left(\|(\bar{z}_{\calR_{s,L}} - \bar{z}_{\calR})\|_2^3 + \|(\bar{z}_{\calR_{s,R}} - \bar{z}_{\calR})\|_2^3\right), 
\end{align*}
and we finish the proof.
\hfill\BlackBox

\subsection{Proof of Claim \ref{claim:varApprox}}
\begin{claim}\label{claim:varApprox}
	Assume that in $\calR$, $y$ is supported on the midpoints $\{y_b\}$, then the covariance matrix approximation by Lindsey's method equals $\nabla^2\psi(\hat{\beta}_{\calR})$.
\end{claim}

{\bf Proof}.
	If $y$ is supported on the midpoints $\{y_b\}$, then the discretization in Lindsey's method is accurate. As a result,  the estimator of Lindsey's method is the exact log-likelihood maximizer $\hat{\beta}_{\calR}$.
	Furthermore, the multinomial cell probabilities based on the Linsey's method's estimator is indeed the response distribution indexed by $\hat{\beta}_{\calR}$. Next, by \citet{lehmann2006testing}, $\nabla^2\psi(\hat{\beta}_{\calR})$ equals the population covariance matrix of the sufficient statistics generated from the distribution indexed by $\hat{\beta}_{\calR}$. Therefore, the covariance approximation by Lindsey's method equals $\nabla^2\psi(\hat{\beta}_{\calR})$.
\hfill\BlackBox

\subsection{Proof of Proposition \ref{prop:computation}}
{\bf Proof}.
We order the observations and count responses in each bin ($\tilde{O}(n_{\calR})$). Next, we run Newton-Rapson algorithm. 
In each iteration, the computation of the gradient vector
\begin{align*}
\sum_{b = 1}^B n_b  \tilde{z}(y_b) (1-e^{z(y_b)^\top \beta + \beta_0}).
\end{align*}
is $O(kB)$, where $\tilde{z}^\top = [z^\top,1]$.
The computation of the Hessian matrix
\begin{align*}
-\sum_{b = 1}^B n_b  e^{z(y_b)^\top \beta + \beta_0} \tilde{z}(y_b) \tilde{z}(y_b)^\top
\end{align*}
takes $O(k^2B)$ operations. 
Finally, one Newton-Raphson update takes $O(k^3)$ operations. Newton-Raphson algorithm is superlinear \citep{boyd2004convex}, thus we can regard the number of Newton-Raphson updates of constant order.  The cell probabilities can be computed in $O(kB)$ time, and the covariance matrix takes $O(k^2B)$ operations. 

To compute the quadratic approximation, $\{n_{\calR_{s,L}}, n_{\calR_{s,R}}, \bar{z}_{\calR_{s,L}}, \bar{z}_{\calR_{s,R}}\}$ can be computed in ${O}(dn_{\calR}k)$ by scanning through the samples once per coordinate. (If observations are not ordered beforehand, we will have $\tilde{O}(dn_{\calR}k)$, where $\tilde{O}$ denotes the order up to $\log$ terms.)
Adding the diagonal matrix $\Omega$ is cheap, and the matrix inversion takes $O(k^3)$ operations.
For a candidate split, the quadratic term in Proposition~\ref{prop:approx} further takes $O(k^2)$ time.
Choosing the optimal split takes $O(S)$.
In summary, the complexity is ${O}(dn_{\calR}k + k^2B + k^3 + Sk^2)$. 
\hfill\BlackBox

\subsection{Proof of Proposition~\ref{prop:approxError}}

The results are built on the expressiveness of logspline density models and tree boosting models. We first develop an intermediate property for unconditional densities, and then combine it with the representability of tree boosting models with arbitrary depth and number of trees. We will use $C$ to denote constants depending on $B$ and the concrete values may vary from line to line.

For unconditional densities, let the true log-density be $h(y) \in S_2^s$ and $\|h^{(s)}(\cdot)\|_2 \le B$. By \citet{barron1991approximation}, there exists a function $h_k \in \Omega_{k, \zeta}$ such that
\begin{align}
	\label{eq:densityApproxError}
	\begin{split}
		\|h(\cdot) - h_k(\cdot)\|_2 &\le C \delta^s \|h^{(s)}(\cdot)\|_2 \le \frac{C}{k^s} ,\\
		\|h(\cdot) - h_k(\cdot)\|_\infty &\le C \delta^{s-1/2} \|h^{(s)}(\cdot)\|_2 \le \frac{C}{k^{s-1/2}},
	\end{split}
\end{align}
where we use $\delta = 1/(k-\zeta+2) \le 2/k$. 
Next, we turn to conditional densities. For $|h(y \mid x) | \in  S_2^s$ and $\|h(\cdot \mid x)\|_\infty \le B$, by \citet[Lemma 3]{barron1991approximation}, there exists a unique vector solution $\beta(x)$ to the minimization problem 
\begin{align*}
	\beta(x) := \argmin_{\tilde{\beta}(x) \in \RR^k} D \left(f_{Y|X}(y \mid X = x) ~\|~ \kappa(y) e^{z(y)^\top \tilde{\beta}(x) - \psi(\tilde{\beta}(x))}\right).
\end{align*}
We next show $\beta(x)$ is measurable on $[0,1]^d$. In fact, define the mapping 
\begin{align*}
	\Phi (\beta) 
	&:= \EE\left[z(Y) \kappa(Y) e^{z(Y)^\top {\beta}- \psi({\beta})}\right].
\end{align*}
Again by \citet[Lemma 3]{barron1991approximation}, $\Phi$ is invertible and thus the inverse $\Phi^{-1}$ is differentiable.
By the optimality of $\beta(x)$, we have
\begin{align*}
	\Phi (\beta(x)) 
	&:= \EE\left[z(Y) \kappa(Y) e^{z(Y)^\top {\beta}(x) - \psi({\beta}(x))}\right] 
	= \EE\left[z(Y) f_{Y|X}(Y \mid X=x)\right].
\end{align*}
Notice that $\log(f_{Y|X}(Y \mid X = x)) \in S_2^s$, then $f_{Y|X}(Y \mid X = x)$ is at least $(s-1)$-th order smooth and as a result $\Phi(\beta(x))$ is differentiable with regard to $x$. Finally, since $\beta(x) = \Phi^{-1} \circ \Phi (\beta(x))$, we conclude $\beta(x)$ is differentiable thus measurable on $[0,1]^d$. Note that $\EE\left[z(Y) f_{Y|X}(Y \mid X=x)\right]$ is bounded since $|h(y \mid x)|\le B$, then $\|\beta(x)\|_\infty \le C_{\beta}$ for some $C_{\beta}$ depending on $B$.

Now we are ready to approximate $\beta(x)$ by tree boosting models. By the simple function approximation theorem, for any bounded measurable function on $[0,1]$, there exists a sequence of simple functions, i.e., linear combinations of indicator functions, such that converge to the target function point-wisely and and in $L_1$.  
Recall the tree boosting models $\{\sum_{i=1}^N a_i \1_{x^{(j)} \in I_j }, N \ge 1, I_j \subseteq[0,1] \}$, then given a measurable function $\beta_j(x)$ supported on the hyper-cube $[0,1]^d$, for any $\varepsilon > 0$, there exists a tree boosting model ${\beta}_{j,\varepsilon}(x) \in \calN_1$ such that
$|{\beta}_{j,\varepsilon}(x)| \le  |\beta(x)| + \varepsilon$ on $[0,1]^d$ and					
\begin{align*}
	|\beta_j(x)  - {\beta}_{j,\varepsilon}(x) | \le \varepsilon, \quad \forall x \in  D_{\varepsilon} \subseteq [0,1]^d, \quad m(D_{\varepsilon}) \ge 1-\varepsilon,
\end{align*}
where $D_{\varepsilon}$ is some measurable set and $m(D_{\varepsilon})$ denotes the Lebesgue measure of the subset $D_{\varepsilon}$. 
By the triangle inequality and Cauchy-Schwarz inequality, 
\begin{align*}
	\left|h(y \mid x) - z(y)^\top {\beta}_{\varepsilon}(x)\right|
	&\le \left|h(y \mid x) - z(y)^\top {\beta}(x)\right|+\left| z(y)^\top {\beta}(x)- z(y)^\top {\beta}_{\varepsilon}(x) \right| \\
	&\le \left|h(y \mid x) - z(y)^\top {\beta}(x) \right|+\|z(y)\|_2 \|{\beta}(x)- {\beta}_{\varepsilon}(x)\|_2.
\end{align*}
On $D_{\varepsilon}$, $\|{\beta}(x)- {\beta}_{\varepsilon}(x)\|_2 \le \sqrt{k}\varepsilon$, and
\begin{align*}
	\|h(\cdot \mid x) - z(\cdot)^\top {\beta}_{\varepsilon}(x)\|_2
	\le C\left(\frac{1}{k^{s}} + \sqrt{k}\varepsilon \|z(\cdot)\|_2\right),
\end{align*}
where we use~\eqref{eq:densityApproxError}.
On $D_\varepsilon^c$, since $|h(y \mid x)| \le B$ and $|{\beta}_{j,\varepsilon}(x)| \le  C_\beta + \varepsilon$, then
\begin{align*}
	\|h(\cdot \mid x) - z(\cdot)^\top {\beta}_{\varepsilon}(x)\|_\infty
	\le \|h(\cdot \mid x)\|_\infty + \| z(\cdot)^\top {\beta}_{\varepsilon}(x)\|_\infty \le C.
\end{align*}
Let $f_\varepsilon(y \mid x) := \kappa(y) e^{z(y)^\top \beta_\varepsilon(x) - \psi(\beta_\varepsilon(x))}$, then by Lemma 1 in \citet{barron1991approximation},
\begin{align}\label{proof:eq:approxError}
	\begin{split}
		&\quad~ \EE\left[D\left(f(y \mid X)~\|~ {f}_\varepsilon(y \mid X)\right)\right]\\
		&= \EE\left[D\left(f(y \mid X)~\|~ {f}_\varepsilon(y \mid X)\right),~ X \in D_\varepsilon\right] \\
		&\quad + \EE\left[D\left(f(y \mid X)~\|~ {f}_\varepsilon(y \mid X)\right),~ X \in D_\varepsilon^c \right] \\
		&\le C \left(\frac{1}{k^{2s}} +   k\varepsilon^2 \|z(\cdot)\|_2^2 + \varepsilon \right).
	\end{split}
\end{align}
Since \eqref{proof:eq:approxError} is valid for arbitrary $\varepsilon > 0$, let $\varepsilon \to 0$ and we finish the proof.
\hfill\BlackBox

\subsection{Proof of Proposition \ref{prop:iterative}}
{\bf Proof}.
Without loss of generality, assume $\calR$ is the covariate space. For simplicity, we omit the superscript of $\beta^t(x)$ and denote the natural parameter functions by $\beta(x)$.

We first show $	\gamma^* = \argmax_{{\gamma}} \ell(\calR; \beta(x) + {\gamma})$.  
For $\lambda = 0$, if $\Delta \gamma = 0$, the KKT condition of the last Poisson regression is 
\begin{align}\label{eq:prop:iterative:eq1}
\frac{1}{n}\sum_{b=1}^B  n_b z(y_b) 
= \sum_{b=1}^B \bar{p}_b(\calR; \beta(x) + \gamma) z(y_b),	
\end{align}
If in $\calR$, $Y$ is supported on the midpoints $\{y_b\}$,
\begin{align}\label{eq:prop:iterative:eq2}
\nabla\psi(\beta(x_i) + \gamma)
= \frac{1}{B} \sum_{b=1}^B p_b(\beta(x_i) + \gamma) z(y_b).
\end{align}
Therefore, by Eq.~\eqref{eq:prop:iterative:eq1} and Eq.~\eqref{eq:prop:iterative:eq2},
\begin{align}\label{eq:prop:iterative:eq3}
\begin{split}
\frac{1}{n}\sum_{i=1}^n \nabla\psi(\beta(x_i) + \gamma)
&= \frac{1}{n}\sum_{i=1}^n \sum_{b=1}^B p_b(\beta(x_i) + \gamma) z(y_b) 
= \sum_{b=1}^B z(y_b) \cdot \frac{1}{n}\sum_{i=1}^n p_b(\beta(x_i) + \gamma) \\
&= \sum_{b=1}^B \bar{p}_b(\calR; \beta(x) + \gamma) z(y_b)
= \frac{1}{n}\sum_{b=1}^B  n_b z(y_b) 
= \frac{1}{n}\sum_{i=1}^n  z(y_i),
\end{split}
\end{align}
which implies the KKT condition of maximizing $\ell(\calR; \beta(X) + {\gamma})$ is satisfied at $\gamma^*$. Since the log-likelihood $\ell(\calR; \beta(X) + {\gamma})$ is strictly concave, then $\gamma^*$ is the unique maximizer.

We next prove the algorithm converges. 
Plug in the MLE estimate of the intercept of the Poisson regression into the log-likelihood and we have
\begin{align}\label{eq:prop:iterative:eq4}
\begin{split}
\ell_{\text{poisson}}(\calR; \beta(x)+\gamma + \Delta \gamma) 
=& \ell_{\text{poisson}}(\calR; \beta(x)+\gamma)  + \sum_{b=1}^B n_b z (y_b)^\top \Delta \gamma \\
&- n \underbrace{\log\left(\sum_{b=1}^B \bar{p}_b(\calR; \beta(x) + \gamma) \exp\left\{z(y_b)^\top \Delta \gamma\right\}\right)}_{:=(I)}.
\end{split}
\end{align}
By Jensen's inequality,
\begin{align}\label{eq:prop:iterative:eq5}
\begin{split}
(I) 
&=\log\left(\frac{1}{n}\sum_{i=1}^n\sum_{b=1}^B {p}_b(\beta(x_i) + \gamma) \exp\left\{z^\top(y_b) \Delta \gamma\right\}\right)\\
&\ge \frac{1}{n} \sum_{i=1}^n\log\left(\sum_{b=1}^B {p}_b(\beta(x_i) + \gamma) \exp\left\{z^\top(y_b) \Delta \gamma\right\}\right).
\end{split}
\end{align}
Notice that  
\begin{align}\label{eq:prop:iterative:eq6}
\begin{split}
&~\quad \psi(\beta(x_i) + \gamma + \Delta \gamma) - \psi(\beta(x_i) + \gamma) \\
&= -\log\left(\frac{\sum_{b=1}^B e^{z(y_b)^\top \left(\gamma + \Delta\gamma\right)}}{\sum_{b=1}^B e^{z(y_b)^\top \gamma }}\right)\\
&= -\log\left(\sum_{b=1}^B {p}_b(\beta(x_i) + \gamma) \exp\left\{z(y_b)^\top \Delta \gamma\right\}\right).
\end{split}
\end{align}
Therefore, by Eq.~\eqref{eq:prop:iterative:eq4}, Eq.~\eqref{eq:prop:iterative:eq5} and Eq.~\eqref{eq:prop:iterative:eq6},
\begin{align*}
&~\quad \ell(\calR; \beta(x) + \gamma + \Delta \gamma) - \ell(\calR; \beta(x) + \gamma)\\
&= \sum_{i=1}^n z(y_i)^\top \Delta \gamma -\log\left(\sum_{b=1}^B {p}_b(\beta(x_i) + \gamma) \exp\left\{z(y_b)^\top \Delta \gamma\right\}\right) \\
&\ge \sum_{i=1}^n z(Y_i)^\top \Delta \gamma - n \cdot (I) 
= \sum_{b=1}^B n_b z(y_b)^\top \Delta \gamma - n \cdot (I) \\
&= \ell_{\text{poisson}}(\calR; \beta(x)+\gamma + \Delta \gamma) - \ell_{\text{poisson}}(\calR; \beta(x)+\gamma).
\end{align*}  
By the strong concavity of $\ell_{\text{poisson}}$, there exists a universal $\delta > 0$ such that
\begin{align*}
\ell_{\text{poisson}}(\calR; \beta(x)+\gamma + \Delta \gamma) - \ell_{\text{poisson}}(\calR; \beta(x)+\gamma)
\ge \delta \|\Delta \gamma\|_2^2.
\end{align*}
Thus,
\begin{align*}
&~\quad\ell(\calR; \beta(x) + \gamma + \Delta \gamma) - \ell(\calR; \beta(x) + \gamma) \\
&\ge \ell_{\text{poisson}}(\calR; \beta(x)+\gamma + \Delta \gamma) - \ell_{\text{poisson}}(\calR; \beta(x)+\gamma)
\ge \delta \|\Delta \gamma\|_2^2.
\end{align*} 	
Since the conditional log-likelihood $\ell(\calR; \beta(x) + \gamma)$ is bounded, thus can not increase linearly forever. Therefore, $\|\Delta \gamma\|_2 \to 0$, i.e. the algorithm converges.
\hfill\BlackBox

\subsection{Proof of Proposition~\ref{prop:approx2}}
{\bf Proof}.
The proof follows that of Proposition~\ref{prop:approx}. We simplify the difference of log-likelihoods,
\begin{align}\label{eq:proof:delta:2}
\begin{split}
\Delta \ell^t(\calR, s)
=& \sum_{x_i \in \calR_L}\left(z(y_i)^\top \gamma^t_{\calR_{s,L}} - \psi(\beta^t(X_i) + \gamma^t_{\calR_{s,L}})\right)  \\
&+ \sum_{x_i \in \calR_R} \left(z(y_i)^\top \gamma^t_{\calR_{s,R}} - \psi(\beta^t(X_i) + \gamma^t_{\calR_{s,R}}) \right) \\
&-  \sum_{x_i \in \calR}\left( z(y_i)^\top \gamma^t_{\calR} - \psi(\beta^t(X_i) + \gamma^t_{\calR})\right) \\
&- \lambda n_{\calR_{s,L}} \gamma^{t \top}_{\calR_{s,L}} \Omega \gamma^t_{\calR_{s,L}} - \lambda n_{\calR_{s,R}} \gamma^{t \top}_{\calR_{s,R}} \Omega \gamma^t_{\calR_{s,R}} +  \lambda n_{\calR} \gamma^{t \top}_{\calR} \Omega \gamma^t_{\calR}\\
=& 	n_{\calR_{s,L}} \bar{z}_{\calR_{s,L}}^\top  (\gamma^t_{\calR_{s,L}} - \gamma^t_{\calR}) + n_{\calR_{s,R}} \bar{z}_{\calR_{s,R}}^\top  (\gamma^t_{\calR_{s,R}} - \gamma^t_{\calR}) \\
&- \sum_{x_i \in \calR_{s,L}}\psi(\beta^t(X_i) + \gamma^t_{\calR_{s,L}})  - \psi(\beta^t(X_i) + \gamma^t_{\calR}) \\
&- \sum_{x_i \in \calR_{s,R}}\psi(\beta^t(X_i) + \gamma^t_{\calR_{s,R}})  - \psi(\beta^t(X_i) + \gamma^t_{\calR}) \\
&- \lambda n_{\calR_{s,L}}\gamma^{t \top}_{\calR_{s,L}} \Omega \gamma^t_{\calR_{s,L}} - \lambda n_{\calR_{s,R}} \gamma^{t \top}_{\calR_{s,R}} \Omega \gamma^t_{\calR_{s,R}} +  \lambda n_{\calR} \gamma^{t \top}_{\calR} \Omega \gamma^t_{\calR}.
\end{split}
\end{align} 	
The score equation of $\gamma^t_{\calR}$ implies
\begin{align}\label{eq:proof:betaL:2}
\bar{z}_{\calR}
= \frac{1}{n_{\calR}} \sum_{i = 1}^n \nabla\psi(\beta^t(X_i) + \gamma^t_{\calR}) + 2 \lambda \Omega \gamma^t_{\calR},
\end{align} 
and similarly for $\gamma^t_{\calR_{s,L}}$, $\gamma^t_{\calR_{s,R}}$.
Plug Eq.~\eqref{eq:proof:betaL:2} into Eq.~\eqref{eq:proof:delta:2}, we obtain
\begin{align}\label{eq:proof:betaL3:2}
\begin{split}
- \Delta \ell(\calR, s) 
=& \sum_{x_i \in \calR_{s,L}} \nabla\psi(\beta^t(X_i) + \gamma^t_{\calR_{s,L}})^\top (\gamma^t_{\calR_{s,L}} - \gamma^t_{\calR}) \\
&+ \sum_{x_i \in \calR_{s,R}} \nabla\psi(\beta^t(X_i) +\gamma^t_{\calR_{s,R}})^\top (\gamma^t_{\calR_{s,R}} - \gamma^t_{\calR}) \\
&- \sum_{x_i \in \calR_{s,L}}\left(\psi(\beta^t(X_i) +\gamma^t_{\calR_{s,L}})  - \psi(\beta^t(X_i) +\gamma^t_{\calR}) \right) \\
&- \sum_{x_i \in \calR_{s,R}}\left(\psi(\beta^t(X_i) +\gamma^t_{\calR_{s,R}})  - \psi(\beta^t(X_i) +\gamma^t_{\calR}) \right)\\
&+ \lambda n_{\calR_{s,L}} \left(\gamma^t_{\calR_{s,L}} - \gamma^t_{\calR}\right)^\top \Omega \left(\gamma^t_{\calR_{s,L}} - \gamma^t_{\calR}\right) \\
&+ \lambda n_{\calR_{s,R}} \left(\gamma^t_{\calR_{s,R}} - \gamma^t_{\calR}\right)^\top \Omega \left(\gamma^t_{\calR_{s,R}} - \gamma^t_{\calR}\right).
\end{split}
\end{align}
By the Taylor expansion of  $\psi(\beta)$ and $\nabla\psi(\beta)$, 
\begin{align}\label{eq:proof:betaL4:2}
\begin{split}
&\nabla\psi(\beta^t(X_i)+\gamma^t_{\calR_{s,L}})^\top (\gamma^t_{\calR_{s,L}} -\gamma^t_{\calR}) - (\psi(\beta^t(X_i)+\gamma^t_{\calR_{s,L}}) - \psi(\beta^t(X_i)+\gamma^t_{\calR}))\\
=& \nabla\psi(\beta^t(X_i)+\gamma^t_{\calR_{s,L}})^\top (\gamma^t_{\calR_{s,L}} -\gamma^t_{\calR}) - \nabla\psi(\beta^t(X_i)+\gamma^t_{\calR})^\top (\gamma^t_{\calR_{s,L}} -\gamma^t_{\calR}) \\
&- \frac{1}{2} (\gamma^t_{\calR_{s,L}} -\gamma^t_{\calR})^\top \nabla^2\psi(\beta^t(X_i)+\gamma^t_{\calR})(\gamma^t_{\calR_{s,L}} -\gamma^t_{\calR}) + O(\|(\gamma^t_{\calR_{s,L}} - \gamma^t_{\calR})\|_2^3)\\
=& \frac{1}{2} (\gamma^t_{\calR_{s,L}} -\gamma^t_{\calR})^\top \nabla^2\psi(\beta^t(X_i)+\gamma^t_{\calR})(\gamma^t_{\calR_{s,L}} -\gamma^t_{\calR}) + O(\|(\gamma^t_{\calR_{s,L}} - \gamma^t_{\calR})\|_2^3).
\end{split}
\end{align} 
Plug Eq.~\eqref{eq:proof:betaL4:2} into Eq.~\eqref{eq:proof:betaL3:2}, we get
\begin{align}\label{eq:proof:betaL5:2}
\begin{split}
&-\Delta \ell(\calR, s) \\
=& \frac{1}{2} \sum_{x_i \in \calR_{s,L}} (\gamma^t_{\calR_{s,L}} -\gamma^t_{\calR})^\top \nabla^2\psi(\beta^t(X_i)+\gamma^t_{\calR})(\gamma^t_{\calR_{s,L}} -\gamma^t_{\calR}) \\
&+ \frac{1}{2} \sum_{x_i \in \calR_{s,R}}  (\gamma^t_{\calR_{s,R}} -\gamma^t_{\calR})^\top \nabla^2\psi(\beta^t(X_i)+\gamma^t_{\calR})(\gamma^t_{\calR_{s,R}} -\gamma^t_{\calR})  \\
&+ \lambda n_{\calR_{s,L}} \left(\gamma^t_{\calR_{s,L}} - \gamma^t_{\calR}\right)^\top \Omega \left(\gamma^t_{\calR_{s,L}} - \gamma^t_{\calR}\right) \\
&+ \lambda n_{\calR_{s,R}} \left(\gamma^t_{\calR_{s,R}} - \gamma^t_{\calR}\right)^\top \Omega \left(\gamma^t_{\calR_{s,R}} - \gamma^t_{\calR}\right) \\
&+ O\left(\|(\gamma^t_{\calR_{s,L}} - \gamma^t_{\calR})\|_2^3 + \|(\gamma^t_{\calR_{s,R}} - \gamma^t_{\calR})\|_2^3\right) \\
=& \frac{1}{2} (\gamma^t_{\calR_{s,L}} -\gamma^t_{\calR})^\top \left(\sum_{x_i \in \calR_{s,L}} \nabla^2\psi(\beta^t(X_i) +\gamma^t_{\calR}) + 2 \lambda n_{\calR_{s,L}} \Omega \right)(\gamma^t_{\calR_{s,L}} -\gamma^t_{\calR}) \\
&+ \frac{1}{2}  (\gamma^t_{\calR_{s,R}} -\gamma^t_{\calR})^\top \left(\sum_{x_i \in \calR_{s,R}} \nabla^2\psi(\beta^t(X_i) + \gamma^t_{\calR}) + 2 \lambda n_{\calR_{s,R}}  \Omega\right) (\gamma^t_{\calR_{s,R}} -\gamma^t_{\calR}) \\
&+ O\left(\|(\gamma^t_{\calR_{s,L}} - \gamma^t_{\calR})\|_2^3 + \|(\gamma^t_{\calR_{s,R}} - \gamma^t_{\calR})\|_2^3\right).
\end{split}
\end{align}
By Eq.~\eqref{eq:proof:betaL:2},
\begin{align*}
&\frac{n_{\calR_{s,R}}}{n_{\calR}}\left(\bar{z}_{\calR_{s,L}}  - \bar{z}_{\calR_{s,R}}\right)
= \bar{z}_{\calR_{z,L}}  - \bar{z}_{\calR} \\
=& \frac{1}{n_{\calR_{s,L}}} \sum_{x_i \in \calR_{s,L}} \nabla^2\psi(\beta^t(X_i) + \lambda_\calR^t) (\lambda_{\calR_{s,L}}^t - \lambda_\calR^t) \\
&+ \frac{n_{\calR_{s,R}}}{n_{\calR}} \left(\bar{z}^t_{\calR_{s,L}} - \bar{z}^t_{\calR_{s,R}}\right) + O\left(\|(\gamma^t_{\calR_{s,L}} - \gamma^t_{\calR})\|_2^2\right),
\end{align*}
where $\bar{z}_{\calR_{s,L}}^t := \sum_{x_i \in \calR_{s,L}} \nabla\psi(\beta^t(X_i) + \lambda_\calR^t)/n_{\calR_{s,L}}$, and similarly for $\bar{z}_{\calR_{s,R}}^t$.
Finally, by the assumption of invertibility,
\begin{align}\label{eq:proof:betaL6:2}
\begin{split}
&\quad~\gamma^t_{\calR_{s,L}} - \gamma^t_{\calR}\\
&= \left(\frac{1}{n_{\calR_{s,L}}} \sum_{x_i \in \calR_{s,L}}\nabla^2\psi(\beta^t(X_i) + \gamma^t_{\calR}) + 2 \lambda \Omega\right)^{-1}\frac{n_{\calR_{s,R}}}{n_{\calR}}\\
&\quad~\left(\left(\bar{z}_{\calR_{s,L}}  - \bar{z}_{\calR_{s,R}}\right) -  \left(\bar{z}^t_{\calR_{s,L}} - \bar{z}^t_{\calR_{s,R}}\right)\right) 
+ O\left(\|(\gamma^t_{\calR_{s,L}} - \gamma^t_{\calR})\|_2^2\right) \\
&= \frac{n_{\calR_{s,R}}}{n_{\calR}} \left(\frac{1}{n_{\calR_{s,L}}} \sum_{x_i \in \calR_{s,L}}\nabla^2\psi(\beta^t(X_i) + \gamma^t_{\calR}) + 2 \lambda \Omega\right)^{-1}\left(\bar{r}^t_{\calR_{z,L}} - \bar{r}^t_{\calR_{z,R}}\right) \\ 
&~\quad + O\left(\|(\gamma^t_{\calR_{s,L}} - \gamma^t_{\calR})\|_2^2\right).
\end{split}
\end{align} 
Plug Eq.~\eqref{eq:proof:betaL6:2} into Eq.~\eqref{eq:proof:betaL5:2},
\begin{align*}
&~-\Delta \ell(\calR, s) \\
&= \frac{n_{\calR_{s,L}}n_{\calR_{s,R}}^2}{2n_{\calR}^2} (\bar{r}^t_{\calR_{s,L}} -\bar{r}^t_{\calR_{s,R}})^\top \left(\frac{1}{n_{\calR_{s,L}}}\sum_{x_i \in \calR_{s,L}}\nabla^2\psi(\beta^t(X_i) +\gamma^t_{\calR}) + 2 \lambda \Omega \right)^{-1}\\
&\quad~(\bar{r}^t_{\calR_{s,L}} - \bar{r}^t_{\calR_{s,R}}) 
+ \frac{n_{\calR_{s,L}}^2n_{\calR_{s,R}}}{2n_{\calR}^2} (\bar{r}^t_{\calR_{s,L}} -\bar{r}^t_{\calR_{s,R}})^\top \\
&\quad~ \left(\frac{1}{n_{\calR_{s,R}}}\sum_{x_i \in \calR_{s,R}}\nabla^2\psi(\beta^t(X_i) +\gamma^t_{\calR}) + 2 \lambda \Omega \right)^{-1}(\bar{r}^t_{\calR_{s,L}} - \bar{r}^t_{\calR_{s,R}}) \\
&\quad~+ O\left(\|(\bar{r}^t_{\calR_{s,L}} - \bar{r}^t_{\calR})\|_2^3 + \|(\bar{r}^t_{\calR_{s,R}} - \bar{r}^t_{\calR})\|_2^3\right), 
\end{align*}
and we finish the proof.
\hfill\BlackBox

\subsection{Proof of  Proposition \ref{prop:computation2}}
{\bf Proof}.
We first evaluate the complexity of the fitting step of LinCDE boosting. The computation of offset is $O(n_{\calR}kB)$, and we store the cell probabilities $p_b(\beta^t(x_i))$. The first step of the fitting runs a penalized Lindsey's method and takes $O(n_{\calR} + k^2B + k^3)$ as shown in the proof of \ref{prop:computation}. The second step in the fitting takes $O(n_{\calR} B + kB)$, and we update the cell probabilities to $p_b(\beta^t(x_i) + \gamma^t_{\calR})$. 

It takes $O(n_{\calR}k^2B)$ to compute the surrogate normalization matrix $\tilde{\Psi}^t(\gamma^t_\calR)$ and an extra $O(k^3)$ for matrix inversion. It takes $\tilde{O}(dn_{\calR}kB)$ to compute all average residuals $\bar{r}_{\calR}^t$.
Finally, the quadratic approximation for all candidate splits and choosing the optimal one takes $O(S k^2)$.
In summary, the time complexity is $\tilde{O}(dn_{\calR}kB + n_{\calR}k^2B + k^3 + Sk^2)$. 
\hfill\BlackBox

\section{Additional Figures}\label{appn:sec:figure}
In this section, we provide additional numerical results. 

\subsection{Additional Simulations}

\begin{figure}[bt]
	\centering
	\begin{minipage}{6.5cm}
		\centering  
		\includegraphics[scale = 0.7]{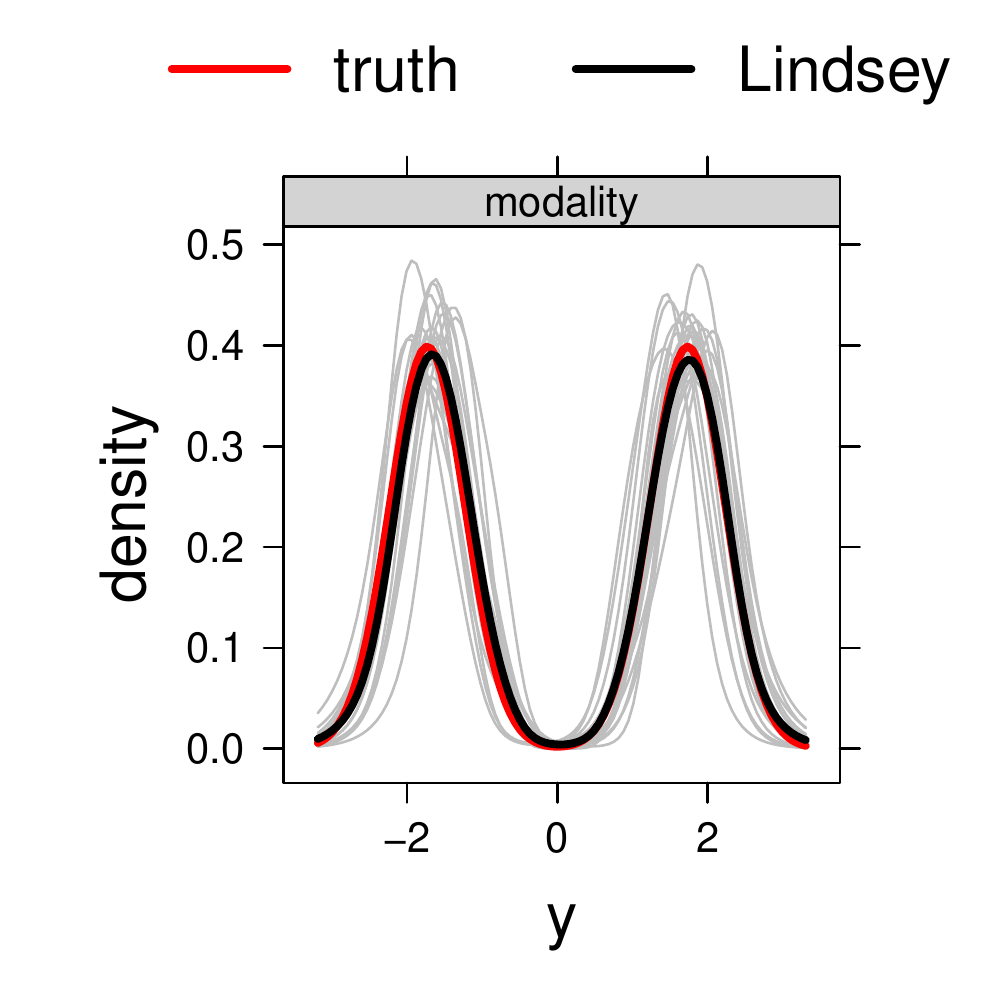}  
	\end{minipage}
	\begin{minipage}{6cm}
		\centering  
		\includegraphics[scale = 0.7]{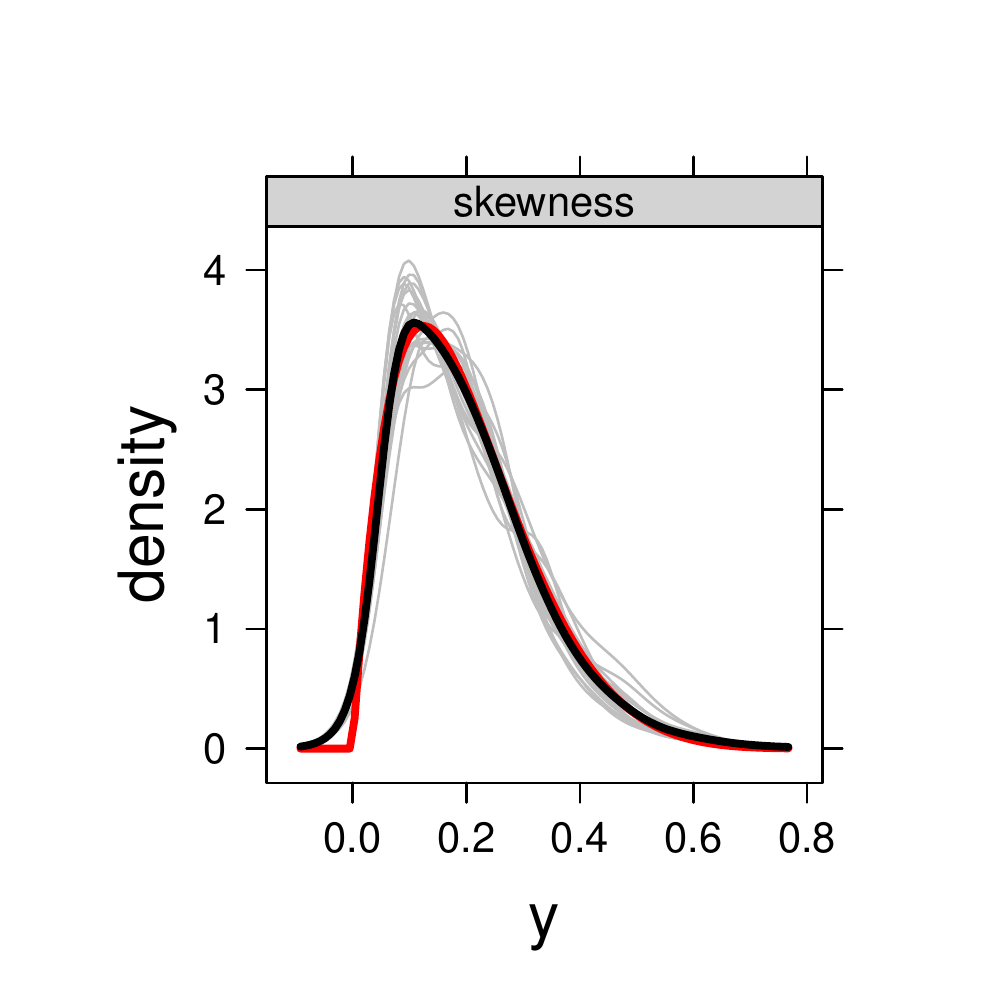}  
	\end{minipage}
	\caption{{Estimation of bimodal and skewed densities
			using our regularized Lindsey's method. In the left panel, the true density is a Gaussian mixture, and in the right panel, the true density is a beta distribution. In each trial, we sample $400$ observations. We generate $10$ natural cubic splines with knots equally spread across the range of observations, 
			and tune the penalty parameter $\lambda$ to
			achieve an effective $5$ degrees of freedom. 
			We repeat each setting $100$ times and plot the average fits against the true densities. }}
	\label{fig:LindseyExample}
\end{figure}

Figure~\ref{fig:LindseyExample} displays the performance of Lindsey's
method in two toy examples. One target density is bimodal, and the other is skewed. In Lindsey's method, we use natural cubic splines, which are arguably the most commonly used splines because they provide good and seamless fits and are easy to implement, and transform them as discussed. 
In both examples, the estimated densities of Lindsey's method match
the true densities quite closely, except for tiny gaps at boundaries and peaks.

\begin{figure}[bt]
	\centering
	\begin{minipage}{7cm}
		\centering  
		\includegraphics[width  = 7cm, height = 7cm]{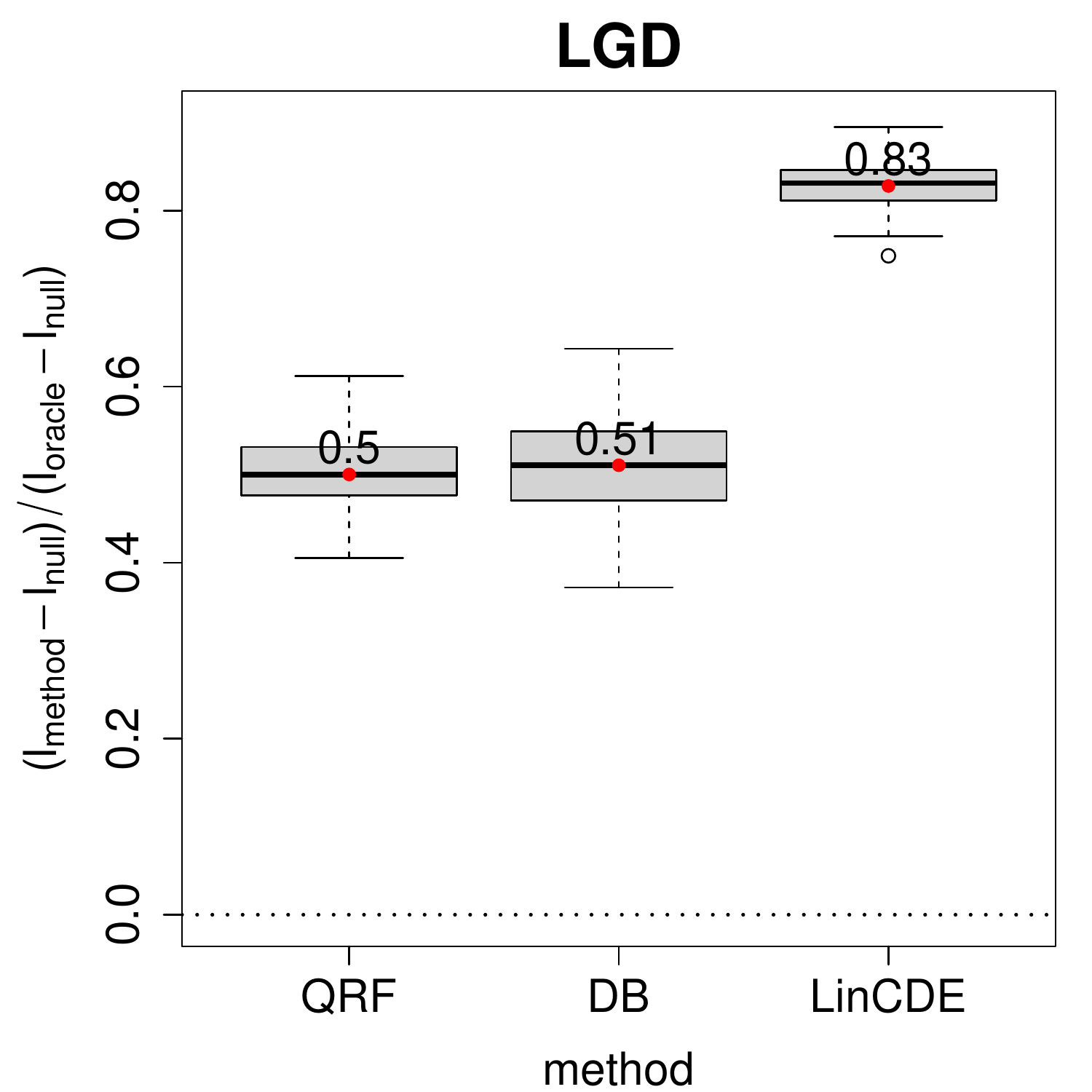} 
	\end{minipage}
	\begin{minipage}{7cm}
		\centering  
		\includegraphics[width  = 7cm, height = 7cm]{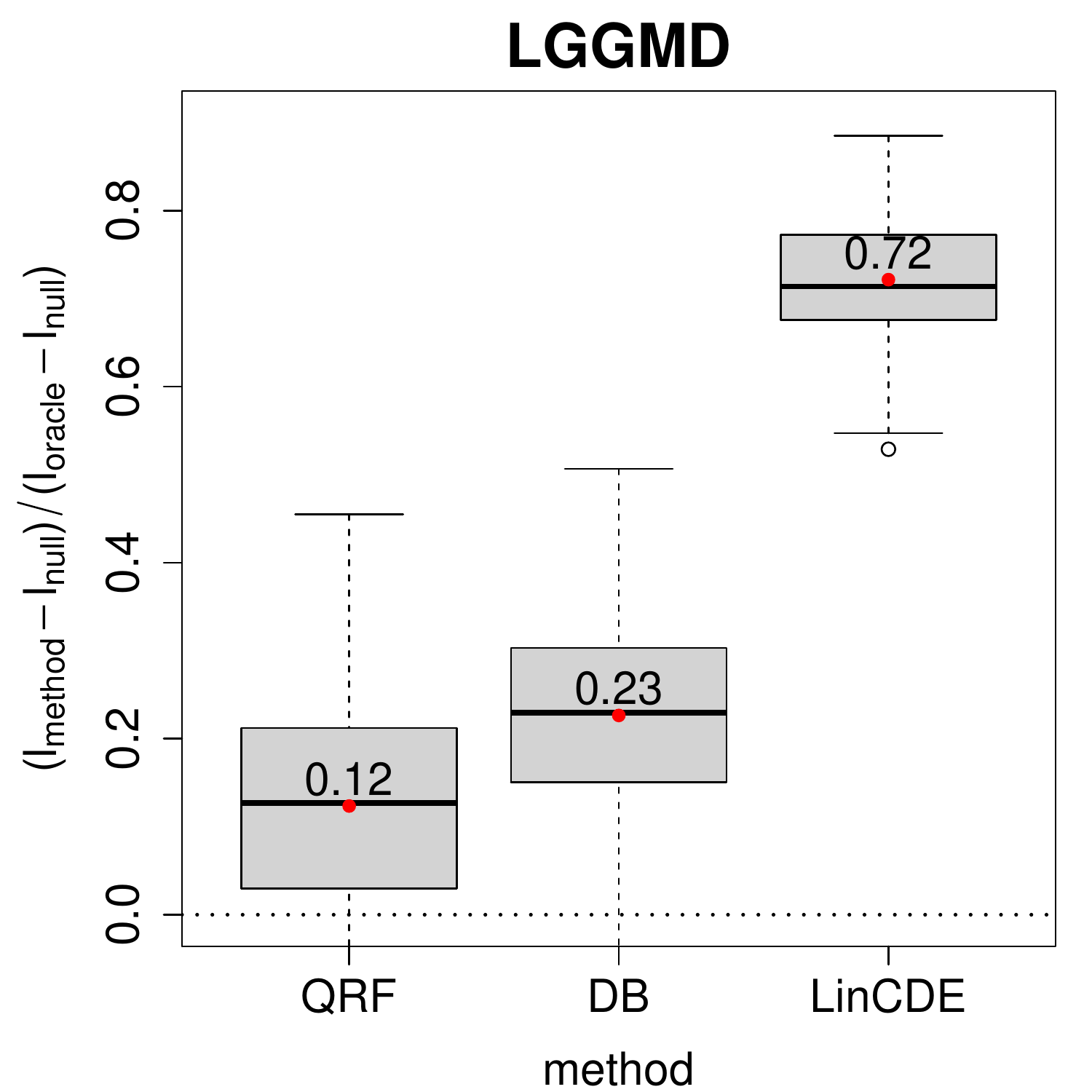}  
	\end{minipage}
	\caption{{Box plots of goodness-of-fit measure \eqref{eq:R2} in the \textit{LGD} (left panel) and \textit{LGGMD} (right panel) settings.  The densities are computed as \eqref{eq:quantileToPdf2} with $50$ bins. QRF stands for quantile regression forest, DB stands for distribution boosting, LinCDE stands for LinCDE boosting respectively.}}
	\label{fig:R2Bin50}
\end{figure}

Figure~\ref{fig:R2Bin50} presents the goodness-of-fit measure \eqref{eq:R2} of the three methods under the \textit{LGD} and \textit{LGGMD} settings with $50$ bins. LinCDE boosting benefits from finer grids. In contrast, distribution boosting and quantile regression forest are hurt by larger numbers of bins due to higher variances, especially in the \textit{LGGMD} setting where the densities themselves are bumpy. 

\begin{figure}[bt]
	\centering
	\begin{minipage}{7cm}
		\centering  
		\includegraphics[width  = 7cm, height = 7cm]{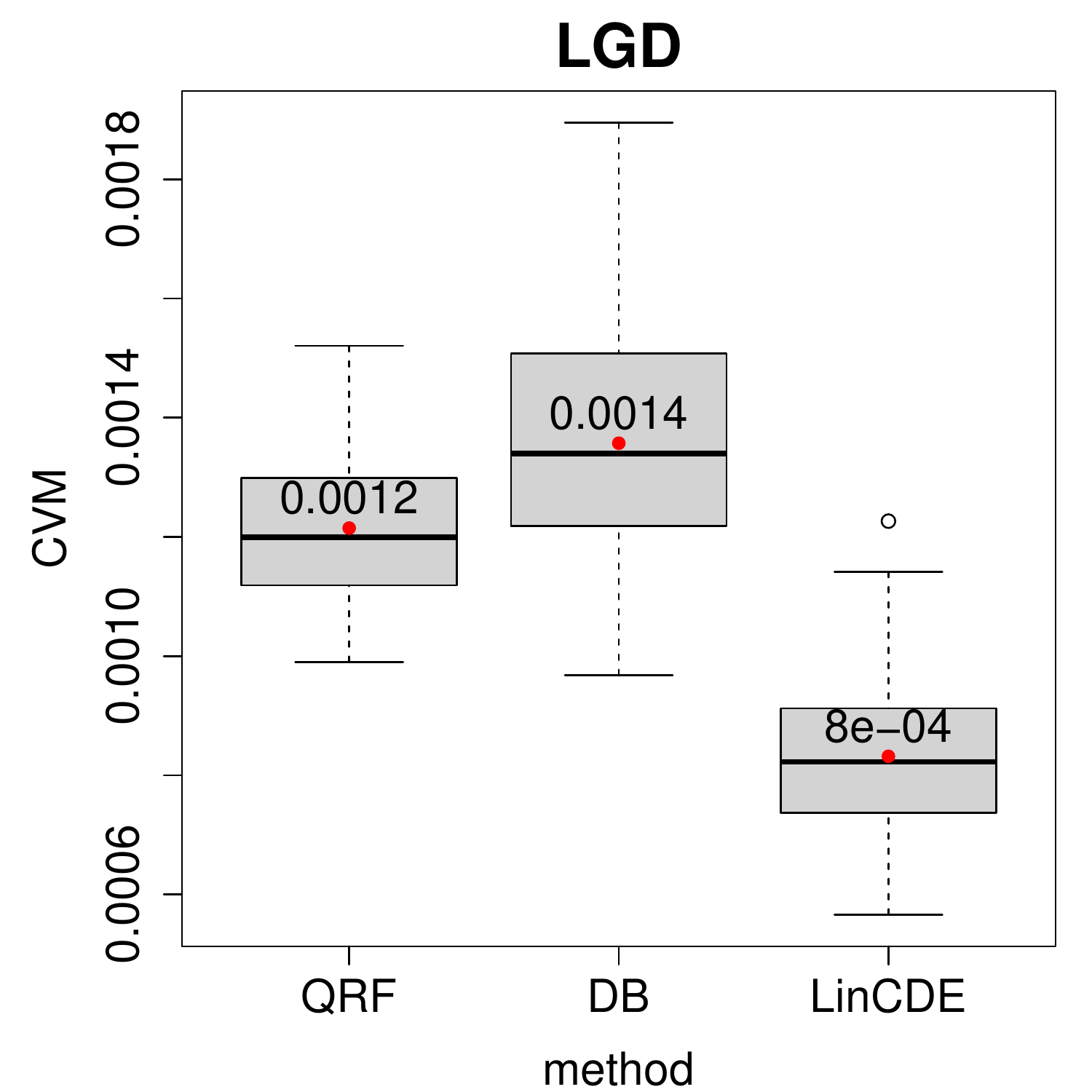} 
	\end{minipage} 
	\begin{minipage}{7cm}
		\centering  
		\includegraphics[width  = 7cm, height = 7cm]{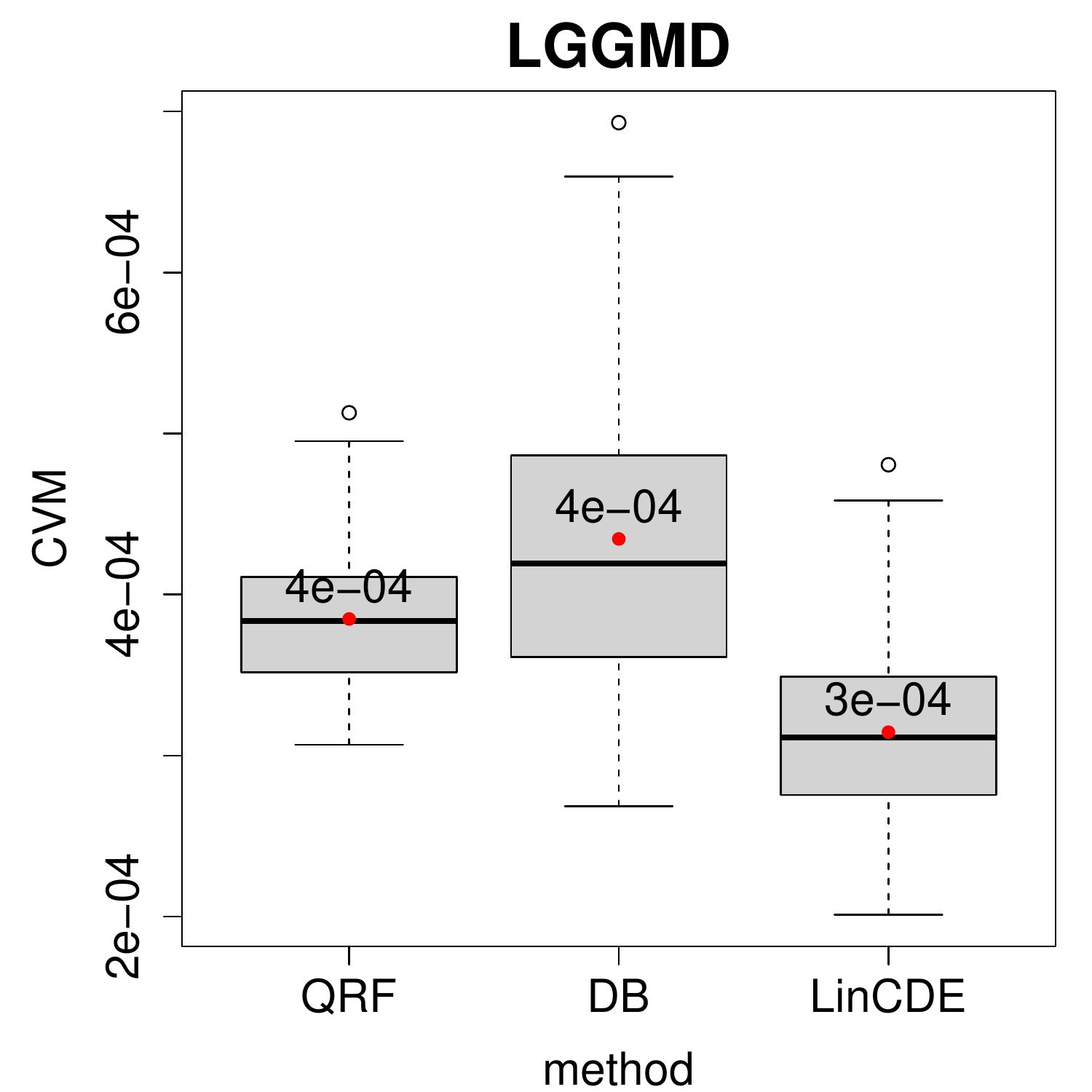}  
	\end{minipage}
	\caption{{Box plots of CVM distance \eqref{eq:CVM} in the \textit{LGD} (left panel) and \textit{LGGMD} (right panel) settings. CVM is a metric naturally defined for conditional CDF estimates \citep{friedman2019contrast}, and a smaller value indicates a better estimate.}}
	\label{fig:CVMDist}
\end{figure}

Similar to AAE, another commonly used CDF-based metric is the cram/'er-von Mises distance \citep{friedman2019contrast}:
\begin{align}\label{eq:CVM}
\text{CVM} = \frac{1}{n} \sum_{i=1}^n \frac{1}{m} \sum_{j=1}^m 
\left(\hat{F}(q(u_j \mid x_i) \mid x_i) - F(q(u_j\mid x_i) \mid x_i)\right)^2,
\end{align}
where $\{u_j\}$ is an evenly spaced grid on $[0,1]$, and $q(u \mid x)$ denotes the $u$ quantile at the covariate value $x$.
Figure~\ref{fig:CVMDist} depicts the cram/'er-von Mises distance. In both settings, LinCDE boosting outperforms the other two. The results are consistent with those of AAE.

\begin{figure}[bt]
	\centering
	\begin{minipage}{7cm}
		\centering  
		\includegraphics[width  = 7cm, height = 7cm]{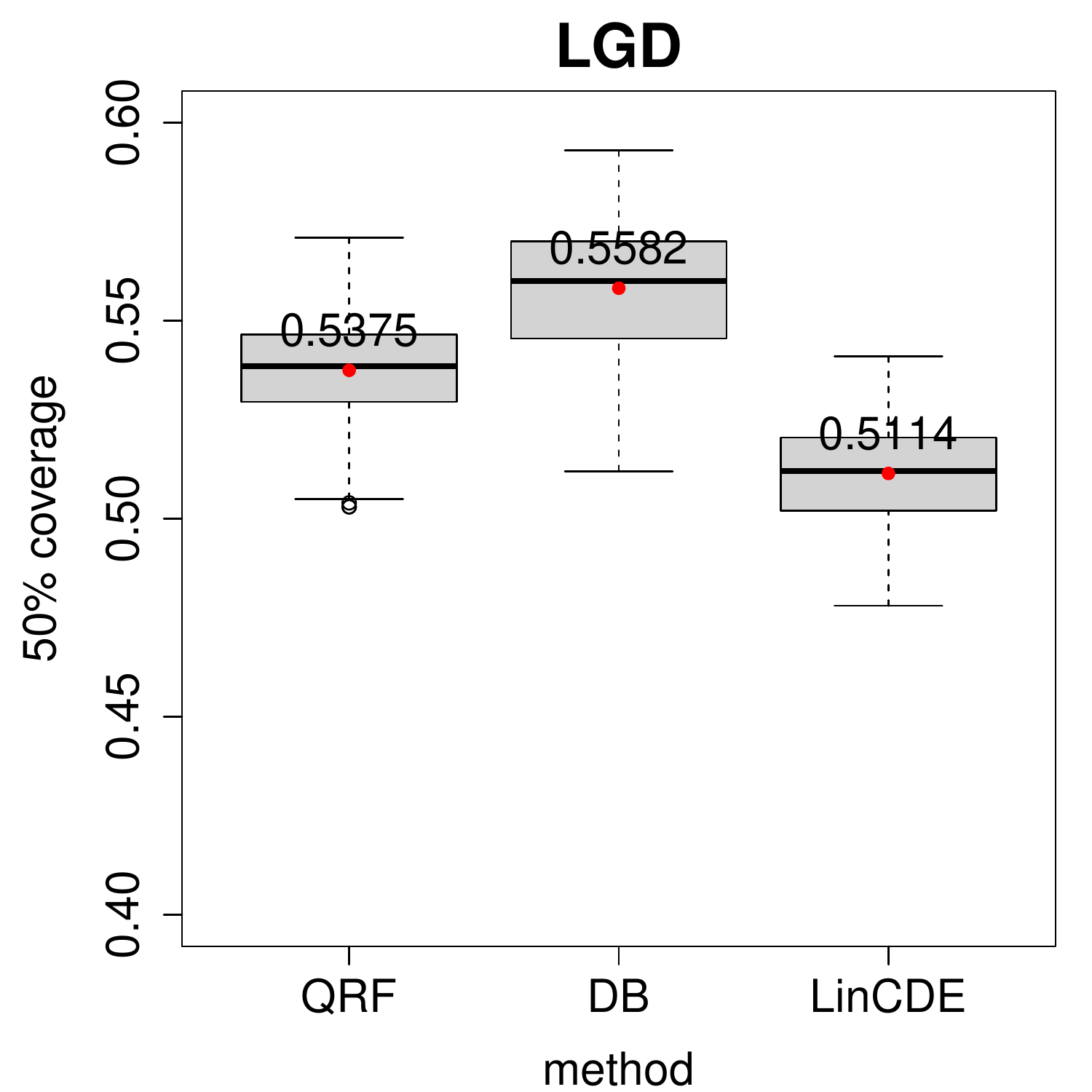} 
	\end{minipage}
	\begin{minipage}{7cm}
		\centering  
		\includegraphics[width  = 7cm, height = 7cm]{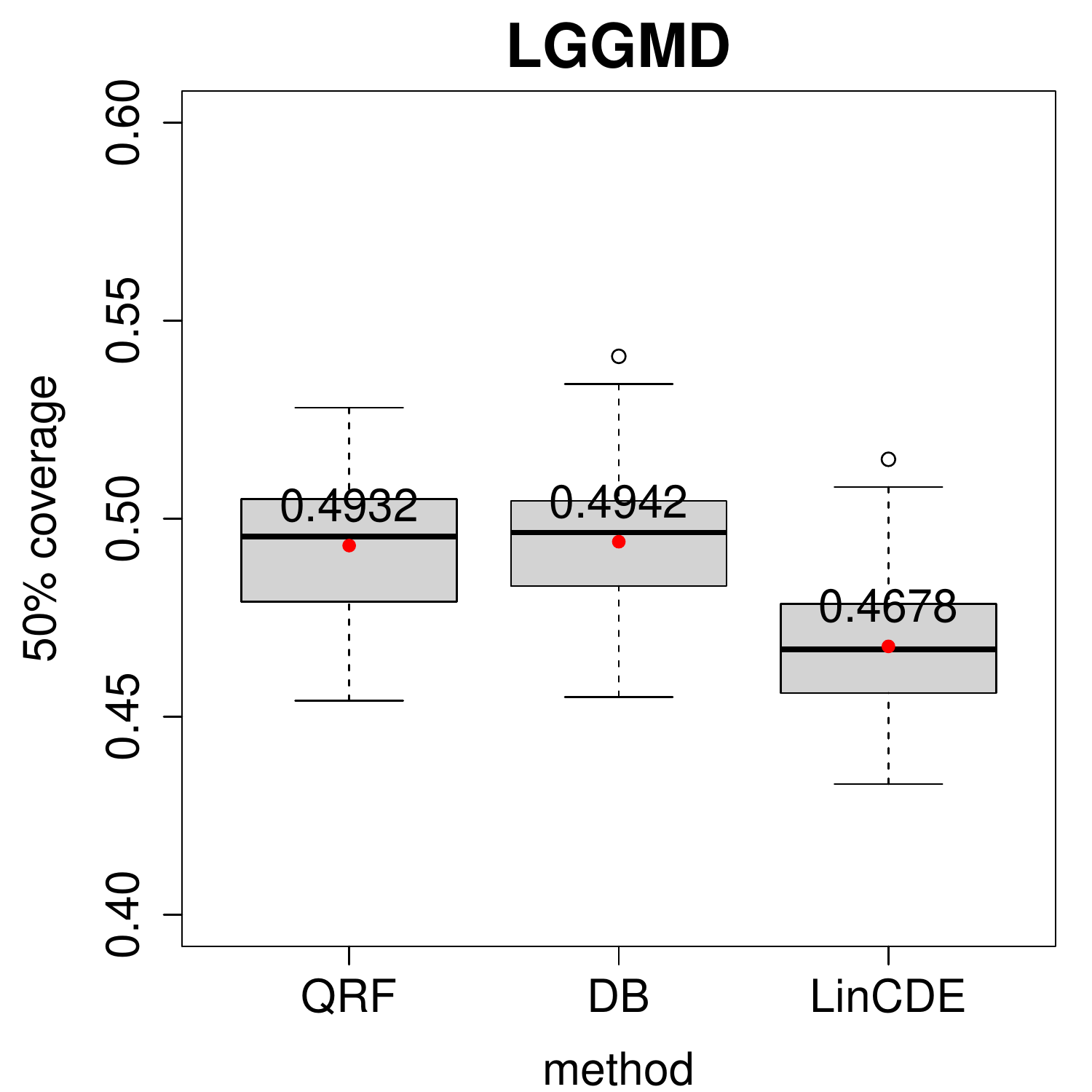}  
	\end{minipage}
	\caption{{Coverage of $50\%$ prediction intervals (ideally $50\%$) in the \textit{LGD} (left panel) and \textit{LGGMD} (right panel) settings.}} 
	\label{fig:CICoverage50}
\end{figure}

In Figure~\ref{fig:CICoverage50}, we plot the coverages of the $50\%$ prediction intervals. The results are consistent with those of $90\%$ prediction intervals.

\subsection{Air Pollution Data}

The air pollution data \citep{wu2020air} focuses on PM2.5 exposures in the United States.\footnote{PM2.5: particulate particles $2.5$ microns or less in diameter. The data represents $98\%$ of the population of the United States.} The responses are $3108$ county-level PM2.5 exposures averaged from $2000$ to $2018$. We incorporate $16$ weather, socio-economic, and demographic covariates, such as winter relative humidity, house value, and proportion of white people. The target is estimating the conditional density of the average PM2.5 exposure. We split the data into training, validation, and test (proportion 2:1:1), and tune on the hold-out validation data. We also identify influential covariates which may help find the culprits of air pollution and manage regional air quality. 

The PM2.5 exposure varies vastly across counties. For example, the
average PM2.5 of west coast counties may soar up to $12 \mu g/m^3$ due
to frequent wildfires, while those of rural areas in Central America
are typically below $8 \mu g/m^3$. The difference in PM2.5 levels
induces the disjoint support issue for LinCDE in
Section~\ref{sec:pretreatment}, and thus we employ the centering. The
comparisons of log-likelihood and quantile losses are summarized in
Table~\ref{tab:realData2}. Centered LinCDE performs the best in log-likelihood and is comparable in quantile losses. 

\begin{figure}[bt]
	\centering
	\begin{minipage}{7cm}
		\centering  
		\includegraphics[scale = 0.45]{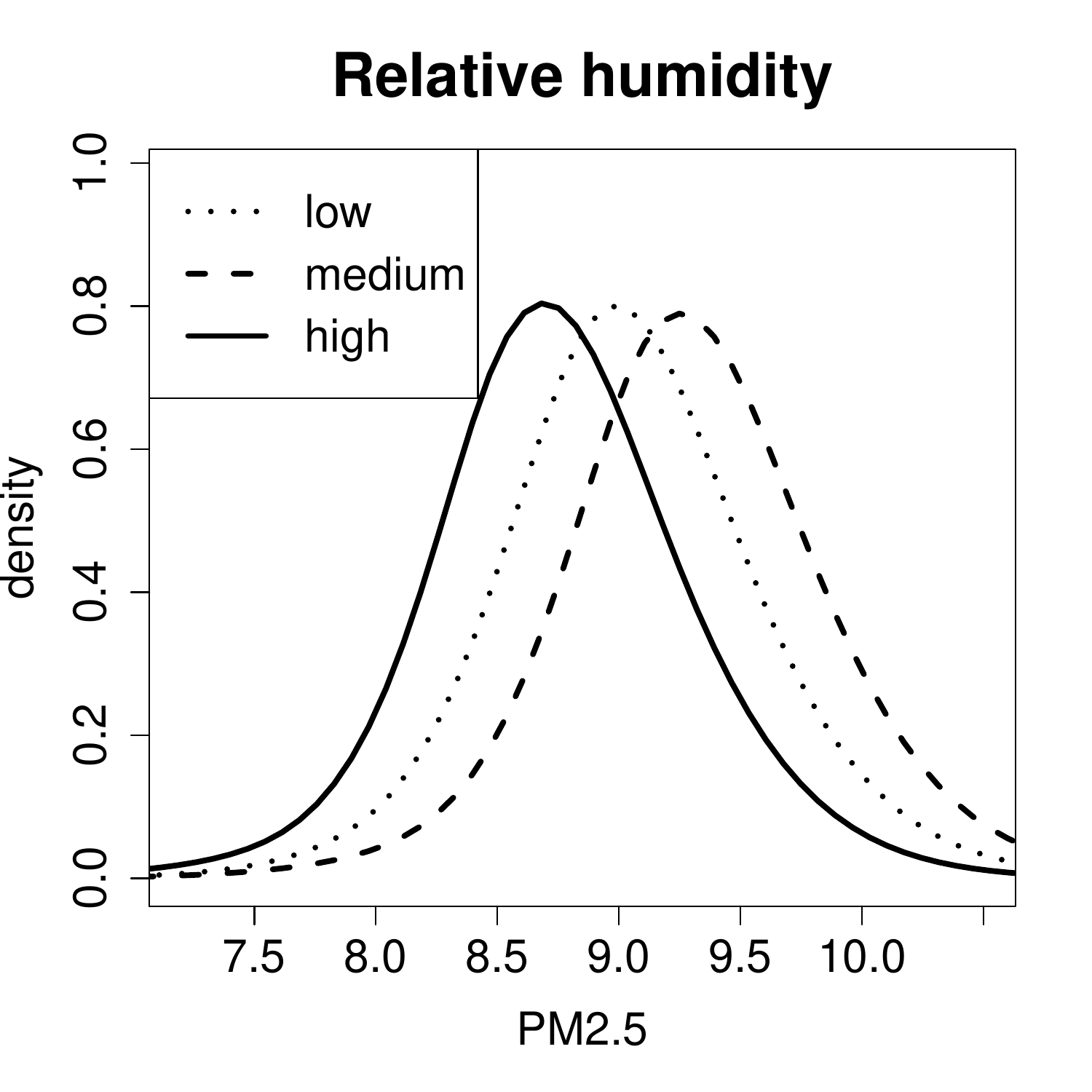}  
	\end{minipage}
	\begin{minipage}{7cm}
		\centering  
		\includegraphics[scale = 0.45]{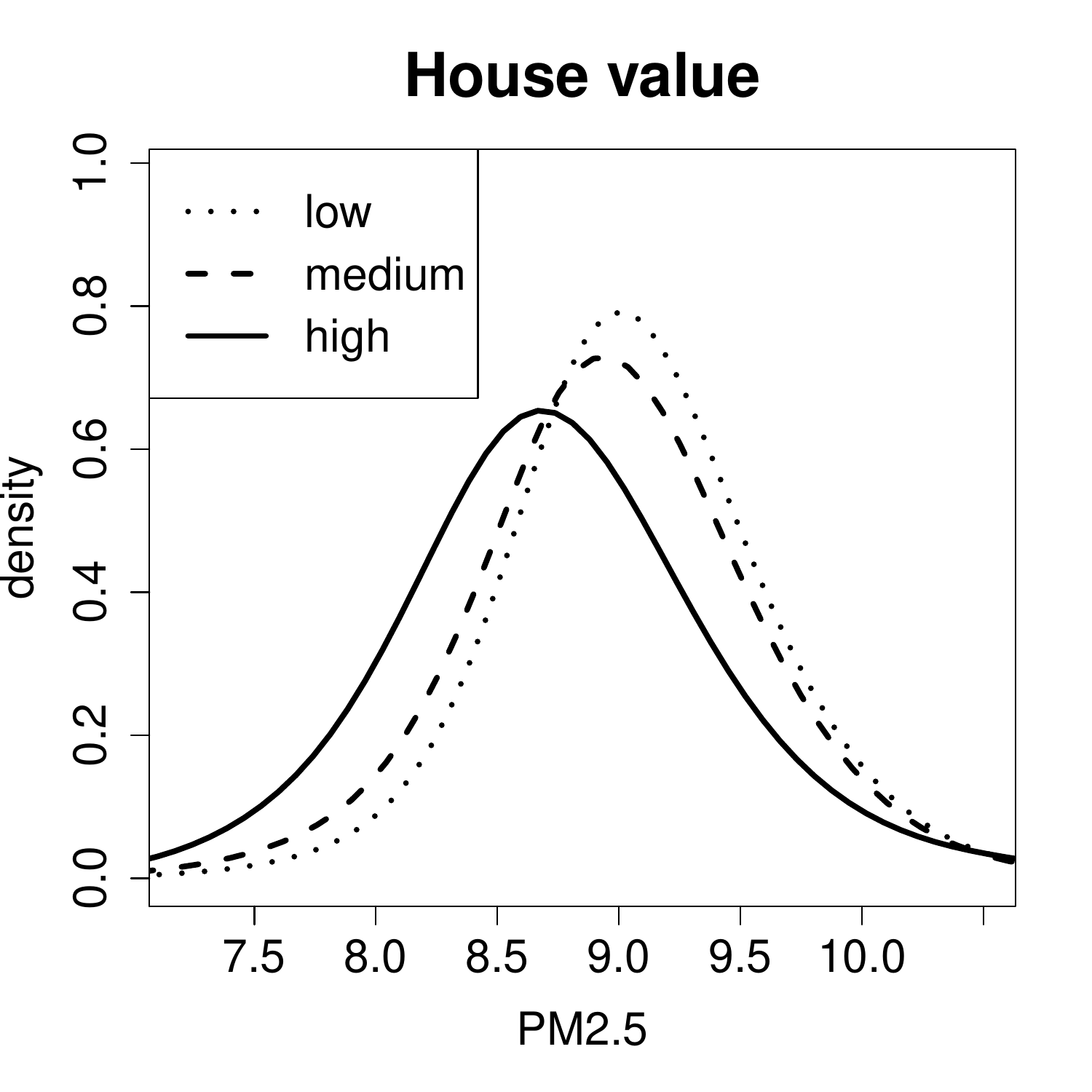}  
	\end{minipage}
	\caption{{LinCDE boosting applied to the air pollution data. We plot the estimated densities of LinCDE boosting across different levels of top influential covariates: winter relative humidity (left panel) and house value (right panel). Other covariates are fixed at the corresponding medians. Winter relative humidity is influential to the locations of the conditional densities: as the humidity goes up, the PM2.5 concentration first increases, then decreases. House value is impactful to the scales of the conditional densities: more expensive houses are associated with more variable PM2.5 exposures.}} 
	\label{fig:airPollutionLocation}
\end{figure}

In Figure~\ref{fig:airPollutionLocation}, we display how the estimated
conditional densities change with respect to winter relative humidity and house value---top influential covariates identified by LinCDE.
\begin{itemize}
	\item Winter relative humidity affects the locations of the
	conditional densities: as the humidity goes up, the PM2.5
	concentration first increases, then decreases. One
	hypothesis of the inverted U-shaped relationship is that in
	dry counties, wind speed that is inversely correlated with the humidity is a powerful factor for PM2.5; in wet counties, moisture particles that accelerate the deposition process of PM2.5 are the driving force.
	\item House value is impactful to the scales of the conditional densities: more expensive houses are associated with more variable PM2.5 exposures. We conjecture that rural areas are generally low in PM2.5 while urban areas vary. Higher house values suggest a higher proportion of urban areas and thus less homogeneous air quality. 
\end{itemize}

\begin{table}[h!]
	\centering
	\begin{tabular}{c|c|c|ccccc}
		\hline\hline
		\multirow{2}{*}{data}                & \multirow{2}{*}{method} & \multirow{2}{*}{-log-like} & \multicolumn{5}{c}{quantile loss}                     \\ \cline{4-8} 
		&                         &                            & 5\%       & 25 \%     & 50\%      & 75 \%     & 95 \%     \\ \hline
				\multirow{6}{*}{Air pollution}              & \multirow{2}{*}{QRF}   
				& 0.95   & 0.077   & 0.189    & 0.229    & 0.206    & 0.087    \\
				&& (0.02)   & (0.002)  & (0.005)  & (0.007)  & (0.006)  & (0.002)  \\ \cline{2-8} 
				& \multirow{2}{*}{DB}     
				& 1.27 & 0.099    & 0.247    & 0.300    & 0.244    & 0.093    \\
				&  & (0.04)  & (0.007)  & (0.009)  & (0.010)  & (0.008)  & (0.006)  \\ \cline{2-8} 
				& \multirow{2}{*}{LinCDE} 
				& \textbf{0.89} & 0.063    & 0.185    & 0.233    & 0.194    & 0.072   \\
				&  & (0.03) & (0.003)  & (0.006)  & (0.007)  & (0.007)  & (0.004)  \\ \hline
	\end{tabular}
	\caption{{Comparison of LinCDE boosting, QRF, and DB on the air pollution data. We display the negative log-likelihoods and quantile losses at $\{5\%, 25\%,50\%,75\%,95\%\}$ levels. Standard deviations are in the parentheses. }}
	\label{tab:realData2}
\end{table}

\vskip 0.2in

\bibliography{21-0840}

\end{document}